\documentclass{jfm}
\usepackage{xcolor}
\usepackage{amsmath}
\usepackage{soul}
\usepackage{subcaption}
\usepackage[normalem]{ulem}
\DeclareFontFamily{OT1}{pzc}{}
\DeclareFontShape{OT1}{pzc}{m}{it}{<-> s * [1.10] pzcmi7t}{}
\DeclareMathAlphabet{\mathpzc}{OT1}{pzc}{m}{it}

\begin{document}

\newtheorem{lemma}{Lemma}
\newtheorem{corollary}{Corollary}
%Commands definitions
\newcommand{\setbackgroundcolour}{\pagecolor[rgb]{0.19,0.19,.19}}  
\newcommand{\settextcolour}{\color[rgb]{0.77,0.77,0.77}}    
\newcommand{\invertbackgroundtext}{\setbackgroundcolour\settextcolour}

\newcommand{\Bo}{B\kern-0.04em o\,}
\newcommand{\Bos}{B\kern-0.04em o}
\newcommand{\We}{W\kern-0.03em e\,}
\newcommand{\Wes}{W\kern-0.03em e}
\newcommand{\Fr}{F\kern-0.04em r\,}
\newcommand{\Frs}{F\kern-0.04em r}
\newcommand{\Dr}{D\kern-0.04em r\,}
\newcommand{\Drs}{D\kern-0.04em r}

%Command execution. 
%If this line is commented, then the appearance remains as usual.
% \invertbackgroundtext

\shorttitle{Capillary-scale solid rebounds} %for header on odd pages
\shortauthor{Galeano-Rios, Cimpeanu, Bauman,  MacEwen, Milewski and Harris} %for header on even pages

\title{Capillary-scale solid rebounds: experiments, modelling and simulations}

\author
 {
 Carlos A. Galeano-Rios\aff{1,2}
  \corresp{\email{C.A.GaleanoRios@bham.ac.uk}},
 Radu Cimpeanu\aff{3,4,5},
   Isabelle A. Bauman\aff{6},
  Annika MacEwen\aff{6},
  Paul A. Milewski\aff{2}
  \and 
  Daniel M. Harris\aff{6}
  }

\affiliation
{
\aff{1}
School of Mathematics, University of Birmingham, Birmingham B15 2TT, United Kingdom
\aff{2}
Department of Mathematical Sciences, University of Bath, Bath BA2 7AY, UK
\aff{3}
Mathematics Institute, University of Warwick, Coventry CV4 7AL, United Kingdom
\aff{4}
Mathematical Institute, University of Oxford, Oxford OX2 6GG, United Kingdom
\aff{5}
Department of Mathematics, Imperial College London, London SW7 2AZ, United Kingdom
\aff{6}
School of Engineering, Brown University, Providence, Rhode Island 02912, USA
}

\maketitle

\begin{abstract}
A millimetre-size superhydrophobic sphere impacting on the free surface of a quiescent bath can be propelled back into the air by capillary effects and dynamic fluid forces, whilst transferring part of its energy to the fluid. We report the findings of a thorough investigation of this phenomenon, involving different approaches. Over the range from minimum impact velocities required to produce rebounds to impact velocities that cause the sinking of the solid sphere, we focus on the dependence of the coefficient of restitution, contact time and maximum surface deflection on the different physical parameters of the problem. Experiments, simulations and asymptotic analysis reveal trends in the rebound metrics, uncover new phenomena at both ends of the Weber number spectrum, and collapse the data. Direct numerical simulations using a pseudo-solid sphere successfully reproduce experimental data whilst also 
providing insight into flow 
quantities that are challenging to determine from experiments. A model based on matching the motion of a perfectly hydrophobic impactor to a linearised fluid free surface is validated against direct numerical simulations and used in the low Weber number regime. The hierarchical and cross-validated models in this study allow us to explore the entirety of our target parameter space within a challenging multi-scale system.
\end{abstract}

\section{Introduction}
Free-surface impacts have been the subject of rigorous scientific study since the pioneering work of \cite{Worthington1882,Worthington1897}. Interest in this area of study was fuelled by military and engineering applications related to alighting of aeroplanes on water and water entry of projectiles. Consequently, a substantial amount of effort has been devoted to the study of the high-Weber-number limit \citep{Karman1929,Richardson1948,HowisonEtAl1991,HowisonEtAl2002}, for which capillary effects can be safely disregarded. Moreover, several advances in these inertia-dominated regimes followed the introduction of the \emph{Wagner model} \citep{Wagner1932}, which describes the early stages of impact of a blunt body onto the free surface of a bath of incompressible, ideal fluid. 

Studies covering moderate Weber number regimes have focused on cavity formation and cavity pinch-off upon surface penetration of projectiles \citep{DuclauxEtAl2007,AristoffAndBush2009, TruscottEtAl2014}, jet formation at the initial stages of impact \citep{ThoroddsenEtAl2004} and forces in the early stages of impact \citep{MoghisiAndSquire1981}. More recently, the study of regimes for which the impact is dominated by capillary effects has been motivated by biological and biomimicry applications \citep{BushAndHu2006,HuEtAl2010,KohEtAl2015}. In these cases, impacts that do not break through the surface are particularly relevant to the study of water-walking mechanisms \citep{YangEtAl2016}.  Inspired by water-walking insects, numerous biomimetic robots have been proposed for use in autonomous environmental exploration and monitoring \citep{BushAndHu2006,HuEtAl2010,yuan2012bio,zhao2012superhydrophobicity,KohEtAl2015,YangEtAl2016,chen2018controllable,kwak2018locomotion}. Dynamic particle motion with capillary effects is also fundamental to a number of industrial processes including self-assembly of particles at interfaces \citep{whitesides2002beyond,whitesides2002self}, wet scrubbing and deposition for removal of particulates from gasses \citep{jaworek2006wet,wang2015behavior}, mineral flotation for material processing \citep{ueda2010water,liu2016critical}, and particle deposition techniques for rapid manufacturing \citep{haley2019modelling}.

\cite{VellaAndMetcalfe2007} addressed these capillary-dominated impacts and described conditions for the sinking of a cylinder in a two-dimensional fluid. \cite{LeeAndKim2008} considered the axisymmetric case of a superhydrophobic sphere impacting a fluid interface and they developed scaling laws to predict the transitions between the regimes in which the impactors stick to, bounce off and penetrate through the surface. In the same work, they presented a mathematical model which that can capture the initial and final stages of the rebound of a superhydrophobic sphere, though it was not possible to use this model to capture the transition between these two stages. Furthermore, only limited experimental data was provided beyond a regime diagram, rendering comprehensive comparison with more advanced dynamical models inviable.

Since the work of Lee \& Kim, there have been other follow-up works on the topic such as \cite{wang2015behavior,ji2017numerical,GaleanoRiosEtAl2017}, including a 2018 study by \cite{chen2018entrapping}, which extended the bouncing and penetration criteria developed by Lee \& Kim to include the wettability of the particle. \cite{GaleanoRiosEtAl2017} introduced the \emph{kinematic match} (KM) formulation of the impact problem, which they used to capture all stages of impact and rebound of a non-wetting sphere onto the free surface of a bath. Their impact model is based on the linearisation of the free-surface equations and is free of any form of fitting parameters.  In the mentioned article and in \cite{GaleanoRiosEtAl2019}, the method is also used to model sub-millimetre diameter droplets that bounce repeatedly on the free surface of a vibrating bath yielding remarkably good agreement with experimental results.

From a numerical standpoint, the study of impact problems is a highly challenging endeavour due to the multi-scale nature of the events in both time and space. In a recent review, \cite{josserand2016drop} provide a comprehensive discussion into the richness of even the most fundamental of questions.
In both low- and high-speed contexts, sub-micron level details may be pertinent to the dynamics of systems which are centimetre-sized or more. Rapidly changing interfacial locations, which may even result in topological transitions (coalescence, secondary jet formation and splashing), require carefully designed algorithms capable of capturing such changes in an accurate and stable manner. Furthermore, the effect of the ambient gas is non-negligible in many such cases if the full dynamics is to be successfully captured for both qualitative and quantitative assessment. 

Over the past two decades, improvements at the algorithmic level, as well as increases in computing power (parallelisation capabilities in particular), have resulted in a number of success stories in this area. These improvements have lead to insight into the key metrics involved in drop impact onto solid surfaces \citep[such as film thickness, maximum spread and underlying structures][]{eggers2010drop, wildeman2016spreading, philippi2016drop}, to access into new regimes, and have even guided and complemented data retrieval for new experimental techniques \citep{Visser1}. While some of the difficulties, e.g. those related to contact line dynamics, are avoided in liquid-liquid impact scenarios, many of the inherent challenges remain the same. The deformation of the impactor and identification of thresholds for splash jet formation has been the subject of much attention \citep{Josserand03,josserand2016jfm}, while the dynamics inside the impinging liquids gives rise to exciting structure, as indicated by initial numerical investigations \citep{Thoraval12}. Finally, employing direct numerical simulations has recently allowed comparisons to Wagner theory in suitable regimes \citep{cimpeanu2018early, moore2020boundary}, providing a strong toolkit for establishing predictive capabilities of analytical formulations and bridging the gap towards direct experimental comparisons and applications.

One highly relevant detail in the present context is the nature of the sphere surface. The superhydrophobic coating is desirable in terms of producing solid rebound behaviour over the largest parameter space.
The ``converse'' problem of liquid droplets impacting superhydrophobic surfaces has been widely studied from the fundamental perspective in order to understand both bouncing and splashing-related effects \citep{richard2000bouncing, reyssat2006bouncing, bartolo2006bouncing, BianceEtAl2006}. Many of these studies on droplet impacts have been motivated by elucidating the underlying physics and guiding designs in applications pertaining to self-cleaning \citep{liu2014pancake}, structure-induced patterning \citep{schutzius2014morphing,lee2010drop} and even aerodynamic (icing prevention) contexts \citep{yeong2014drop,peng2018all}. In the context at hand, the superhydrophobic coating around the impacting sphere is used to ensure a large contact angle and low contact-angle hysteresis. Our assumption of perfect hydrophobicity also has the added advantage of (comparatively) simplified contact line dynamics for the associated theoretical investigations.

Studies in the aforementioned scenarios raise valid questions for the case of solid spheres rebounding off the free surface of a bath, considered here. For instance, in \citet{BianceEtAl2006} it has been shown that for droplets bouncing off of a solid, the coefficient of restitution is a non-monotonic function of the Weber number. Specifically, it increases with Weber in the low Weber number regime, and it reverses its behaviour in the moderate to high Weber number range. It is not known whether this behaviour is reproduced in the converse system.
Another question is whether the criterion for bouncing off the surface versus oscillating without detaching from it, and the criterion for sinking that were presented by \citet{LeeAndKim2008} holds for densities and Bond numbers outside the range they reported.
Furthermore, in some related problems  \citep[e.g.][]{GiletAndBush2009JFM}, it has been shown that scaling based on a linear spring model is sufficient to rationalise a collapse of the relevant rebound metrics for a wide range of rebounds. The question of whether a similar collapse, on the basis of a linear model, is possible in the system considered herein is of interest.

In the following sections, we address these and other related questions. We present a combined experimental, numerical, and theoretical investigation focusing on the dependence of contact time, maximum penetration depth and coefficient of restitution on the different impact parameters. We show that direct numerical simulations (DNS) of pseudo-solid spheres impacting a fluid bath are able to accurately capture all features observed in our experimental studies and
act as a bridge between experiments and modelling efforts.
In view of the above, we show that the kinematic match method produces results that are in full agreement with data obtained via DNS for impacts in which the modelling assumptions remain valid. Furthermore, we use the kinematic match to explore the low Weber number limits in which we identify impact velocities that maximise the fraction of the initial energy that is recovered by the impactor. Finally we use asymptotic analysis to produce a non-linear spring model, which we use to rationalise and interpret the maximum penetration depth and contact time data amalgamated from the three approaches.

\section{Experimental Methods}\label{sec.ExperimentalSetUP}

\subsection{Experimental Setup}
\begin{figure}
    \centering
    \includegraphics[width=1\textwidth]{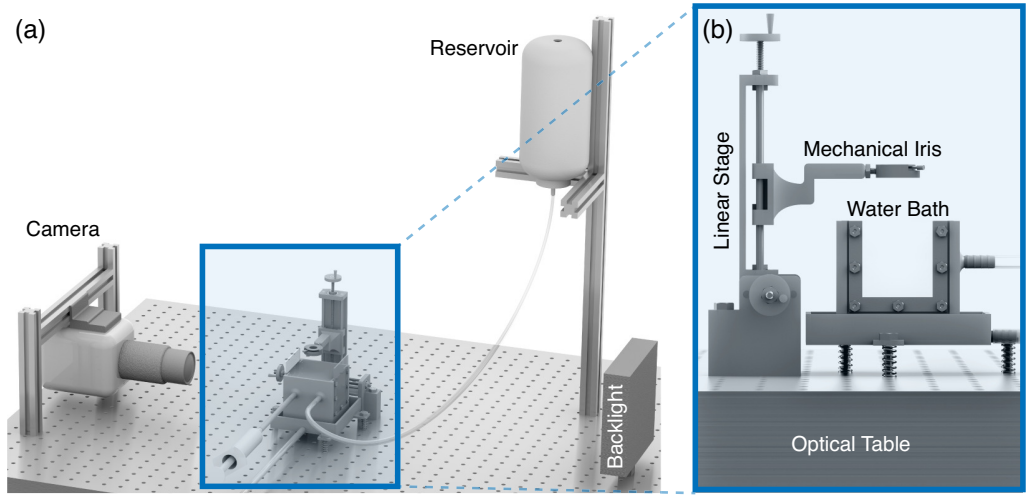}
    \caption{(a) Rendering of the complete experimental setup, including the high-speed camera, water bath, linear stages, water reservoir, and backlight. (b) Closer view of the water bath and linear-stage system, viewed from the perspective of the camera.}
    \label{fig:setup}
\end{figure}

The experimental setup is depicted in Figure \ref{fig:setup}.  In each trial, spheres were dropped from a mechanical iris that could be height-adjusted by a system of custom linear stages.  A two-stage system was custom designed and fabricated to allow for three degrees of freedom for the iris position above the water bath (vertical and horizontal stages provide two degrees of freedom, and the threaded rod that held the iris provided a third). 
The sphere was dropped approximately 2 cm from the panel closer to the camera, 3.5 cm from each of the side walls of the bath, 5 cm from the back wall panel, and 7 cm from the bottom of the bath. These distances were chosen such that the boundaries of the bath would not interfere with the dynamics near impact. This was confirmed in experiment by increasing the distance of impact from the front panel until the rebound metrics were not affected in a statistically significant manner. In many cases, the influence of the reflected waves during impact was also that the sphere did not rebound vertically (and moved in or out of the narrow focal plane).

The water bath itself was designed to be easily filled, flushed, and drained to minimise contamination of the free surface \citep{kou2008method}.  There are two tubes connected directly to the bath; one that connects to a water reservoir filled with deionized water, and the other to a syringe for fine water-height adjustments. Overflow from flushing the bath is caught by a lip at the base of the bath and then drained by gravity through an outlet to a waste container beneath the optical table. The 3D-printed bath can be precisely levelled using three levelling springs and is mounted directly to an optical table.  The vibration isolation provided by the optical table ensured minimal disturbances on the free surface prior to impact, which could interfere with the results. 
The bath panels were laser cut from clear polystyrene, a material with a contact angle of approximately $90^{\circ}$ \citep{ellison1954wettability}, such that a pronounced meniscus would not form and interfere with imaging at impact. The panels were laser cut to have a line of etched dots (0.2 mm in diameter, spaced 10 mm apart so as to not interfere with the visualisation at the impact location) at the desired water level as a visual indicator to ensure that the water level remained consistent between trials. 

A Phantom Micro LC311 camera with a Nikon Micro 200 mm lens was used for the video capture.  The camera was mounted on a system of linear stages with three degrees of freedom to allow for fine positioning. In particular, the camera position could be finely adjusted along its axis such that the focus was fixed at the minimal working distance for all experiments in order to ensure consistent (and maximal) image spatial resolution. The lens was set at its minimal aperture (f/32), which allowed for the focus to be satisfactory when the sphere was both above and below the water surface.  The camera captured images at 10,000 frames per second at an exposure time of 99.6 microseconds. The window size of images was approximately 5 mm by 10 mm, which was captured in 256 by 512 pixels. The images were uniformly back-lit with a 100 mm by 100 mm Phylox LED light panel. Sample image data is shown in Figure \ref{fig:traj}(a-c). 

In order to maintain a large equilibrium contact angle ($\theta_c$) of approximately $160^{\circ}$ with very low contact angle hysteresis \citep{weisensee2017droplet}, the spheres were coated with a commercially-available 2-part (henceforth referred to as `Step 1' and `Step 2') superhydrophobic spray (NeverWet). Spheres that are not coated or have a damaged coating have significantly reduced propensity to bounce under most impact conditions. The protocol that allowed for the application of an uniform coating is described in detail in what follows.  Approximately 10 spheres were initially distributed in a clean Petri dish, arranged so that none of the spheres were in contact with each other or the side walls of the container.  Then two rapid sprays of Step 1 were applied to the spheres from approximately 30 cm away.  The spheres were then left to sit for 1 minute.  At this point, the spheres were redistributed on the petri dish using a small toothpick.  This procedure for Step 1 was then repeated 5 times.  The spheres were then left in a fume hood for 15 minutes for the Step 1 coating to dry.  They were then moved to a clean Petri dish and left in the fume hood for at least another 15 minutes.  Following this procedure, the Step 2 coating was applied.  Ten rapid sprays of Step 2 were applied in succession.  The spheres were redistributed between each spray without external contact by gently tilting the Petri dish and allowing them to roll to new positions and orientations.  They were then left to sit for approximately 5 minutes, and 10 more sprays were applied in the same manner.  The spheres were then left to dry in the fume hood for at least 12 hours before being used in any experiment.  This protocol allowed for the millimetric spheres to be coated uniformly, which proved an essential step for obtaining repeatable results.  Note that applying too much coating in any given step led to the spheres become overly saturated, and the resulting fluid meniscus bridging the sphere to the base of the Petri dish would dry and leave the surface with visible defects.  Any such spheres were discarded.

\subsection{Experimental Parameters}

All spheres tested were denser than water (see table \ref{tab:parameters}), with density ratios $\Dr = \rho_s/\rho$ ranging from 1.2 to 3.2. These ratios were obtained by using nylon ($\Dr$ = 1.2), polytetrafluoroethylene (PTFE) ($\Dr$ = 2.2), and ceramic spheres ($\Dr$ = 3.2), each coated with the superhydrophobic NeverWet spray. Spheres of radius 0.83 mm were tested for all three densities. Three sizes of nylon spheres were also tested, with radii 0.83 mm, 1.24 mm, and 1.64 mm.  Release heights were varied to achieve impact velocities from 30 to 110 cm/s.  All values and non-dimensional parameters associated with the experiments are listed in Table \ref{tab:parameters}.  Note that the variable subscript `$s$' delineates parameters that correspond to the sphere properties.

\begin{table}
\centering
\caption{Relevant parameters and their characteristic values in our experimental study.}

\begin{tabular}{lccc}
\textbf{Parameter}       
& \textbf{Symbol} 
& \textbf{Definition} 
& \textbf{Value(CGS)}
\\
Impact Speed          
& $V_0$          
&
& $30-110\,$cm/s
\\
Sphere Radius  
& $R_s$
&
& $0.083-0.164\,$cm
\\
Sphere Density      
& $\rho_s$ 
&
& $1.2-3.2\,$g/cm$^3$
\\
Water Density         
& $\rho$        
& 
& $1.00\,$g/cm$^3$
\\
Equilibrium Contact Angle
& $\theta_c$      
&
& $160^{\circ}$ 
\\
Surface Tension 
& $\sigma$        
&                               
& $72\,$dynes/cm
\\
Kinematic Viscosity of Water 
& $\nu$ 
&
& $0.978\,$cSt
\\
Gravity Constant 
& $g$
& 
& $980\,$cm/s$^2$
\\
Capillary Length 
& $l_\sigma$
& $\sqrt{\sigma/(\rho g)}$
& $0.271\,$cm
\\
Capillary Time 
& $t_\sigma$
& $\sqrt{\rho R_s^3/\sigma}$
& $2.82\,$ms
\\
Reynolds Number 
& \Rey
& $V_0 R_s/\nu$
& $250-1225$
\\
Weber Number
& $\We$     
& $\rho V_0^2 R_s/\sigma$ 
& $1.0-14.0$
\\
Bond Number 
& $\Bo$
& $\rho g R_s^2/\sigma$
& $0.09-0.37$
\\   
Froude Number
& $\Fr$ 
& $V_0^2/(g\,R_s)$  
& $5.60-148.76$
\\
Density Ratio        
& $\Dr$          
& $\rho_s/\rho$  
& $1.2-3.2$
\end{tabular}
\label{tab:parameters}
\end{table}

\subsection{Experimental Procedure}\label{Section:ExpProc}

Spheres were released from the mechanical iris at a range of heights, beginning at approximately one centimetre above the water bath, and gradually increased until the spheres sunk upon impact. 
Five spheres for each radius and density combination were tested at each height, with three trials for each sphere, for a total of fifteen trials per height. The water bath was flushed each time a new sphere was used (every three trials or approximately every 5 minutes). If a sphere showed indications of a damaged coating or was noticeably non-spherical due to uneven coating, the sphere was discarded immediately and any associated trajectories were also disregarded.

High-speed video footage of each bounce were recorded and directly imported into MATLAB. Custom image-processing software in MATLAB was used to determine the vertical trajectory of the sphere as described in what follows.
First, the video data was processed using a built-in Canny edge detection in MATLAB.  The top (highest) and bottom (lowest) edges in the image were then recorded.  During the initial free fall, the top edge corresponded to the top of the sphere, and the bottom edge corresponded to the water surface. For the cases where the sphere passes entirely below the still air-water interface level, the top edge in the frame became the water's surface, and the bottom edge corresponds to the bottom of the sphere. When the sphere then resurfaced and bounced above the interface, the top edge corresponded again to the top of the sphere and the bottom edge is on the disturbed air-water interface. Once the sphere landed and stoped oscillating on the surface of the water, the top and bottom edges correspond to the top and bottom of the sphere.  In summary, the top of the sphere was tracked during the initial free fall, the bottom was tracked when the sphere is submerged, and top of the sphere was tracked from the rebound onward. The equilibrium resting state after the bounce was used to define the difference between the top and bottom trajectories (i.e. the sphere diameter in pixels). This value was then subtracted from all points in the trajectory that corresponded to the top of the sphere, thus generating a smooth curve representing the trajectory of the bottom point of the sphere, with $z=0$ corresponding to the height of the undisturbed air-water interface. The final trajectories were then used to generate our variables of interest in the present work, including impact speed, $V_0$; penetration depth $\delta$, contact time, $t_c$; and coefficient of restitution, $\alpha$. Sample trajectories are shown in Figure \ref{fig:traj}(d). The complete set of the experimental trajectories is provided in appendix \ref{app:trajects}.

\begin{figure}
    \centering
    \includegraphics[width=1\textwidth]{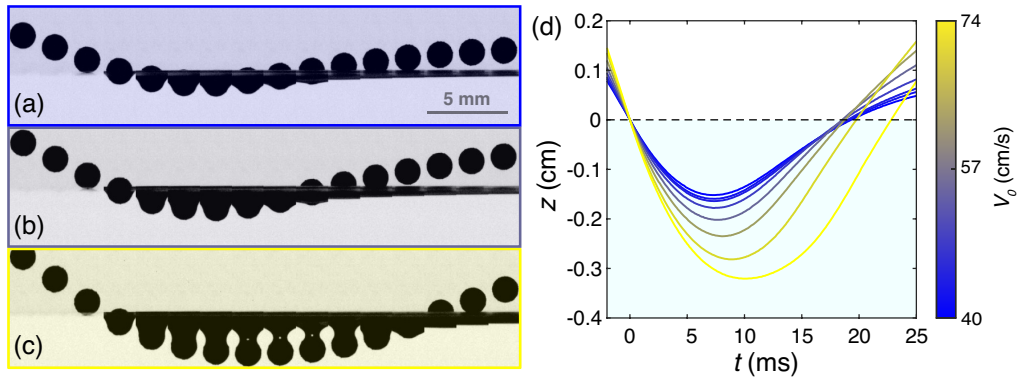}
    \caption{Rebound data for superhydrophobic spheres with radius $R_s=0.083$ cm and density $\rho_s=2.2$ g/cm$^3$.  (a-c) Sequence of images with different impact speeds $V_0$: (a) $40.2\pm 0.7$ cm/s, (b) $53.7\pm 0.7$ cm/s, (c) $73.6\pm 0.6$ cm/s. Images are evenly spaced in time by 2 ms, corresponding to 20 frames. (d) Trajectories of the bottom of the spheres (relative to the undisturbed free surface height) measured for 8 different impact velocities.  Shown are the average trajectories over all trials at a fixed release height with outliers removed, as described in the text. Videos corresponding to the trials shown in (a)-(c) are available as supplementary material.}
    \label{fig:traj}
\end{figure}

There are several parameters of interest in our study, which we define in what follows. The maximum penetration depth, $\delta$, of a bounce is defined as the position of the bottom of the sphere at the lowest point in the trajectory (computed relative to the undisturbed interface height).  In order to determine the contact time, $t_c$, and coefficient of restitution, $\alpha$, a parabola was fit using a least-squares method to the incoming and outgoing trajectories, separately, with at least 10 data points prior to impact and at least 20 data points following rebound.  The analytical form of the parabolic fit was then used to extrapolate the time at which the sphere crosses the still air-water interface height (which corresponds to a root of the parabolic function).  The derivative of the parabolic fit function was then computed analytically and its value evaluated at these times in order to calculate the impact speed, $V_0$, and exit speed, $V_e$.  

In the present work, contact time, $t_c$, is defined as the time duration from which the bottom of the sphere crosses the still air-water interface to the time the bottom of the sphere next reaches that height.  Note that, due to the nature of visualisation setup, it was impossible to determine precisely when the spheres lost physical contact with the fluid; however, this always occurred before the sphere returned to the level of the free surface.  Each bounce was also characterised by its coefficient of restitution, $\alpha$, which is defined here as the negative of the normal exit velocity, $V_e$, divided by the normal impact velocity, $V_0$. This parameter ranges between 0 and 1, and is related to the momentum transfer to the fluid bath.  Outliers within each data set (generally due to accidental damage to the sphere coating) were identified using a modified 0.02-level two-sided Thomson T-test to determine a suitable rejection region of $\alpha$ \citep{wackerly2014mathematical}.  In each set of fifteen trials (five spheres, with three bounces each), this test identified at most 2 outliers. 

\section{Direct numerical simulations}\label{sec.ModellingAndSimulations}

In the present section we describe the construction of a computational framework capable of resolving the complex bouncing dynamics in this multi-scale context. Our implementation is built as an extension of the well-known, open-source, volume-of-fluid package $\mathpzc{Gerris}$ (see \cite{Popinet03, Popinet09}), which has been proven to be one of the most successful tools in multi-phase computational fluid dynamics studies in recent years. As described in the previous section, the physical process we are aiming to elucidate is highly non-trivial due, in no small part, to significant nonlinear effects and liquid surface deformations.

\begin{figure}
    \centering
    \includegraphics[width = 13.0cm]{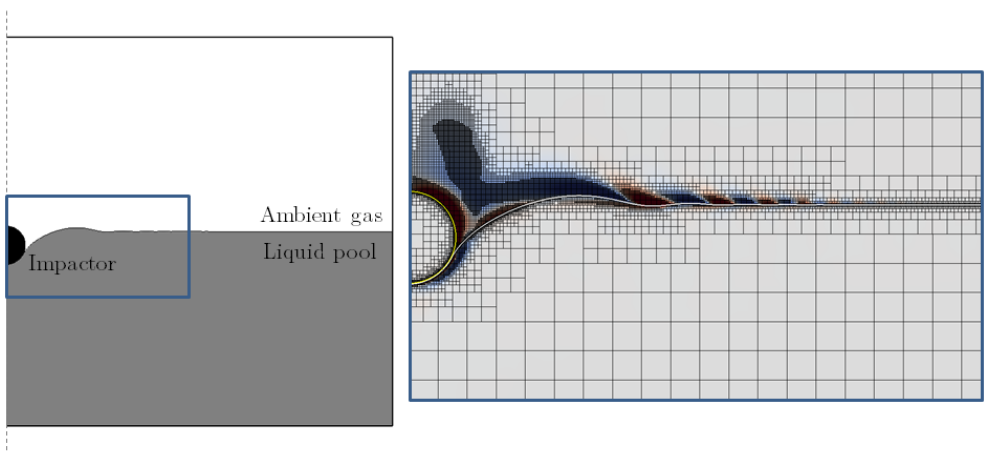}
    \caption{Axisymmetric simulation domain of size $20R_s \times 20R_s$, with $R_s$ denoting the impactor radius. The inset illustrates the adaptive mesh refinement strategy, with changes in vorticity (shown as background colour) and interfacial locations used as primary criteria. A video corresponding to this particular case (expanded on in Figures~\ref{fig:DNSBehaviour}-\ref{fig:DNSDeformation} as well) is available as supplementary material.}
    \label{fig:DNSDomain}
\end{figure}

Before outlining the numerical setup as a whole, a particularly meaningful detail relates to our treatment of the coated spheres. The specific surface features on the sphere present in the experiment pose a formidable challenge and require much finer resolution than a full DNS framework is capable of resolving, even with the very high end of modern day computing resources. Resolution of these fine scale features would arguably also require additional physics and the formulation of a hybrid model containing sub-continuum effects (see, e.g., \cite{SprittlesImpact2020}),
which are beyond the scope of the present work. Furthermore, the quadtree/octree multi-grid setting in $\mathpzc{Gerris}$ makes true fluid-structure interaction very difficult to embed accurately. Therefore a simplification was adopted instead: the solid spheres are computationally modelled as highly viscous liquid drops ($250$ times the viscosity of water at room temperature) with very large surface tension coefficients ($20$ times the air-water value). These simulations are implemented using two distinct height functions (level set definitons of interfaces) to avoid coalescence, and were found to represent a viable compromise from both numerical stiffness and physical behaviour perspectives. We have studied this approximation extensively (see also Figure~\ref{fig:DNSDeformation}) and have pushed the setup as close to a true solid as possible, whilst retaining reasonable run times given that the disparity in physical properties causes significant slowdown in terms of convergence. A quantitative study on the deformation of this ``pseudo-solid'' has revealed deviations of less than $5\%$ in even the most challenging of scenarios. As is to be anticipated, a flattening of the sphere occurs on impact in the vertical direction, with mass conservation thus leading to an elongation of the impactor at the equator into an oblate ellipsoidal shape. Given the large imposed surface tension coefficient, the pseudo-solid relaxes to a spherical shape as soon as the impactor has left the pool surface. Whilst this effect is consistently observed across all DNS realisations, we have made significant efforts in ensuring that variations in the mentioned pseudo-solid geometrical parameters no longer affect the dynamics at the prescribed resolution levels,
and are thus a viable platform for understanding mechanistic features of the studied system. Apart from the observed qualitative behaviour and comprehensive validation studies performed, this approach is also confirmed quantitatively versus another model and experimental data in Section~\ref{sec:Results}. 
We thus underline the rather remarkable feature that the behaviour we describe appears to be independent of the microscopic details of contact with the superhydrophobic surface. This pseudo-solid approach however is not a good model for experiments on spheres with smaller contact angles, which exhibit notably different behaviour in the experiment.  This experimentally observed sensitivity to the wetting behaviour suggests that a contact line exists during impact, and that a continuous air layer is not maintained during impact as is the case for rebounding droplets \citep{CouderEtAl2005PRL}.

Our setup for investigating this challenging multi-fluid system is shown in Figure~\ref{fig:DNSDomain}. Together with second-order accuracy in both time and space, the adaptive mesh refinement and parallelisation capabilities make a difficult setting tractable. We assume axisymmetry and build a domain sufficiently large to avoid reflections and artifacts from the side walls. This constraint sets the maximum length scale captured, which is fixed at $20$ impactor radii (typically $R_s \approx 1$ mm) for all realisations that follow. The smallest length scale to capture is arguably the variation in physical quantities across the gas film between the impacting body and the liquid pool, which in the past has been reported at $\mathcal{O}(10)\ \mu$m for droplet impacts \citep{CouderEtAl2005PRL}. This translates to at least three orders of magnitude being required, thus leading to a maximum grid refinement of level $12$ (translating to $2^{12}$ cells per dimension), with the minimum cell size spanning approximately $4\ \mu$m. This means that there are at least $200$ grid cells per impactor radius and that quantities across the gas film are allowed at least $3-4$ cells to manifest any meaningful variation. 

The mesh adaptivity criteria used are stringent and based on changes in the magnitude of velocity components, vorticity and interfacial locations in the domain. The strategy was developed to ensure sufficient accuracy, as well as an accessible run time for extensive parameter studies for future comparisons. This resolution translates to $\mathcal{O}(10^5)$ cells for the most challenging settings and a typical runtime of $500$ CPU hours per run, with local high performance computing facilities equipped to handle realisations on $1-16$ CPUs. We have conducted extensive validation studies, using metrics related to interfacial shapes (in particular; tracking maximum depth, gas film thickness and impactor radii) to establish convergence before any direct comparisons with our other approaches.

\begin{figure}
    \centering
    \includegraphics[height = 4.0cm]{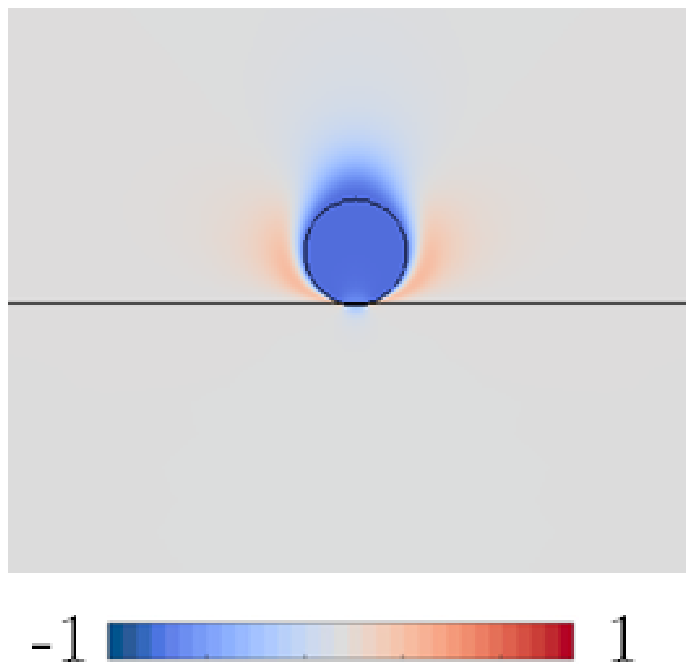}
    \includegraphics[height = 4.0cm]{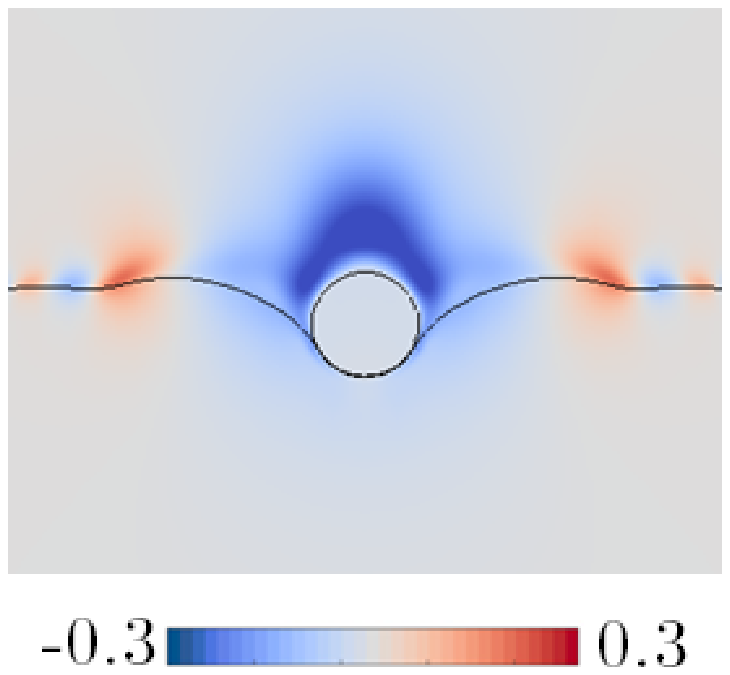}
    \includegraphics[height = 4.0cm]{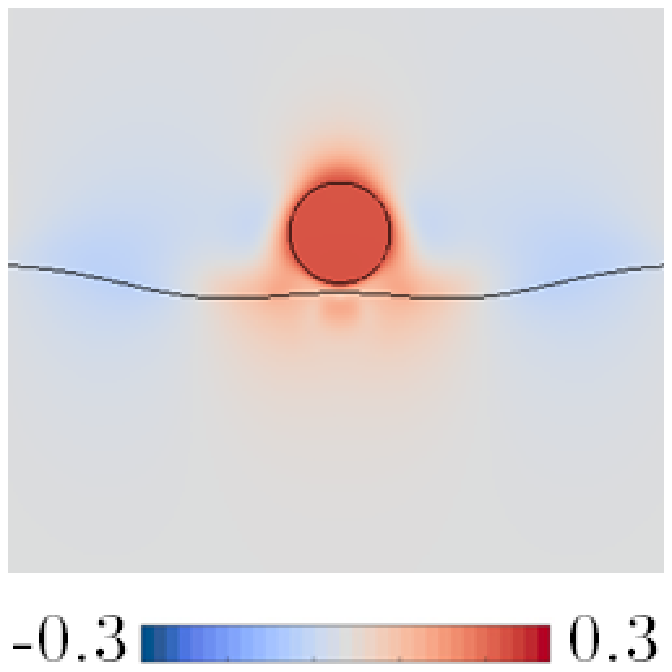}    
    \caption{Typical bouncing behaviour as observed in the direct numerical simulations for a case described by sphere radius $R_s=0.083$ cm, density $\rho_s=2.2$ g/cm$^3$ and impact speed $V_0 = 56.67$ cm/s. The background colour represents the dimensionless vertical velocity field, with the relevant interfaces also highlighted in black. The three illustrated instances represent, in dimensionless time units: (left) $t \approx 1.0$, as the impactor touches the surface, (middle) $t \approx 4.5$, as the impactor reaches its maximum depth and (right) $t \approx 10.0$, as the impactor leaves the surface for its first bounce.}
    \label{fig:DNSBehaviour}
\end{figure}

\begin{figure}
    \centering
    \includegraphics[width = 3.0cm]{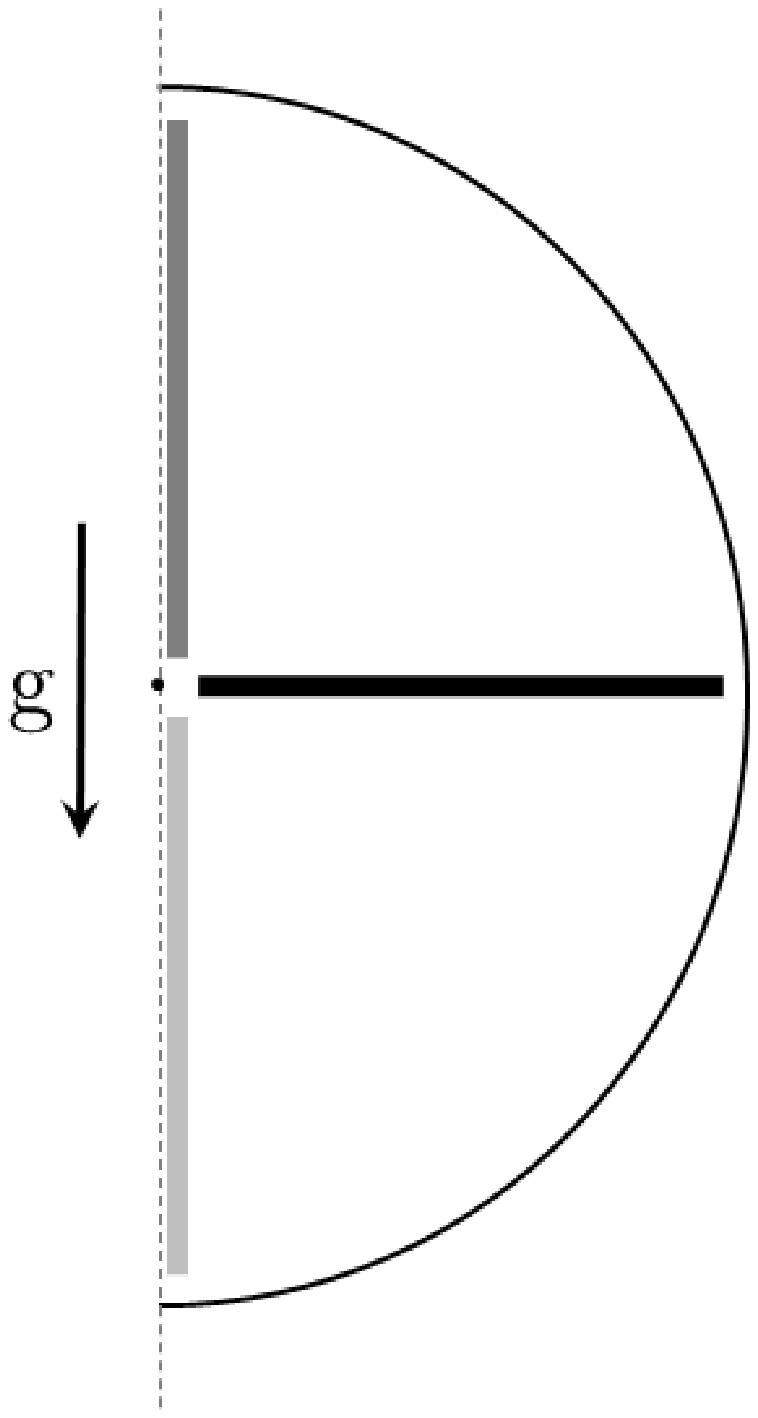}
    \includegraphics[width = 7.0cm]{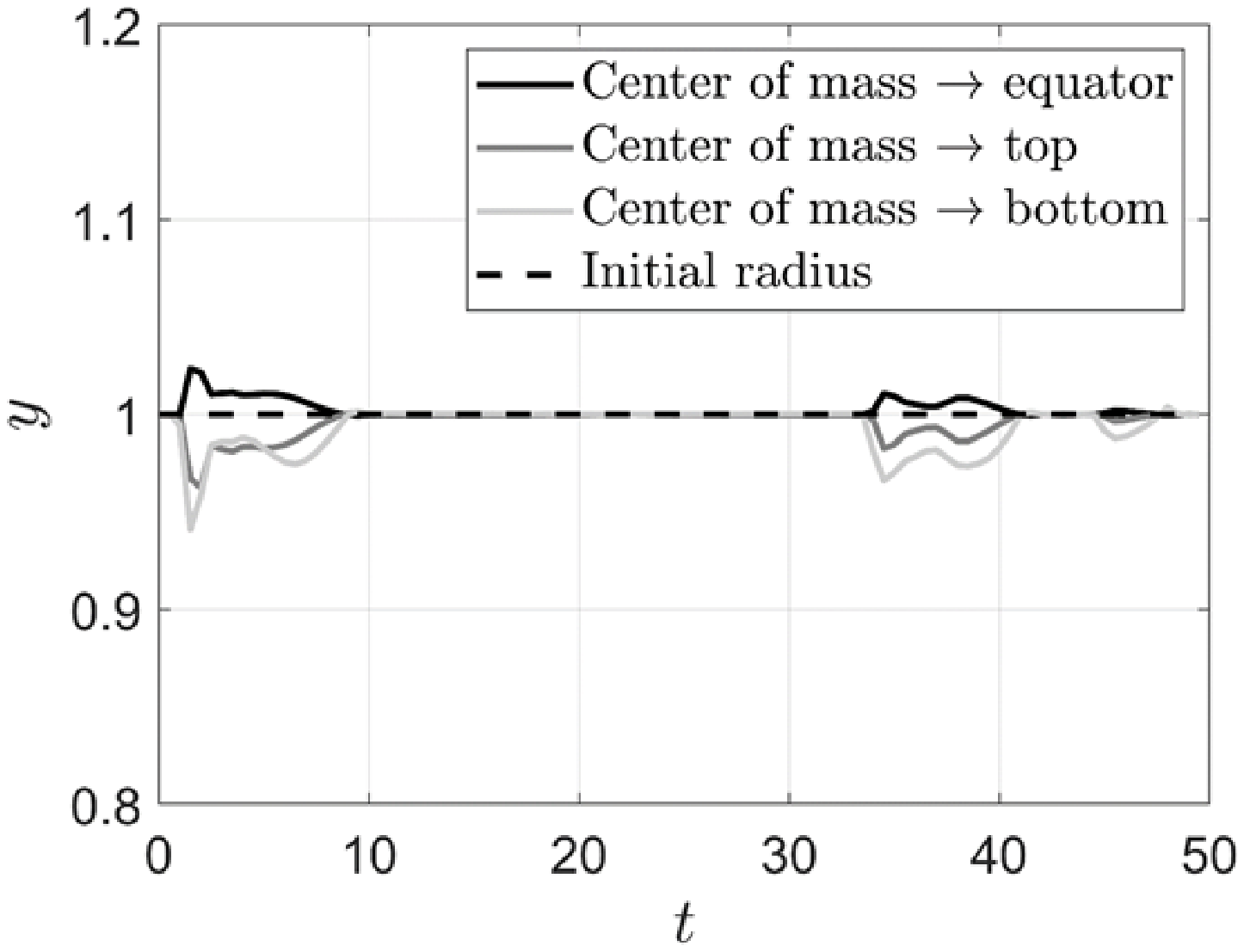}
    \caption{Pseudo-solid deformation study for a representative test case described by an impacting sphere of radius $R_s = 0.83\,$mm, $\rho_s = 2.2\,$g$/$cm$^3$ and $V_0 = 56.67\,$cm$/$s. (Left) Sketch of measured segments as distances from the centre of mass of the impactor to its relevant extremities. (Right) Segment size evolution as a function of dimensionless time, compared to a reference undeformed $y = 1$ radius, indicated here with a dashed line.} 
    \label{fig:DNSDeformation}
\end{figure}

Using a non-dimensionalisation based on the sphere radius and initial impact velocity as reference scales, with the arising dimensionless groups presented further on as equation~\eqref{eq:dimensionless}, we consider $50$ time units (the equivalent of $\mathcal{O}(0.1)$ s), which has proven sufficient to capture $2-3$ rebounds for each parameter setting. The example expanded upon in the present section underlying each of Figures~\ref{fig:DNSDomain}-\ref{fig:DNSDeformation} is described by sphere radius $R_s=0.083$ cm, density $\rho_s=2.2$ g/cm$^3$ and impact speed $V_0 = 56.67$ cm/s and represents a typical test scenario in this context, as illustrated in Figure~\ref{fig:DNSBehaviour}. 
Part of its evolution (concentrating on the first bounce) is also presented as video supplementary material.

The developed computational framework is used to study regimes and uncover a host of details at length- and timescales beyond the reach of other approaches. The inclusion of the effect of the ambient gas and fully nonlinear formulation provides a comprehensive resolution of the studied dynamics, while the ability to inspect the flow field in a precise manner leads to a constructive interplay with other methodologies. However such an approach, even with considerable efforts in terms of parallelisation and overall efficiency, is nevertheless extremely expensive. The resources required (computing power and ultimately time) make the usage of carefully resolved numerical simulations prohibitive for certain applications; such as many body impacts or longer time dynamics (as in the case of periodic bouncing).
In what follows we elaborate on a simpler model, which in the low Weber number regime provides an efficient alternative while also resolving the impact and the wave motion in the bath.

\section{Linearised quasi-potential fluid model}
An alternative fluid model that is considerably less computationally intensive is now described. The model forgoes the gas layer, assumes a near inviscid bath, small free-surface slopes, and hydrophobicity \emph{ab initio}. What follows is a brief summary of the method in \citet{GaleanoRiosEtAl2017}.
Consider a bath of incompressible fluid of infinite depth and unbounded lateral extension. The fluid has density $\rho$, kinematic viscosity $\nu$ and surface tension coefficient $\sigma$. Imposing axisymmetry, we introduce cylindrical coordinates $(r,\theta,z)$, with the origin at the point of first contact of the sphere with the free surface, and the $z$ axis pointing vertically upwards. We define functions $\eta(r,t)$, $\varphi(r,z,t)$ and $p_s(r,t)$ as the free surface elevation, a velocity potential and the pressure on the free surface, respectively. 
The impacting sphere has a density $\rho_s$ and at time $t=0$ is in imminent contact with the free surface whilst moving downward with speed $V_0$.  

Taking $R_s$, $V_0$ and $\rho$ as the characteristic length, velocity and density, respectively, results in the following dimensionless numbers:
\begin{equation}\label{eq:dimensionless}
\Rey = R_sV_0/\nu,\ \Fr = V_0^2/(g\,R_s),\ \We = \rho\,V_0^2\,R_s/\sigma,\ \Dr = \rho_s/\rho;
\end{equation}
i.e. Reynolds number, Froude number, Weber number and density ratio. We note that $\Fr = \We/\Bo$, where $\Bo = \rho g R_s^2/\sigma$ is the Bond number.

Defining $\phi(r,t) = \varphi(r,0,t)$ and using the linearised quasi-potential formulation for the fluid flow:
\begin{equation}\label{Laplace}
\Delta\varphi = 0,\ z\leq 0;
\end{equation}
\begin{equation}\label{KBC}
\partial_t\eta = \frac{2}{\Rey}\Delta_H \eta + \partial_z\varphi,\ z = 0;
\end{equation}
\begin{equation}\label{DBC}
\partial_t\phi = -\frac{1}{Fr}\eta+\frac{1}{We}\kappa \left[\eta\right]+\frac{2}{\Rey}\Delta_H \phi - p_s,\ z = 0;
\end{equation}
subject to 
\begin{equation}
\eta\to 0, \text{ when } r\to\infty; \ \ \ \ \varphi\to 0, |\nabla\varphi|\to 0, \text{ when } (r,z)\to\infty,
\end{equation}
where $\Delta_H = \partial_{rr} +\frac{1}{r}\partial_{r}$ and $\kappa$ is twice the mean curvature operator with the convention that convex functions have positive curvature. The system given by (\ref{Laplace}), (\ref{KBC}) and (\ref{DBC}) can be reduced to two equations defined on the free surface by the introduction of a Dirichlet-to-Neumann transform, which is denoted by $N$ and defined on a set of suitably smooth functions of the plane. It is given by the singular integral representation detailed in \citet{GaleanoRiosEtAl2017} and is such that, for any given time $t$,
\begin{equation}
N(\phi)(r,t) = \partial_z\varphi(r,z = 0,t).
\end{equation}

The free-surface evolution is thus given by:
\begin{equation}\label{KBC1}
\partial_t\eta = \frac{2}{\Rey}\Delta_H \eta + N\phi,\ z = 0;
\end{equation}
\begin{equation}\label{DBC1}
\partial_t\phi = -\frac{1}{Fr}\eta+\frac{1}{We}\kappa \left[\eta\right]+\frac{2}{\Rey}\Delta_H \phi - p_s,\ z = 0.
\end{equation}
\subsection{Motion of the sphere and the natural constraints}
Defining a contact surface, $S(t)$, on the sphere, where the surface of the fluid bath coincides with that of the solid sphere, we introduce a contact area, $A(t)$, which is the orthogonal projection of $S(t)$ onto the $(r,\theta)$-plane. Assuming that $A(t)$ is a disc of radius $r_c(t)$, we impose that $p_s= 0$ everywhere outside $A(t)$. The motion of the south pole of the sphere $h(t)$ is thus governed by 
\begin{equation}\label{Newton2nd}
\frac{d^2h}{dt^2} = -\frac{1}{Fr}+\frac{3}{4\pi\Dr }\int_{A(t)}{p_s \text{d}A(t)}.
\end{equation}
The function $p_s$ couples equations (\ref{DBC1}) and (\ref{Newton2nd}).

Equation (\ref{KBC1}), (\ref{DBC1}) and (\ref{Newton2nd}) must be solved subject to the following constraints
\begin{equation}\label{contact}
\eta(r,t) = h(t)+z_s(r), \ \ r\leq r_c(t);
\end{equation}
\begin{equation}\label{free}
\eta(r,t) < h(t)+z_s(r), \ \ r> r_c(t);
\end{equation}
where $z_s(r)$ is given by the bottom half of the sphere (whose centre is on the $r = 0$ vertical) for $r \leq R_s$, and $z_s = \infty$ otherwise. Finally, we impose that the solid is perfectly hydrophobic and therefore the contact angle is always of $\pi$, which yields the final constraint
\begin{equation}\label{tangent}
\partial_r \eta(r=r_c(t),t) = \partial_r z_s(r=r_c(t)).
\end{equation}
\subsection{The kinematic match}

The kinematic match (KM) method, presented in \citet{GaleanoRiosEtAl2017} and \citet{GaleanoRiosEtAl2019}, introduces an algorithm to solve all stages of a collision, in which the impactor does not break through the free surface. Moreover, the method predicts the evolution of the contact area, and the pressure distribution within it, whilst imposing only first principles and the natural geometric and kinematic constraints. The algorithm is built on the idea that, when one imposes a given contact area evolution, equations (\ref{KBC1})-(\ref{contact}) form a closed system. One can then iterate on the geometry of the contact area, solving the system (\ref{KBC1})-(\ref{contact}) at each iteration and assessing the iteration result by checking the remaining equations of the system, i.e. (\ref{free}) and (\ref{tangent}). The numerical implementation of the method uses an adaptive time step to satisfy a constraint on the time-step size that is necessary to capture the motion of the boundary of the pressed area. For all simulations here, we adopt the domain $D = \left\{(r,t); 0\leq r \leq 50 R_s, \  0\leq t \leq T\right\}$ and we discretise the spatial domain using a regular mesh of with nodes spaced $R_s/50$.

The domain size was chosen to prevent waves from being reflected off the boundary back toward the impact location during contact, thus ensuring that the rebound is not affected by the finite size of the numerical domain. To find an adequate domain size we ran a preliminary KM simulation to find the contact time and compare it to the time a capillary-gravity wave, whose wavelength is equal to the radius of the sphere, would have returned from the boundary. The domain size in the KM (radius of $50 R_s$) is chosen so as to satisfy this condition for all impact times with a ``safety factor'' of 4. Additionally, we can verify that no waves are observed returning towards the impactor before lift-off. The information from the KM is also used to calibrate the domain size for the DNS, though in that case, due to the computational cost involved, we used a safety factor of 1.6 (a domain radius of $20 R_s$).

The programs needed to produce the linearised free-surface simulations are made available as supplementary material.

\subsection{Small surface gradient regime}

The aforementioned implementation of the KM includes the assumption that the free-surface gradient is small. This approximation significantly simplifies the calculation of a rebound (at the cost of a loss of accuracy in the higher Weber number regime), allowing it to be carried out in the order of tens of minutes in standard current laptop computers. Consequently, the implementation of the KM method here presented is better suited to efficiently study the low Weber number regime. 

The kinematic match is also useful in the study of small rebounds for which the sphere's south pole may not return to the height of the initial contact. In this range, one needs to directly observe lift-off to assess the coefficient of restitution, which is more challenging in experiments. This regime is also accessible to direct numerical simulations, which we use to validate the KM predictions. However, these DNS calculations have run-times of the order of days even when using computer clusters. Moreover, the typically small size of spheres for which these weak rebounds are observed produces very short, and therefore fast, capillary waves that require a considerably extended numerical domain to rule out any influence that waves could otherwise have on the rebound if allowed to reach the boundary of the domain and therefore be reflected back and arrive to the vicinity of the impact point. This requirement further increases the computational cost of the direct numerical simulations. 
In these cases, though the need for a large numerical domain is also present when using the KM, scaling it up is much less costly since the numerical fluid domain is one-dimensional.

In practice, we limit the use of the KM method on the linearised free surface model to the cases where the maximum surface slope (a standard measure of nonlinearity in water waves) is no greater than $1$ ( $\|\nabla\eta\|_{\infty} \leq 1$) over the full simulation of the rebound. 

\section{Experimental results and model predictions}
\label{sec:Results}
\subsection{Trajectories and waves}

\begin{figure}
    \begin{subfigure}[c]{0.495\textwidth}
        \centering
        \includegraphics[width =\textwidth]{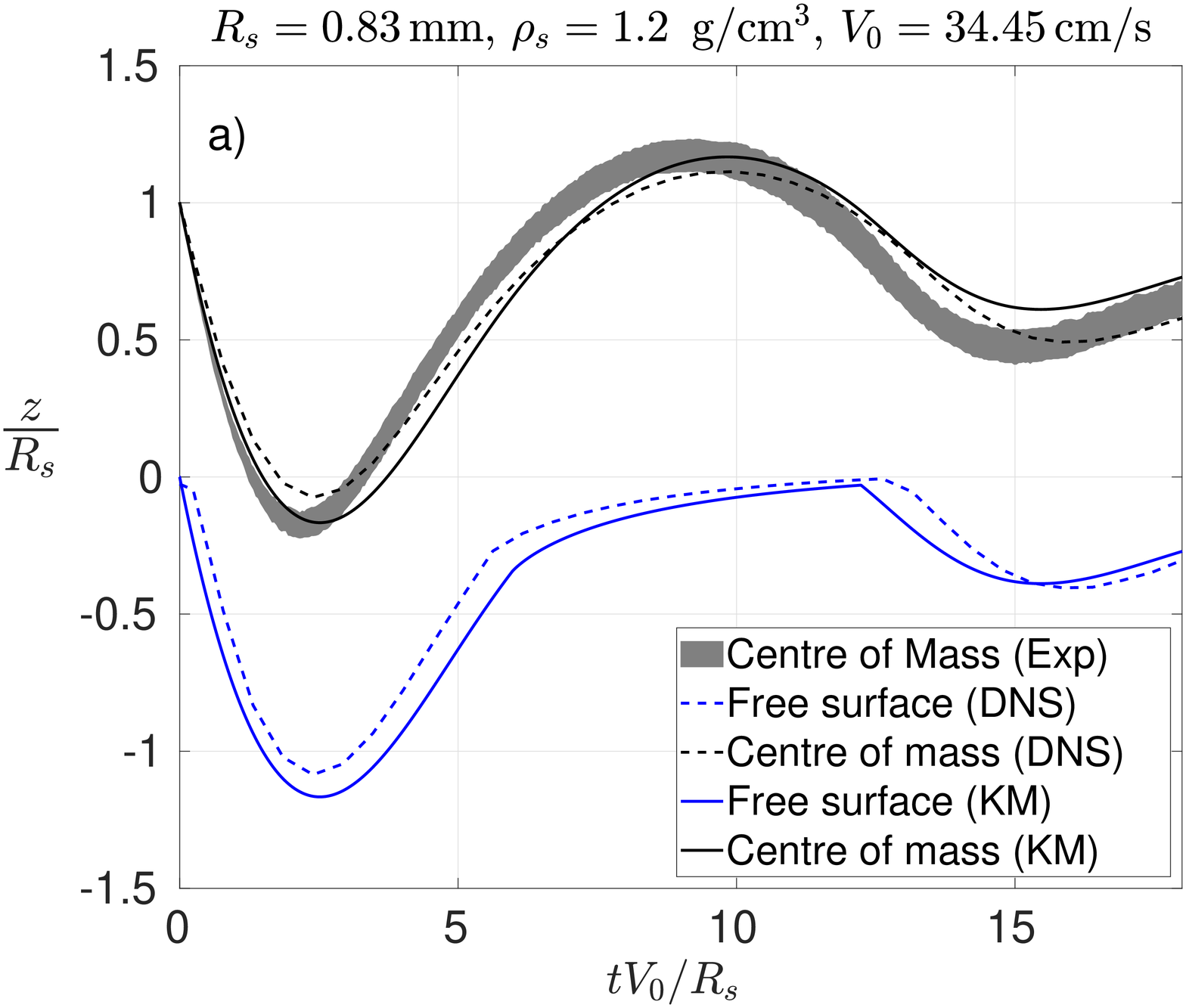}
    \end{subfigure}
    \begin{subfigure}[c]{0.495\textwidth}
    \centering
    \includegraphics[width = \textwidth]{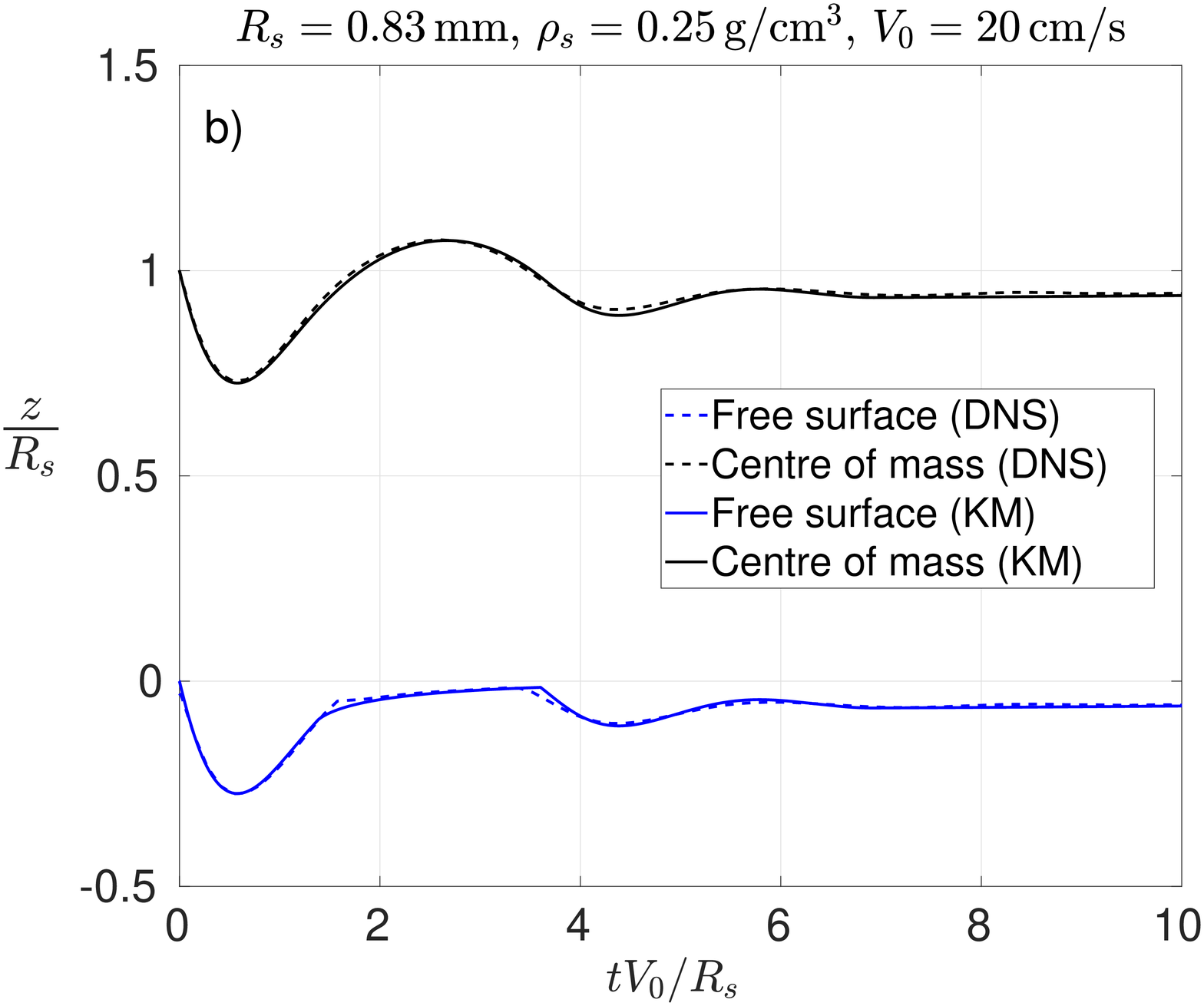}
    \end{subfigure}
    \caption{Comparison of predicted and measured trajectories. Panel (a) shows the resulting trajectory of the centre of mass in experiments (Exp) versus those obtained via direct numerical simulation (DNS) and kinematic match (KM) calculations. The width of the shaded regions that describe the experimental results enclose one standard deviation above and below the mean experimental trajectory. Panel (b) compares the results of the two numerical methods in the low Weber number regime, which is not accessible in the present experiments. Both panels also include the free-surface elevation at the centre of the bath, i.e. directly below the south pole of the sphere (the impact point), as obtained from DNS and KM simulations. Videos of the KM simulation for the case in panel (a) are available as supplementary material.}
    \label{fig:TrajectsLinear}
\end{figure}

A comparison between the trajectories obtained in the small surface gradient regime is presented in figure \ref{fig:TrajectsLinear}. Panel (a) corresponds to one case for which we have experimental, DNS and KM trajectories for the sphere. Panel (b) shows the comparison between DNS and KM for a weaker impact, for which there is no experimental data. We highlight that the disagreement between DNS and KM is of the order of the predicted droplet deformation in the pseudo-solid sphere used in DNS. In this figure, we have exceptionally included the evolution of a second impact, as a way to show that the methods employed here are able to capture successive impacts, though these are not the focus of the present work. The second impact is made evident in the corner that is present in the curve that tracks the free-surface elevation directly below the south pole of the sphere as a function of time. All experimental and DNS trajectories for different spheres and impact velocities  are presented in appendix \ref{app:trajects}.

\begin{figure}
    \centering
    \begin{tabular}{cc}
     \includegraphics[width = .49\textwidth]{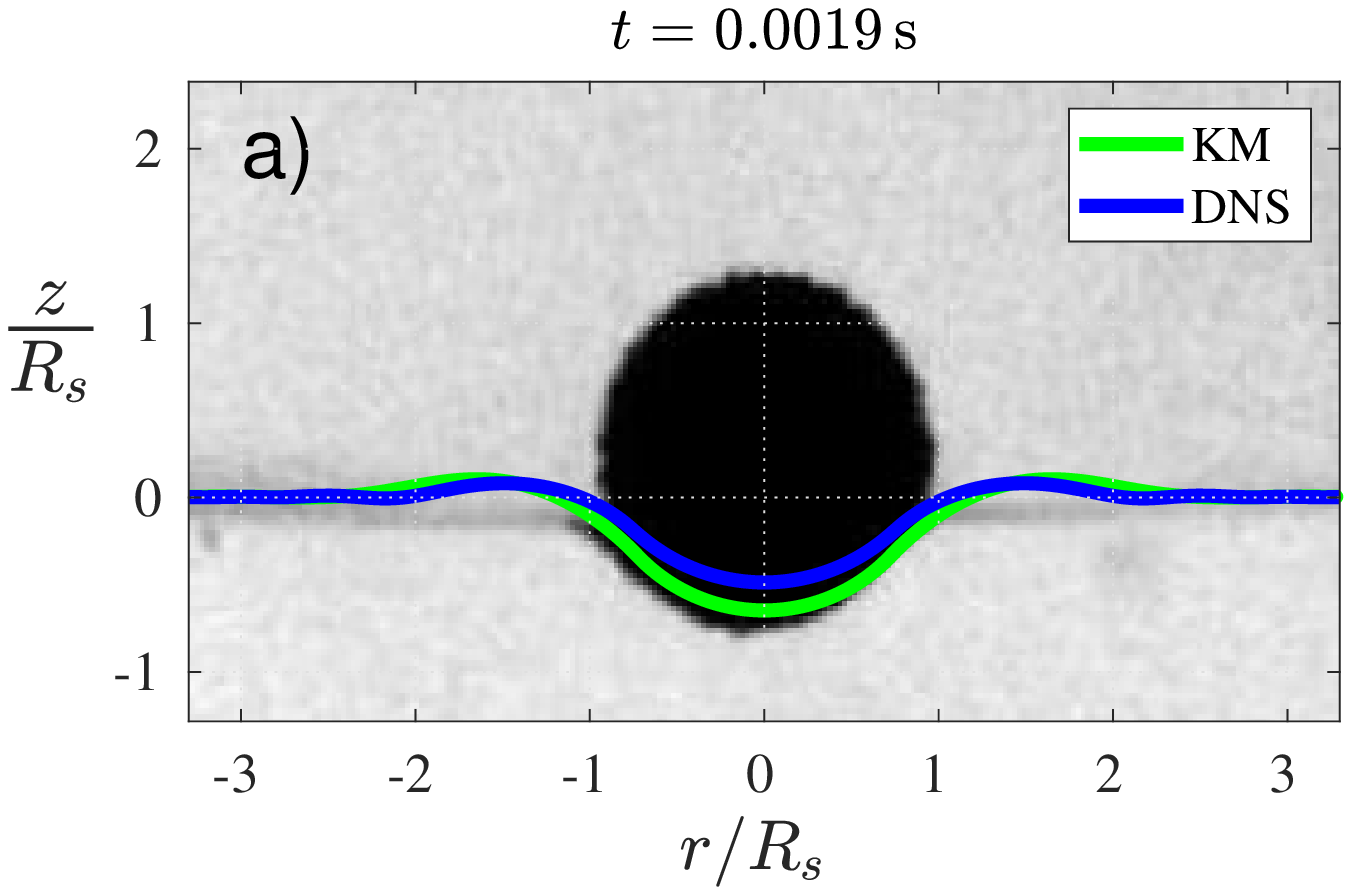}
     &
     \includegraphics[width = .49\textwidth]{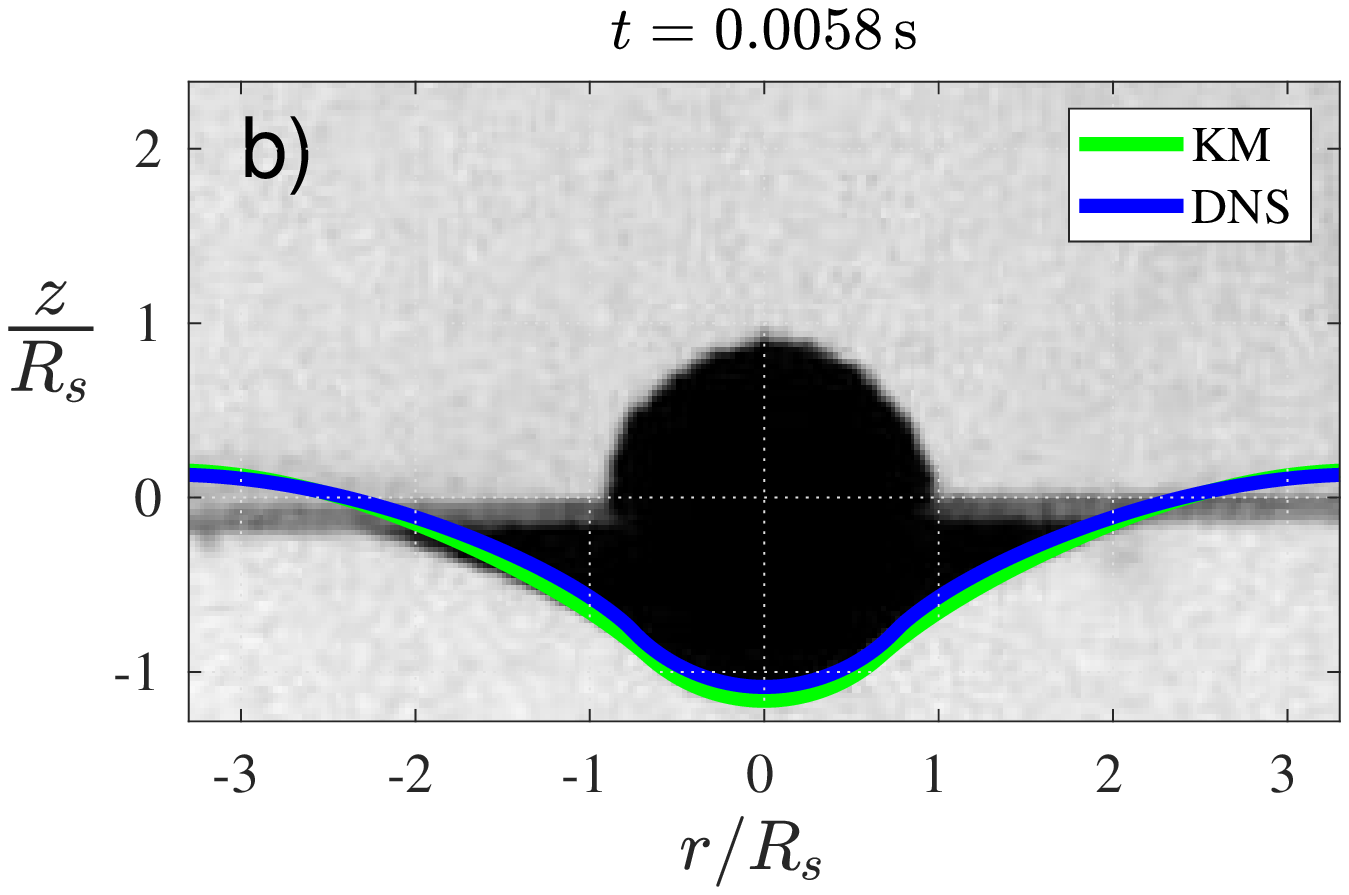}
     \\
     \\
     \includegraphics[width = .49\textwidth]{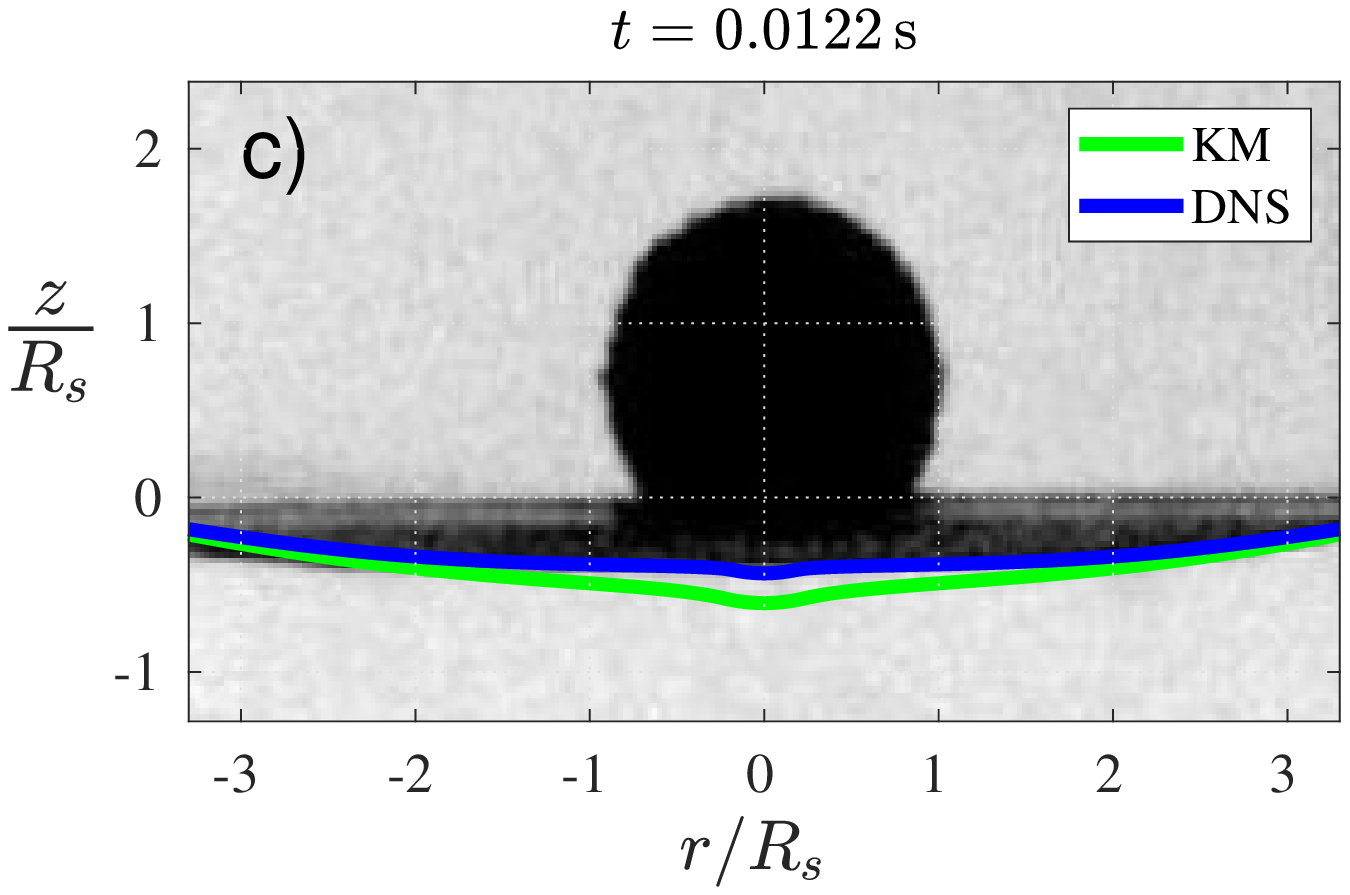}
     &
     \includegraphics[width = .49\textwidth]{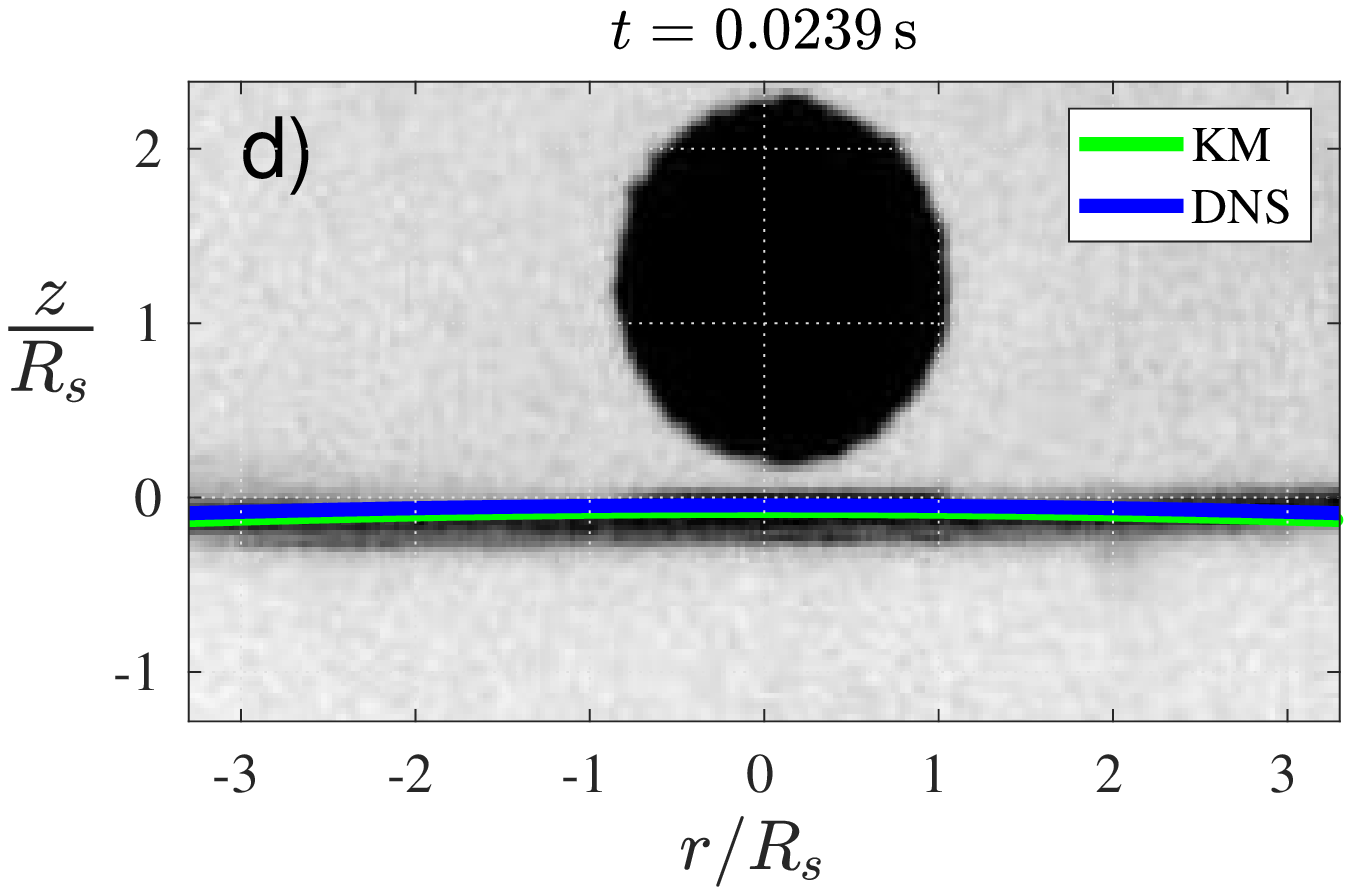}
    \end{tabular}
    \caption{Surface profile predictions superimposed onto experimental high speed camera images for $R_s = 0.83\,$mm, $\rho_s = 1.2\,$gr$/$cm$^3$ and $V_0 = 34.45\,$cm$/$s. }
    \label{fig:photos}
\end{figure}

\begin{figure}
    \centering
    \begin{tabular}{cc}
     \includegraphics[width = .43\textwidth]{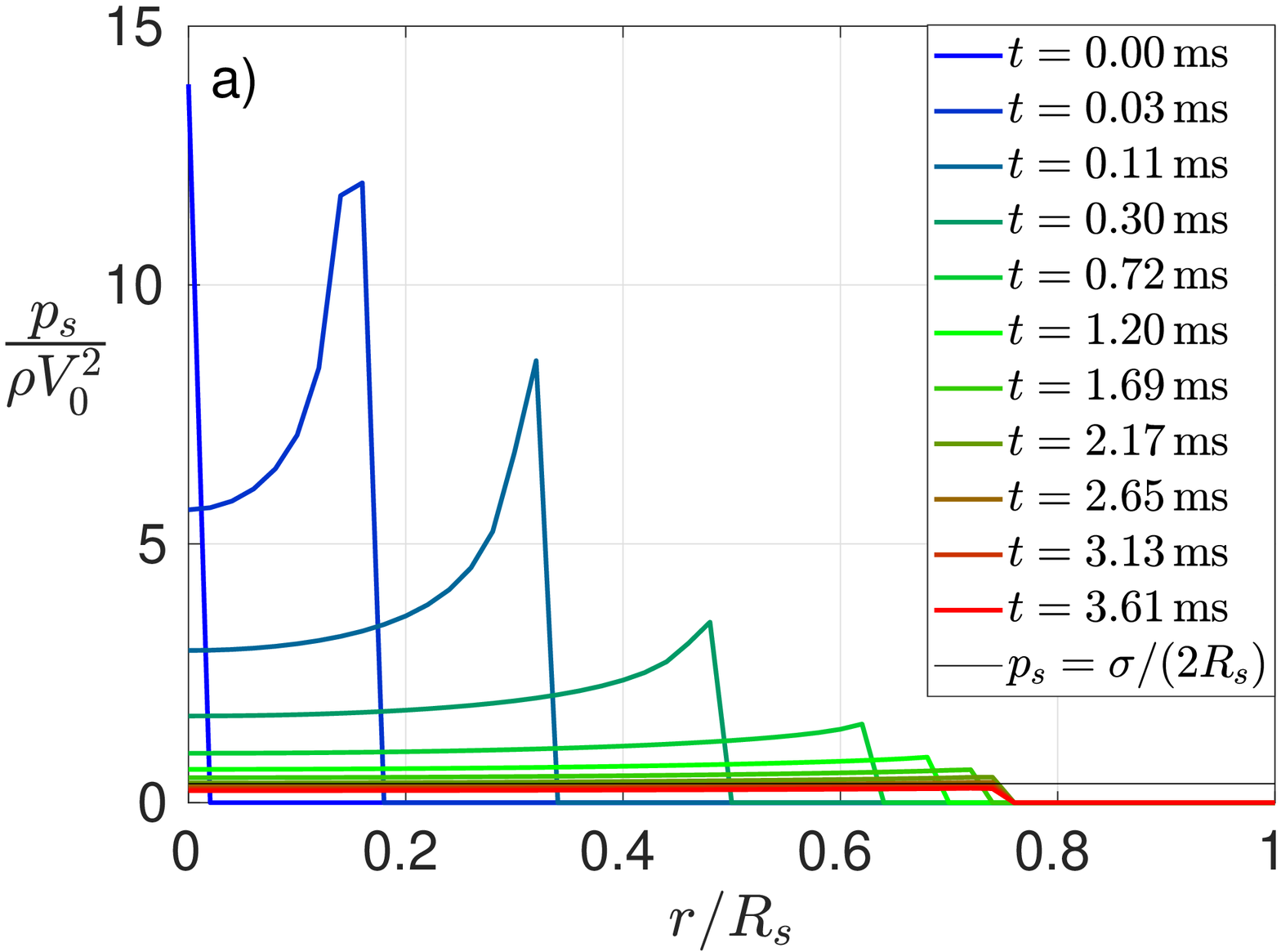}
     &
     \includegraphics[width = .56\textwidth]{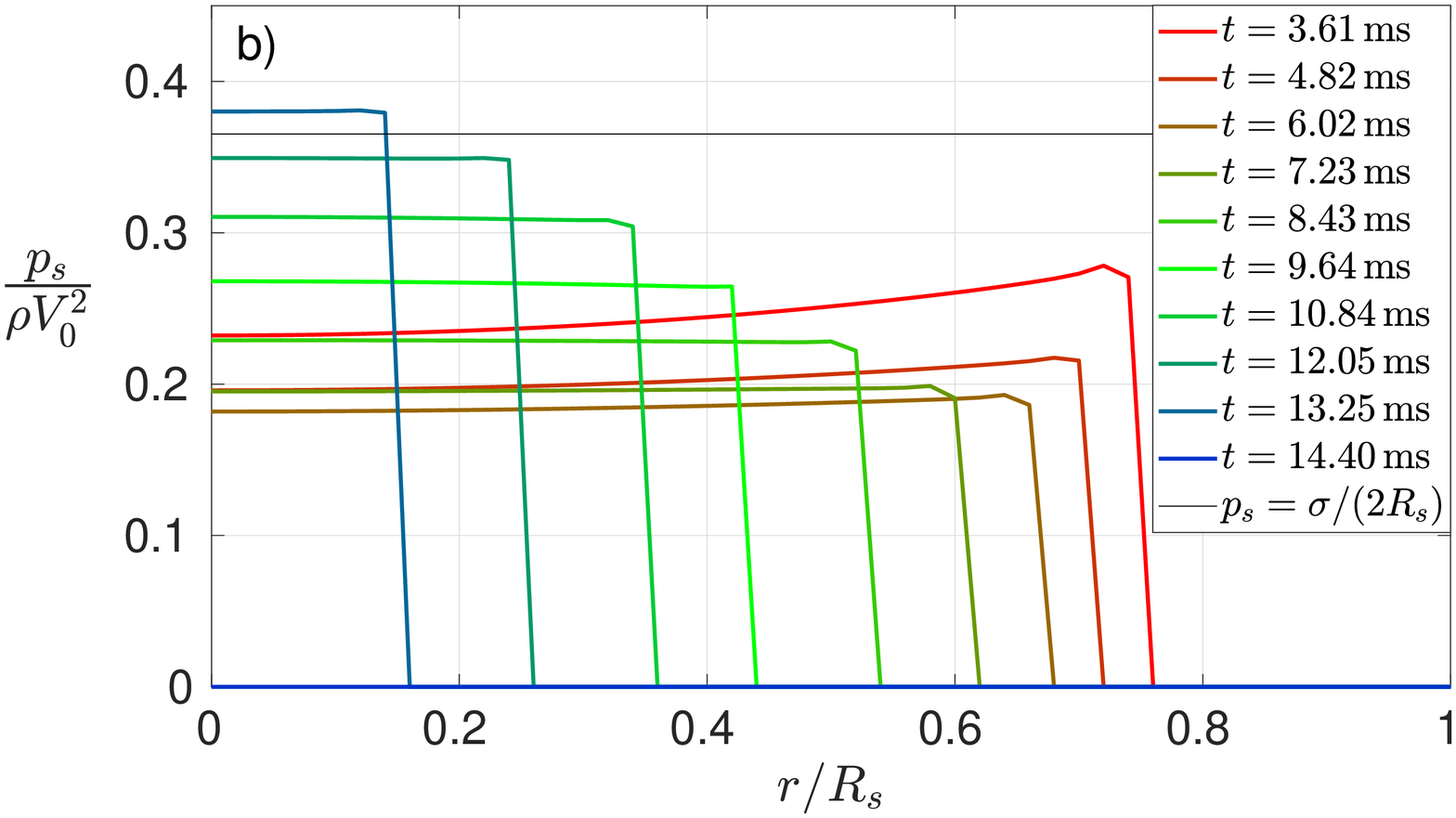}
    \end{tabular}
    \caption{Evolution of pressure distribution as predicted by the linearised model. Panel (a) shows the pressure distributions as the pressed area expands following impact, and panel (b) shows the pressure distribution as the pressed area contracts before lift-off ($R_s = 0.83\,$mm, $\rho_s = 1.2\,$g$/$cm$^3$, $V_0 = 34.45\,$cm$/$s). The black horizontal line  indicates the contribution of surface tension to the pressure distribution and thus serves as a reference level.     } 
    \label{fig:press_prof}
\end{figure}

An example of a comparison between experimental results and model predictions for the interface shape is provided in Figure \ref{fig:photos}. Four snapshots of the impact reported in figure \ref{fig:TrajectsLinear}(a) are chosen. Initial stages of the impact, panel \ref{fig:photos}(a), show a slight better agreement of the kinematic match; this effect is expected as a consequence of the deformability of the pseudo-solid sphere in our DNS simulations. However; in later stages of the rebound, panel \ref{fig:photos}(c), the DNS better captures the interface deflection.

The agreement observed in figures \ref{fig:TrajectsLinear} and \ref{fig:photos} suggest that the role of the flow within the air layer is not dominant in this system for low Weber numbers, as we are able to capture the experimental dynamics with both the air-layer modelling DNS, and the linearised modelling that completely ignores the role of air flow. 

Figure \ref{fig:press_prof} shows the model's predicted evolution of the pressure field as the pressed area expands (panel a), and the subsequent contraction of the pressed area as lift-off approaches (panel b). Note that the time scales of panel (a) are much faster than those of panel (b). The pressure profiles are consistent with those previously observed but unreported in  \citet{GaleanoRiosEtAl2017,GaleanoRiosEtAl2019}, where the initial spike in pressure is followed by an approximately constant pressure distribution with a peak at the boundary of the pressed area. This model predicts that the peak is most pronounced in the early impact times.

\subsection{Rebound metrics}

We consider three different output parameters for the rebounds, namely: contact time ($t_c$), coefficient of restitution ($\alpha$) and maximum surface deflection ($\delta$). As mentioned in section \ref{Section:ExpProc}, given the experimental difficulty to accurately determine the time of surface detachment of the sphere, contact time, $t_c$, is defined as the interval between the two instances when the south pole of the sphere crosses level $z=0$ and the coefficient of restitution, $\alpha$, is defined as minus the ratio of the vertical velocities at those times.

\begin{figure}
    \centering
    \begin{tabular}{rr}
     \includegraphics[width = .45\textwidth]{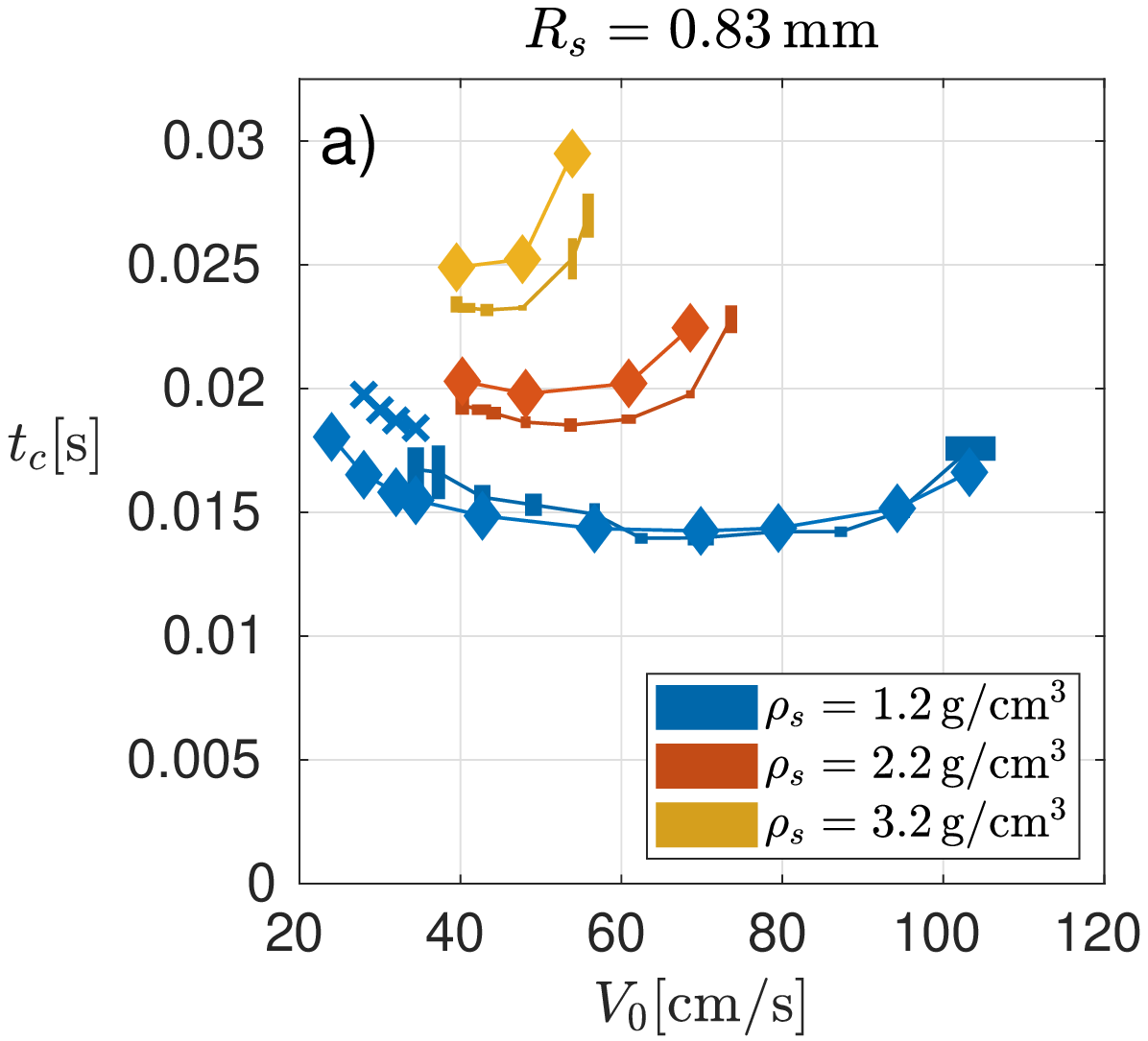}
     &
     \includegraphics[width = .44\textwidth]{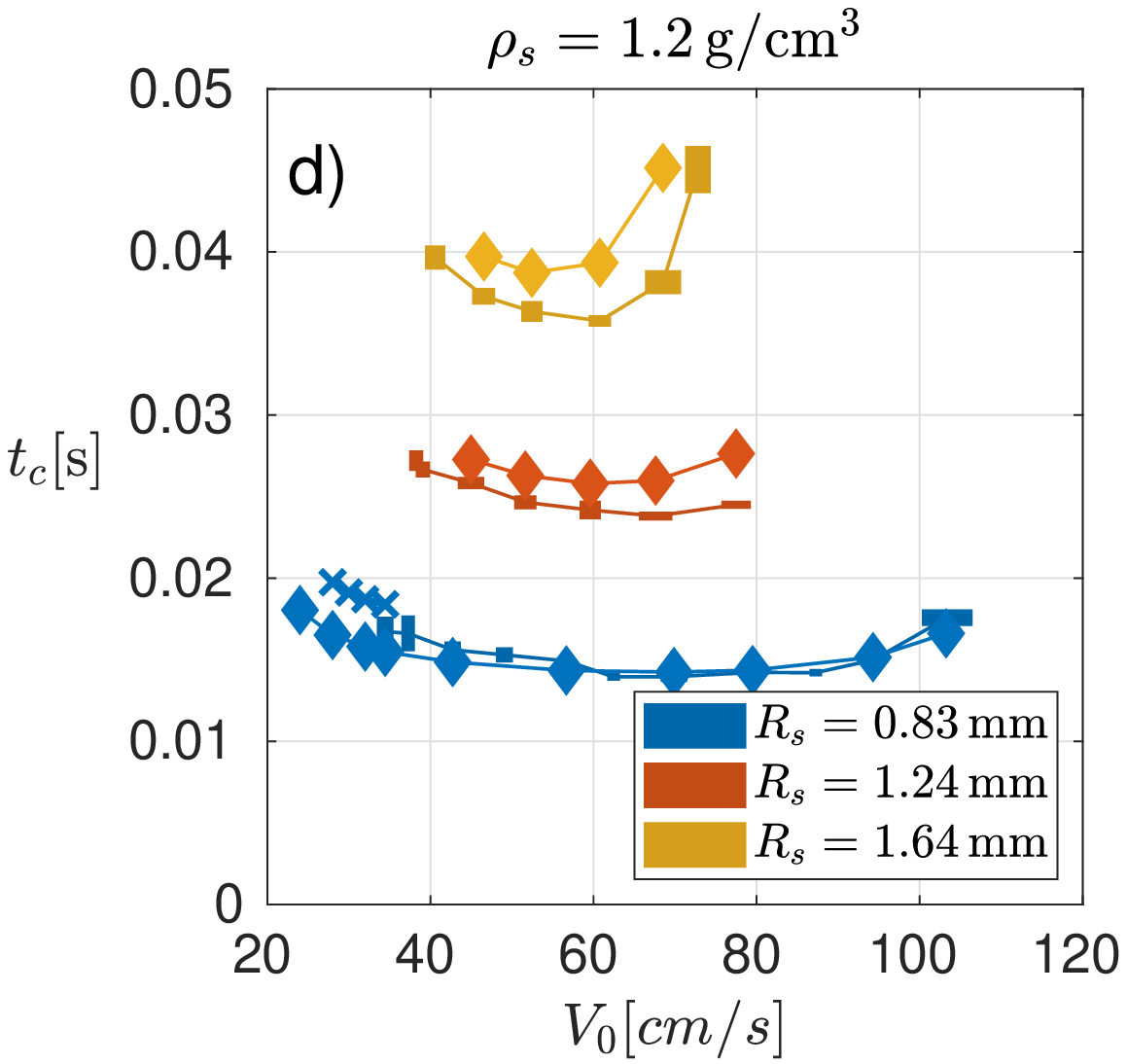}
     \\
     \\
     \includegraphics[width = .4\textwidth]{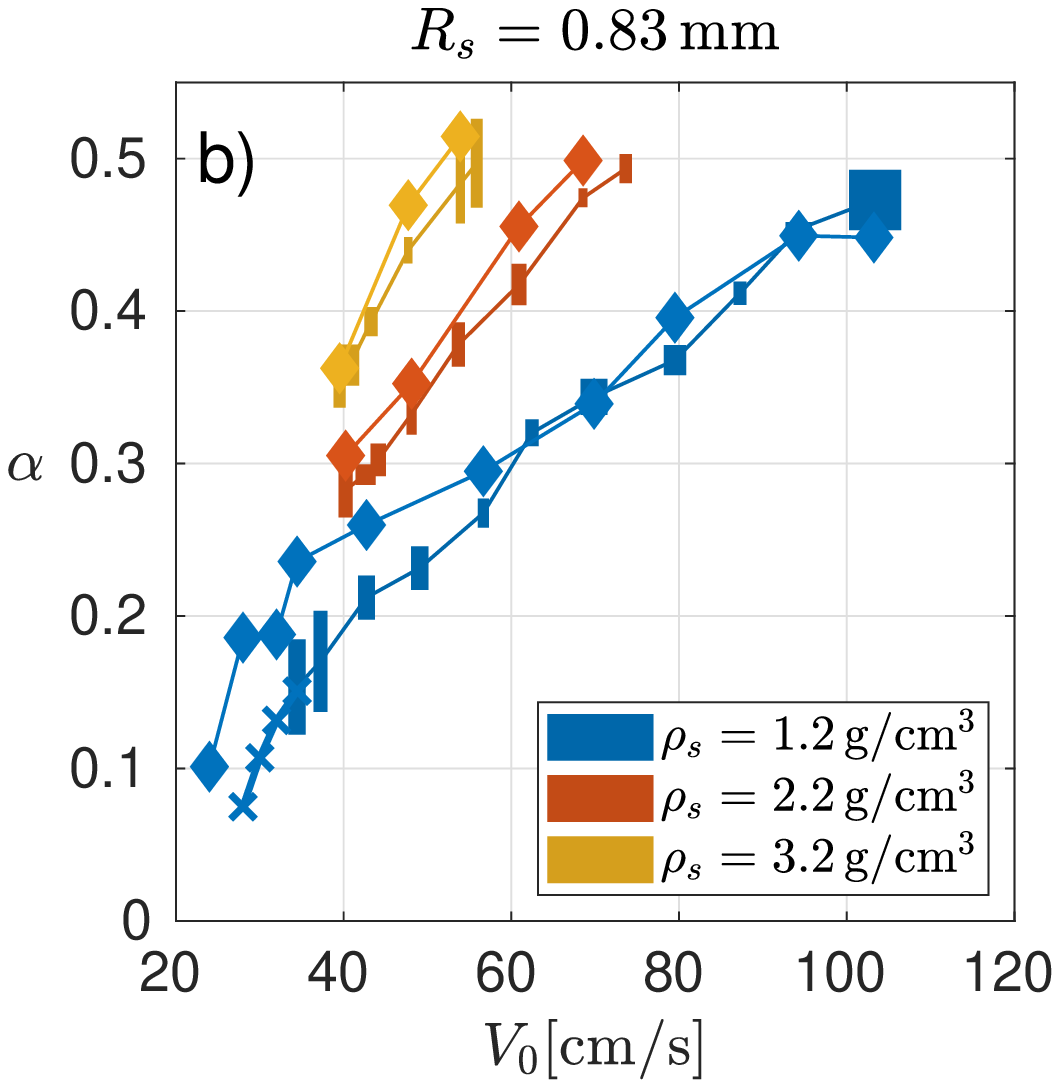}
     &
     \includegraphics[width = .4\textwidth]{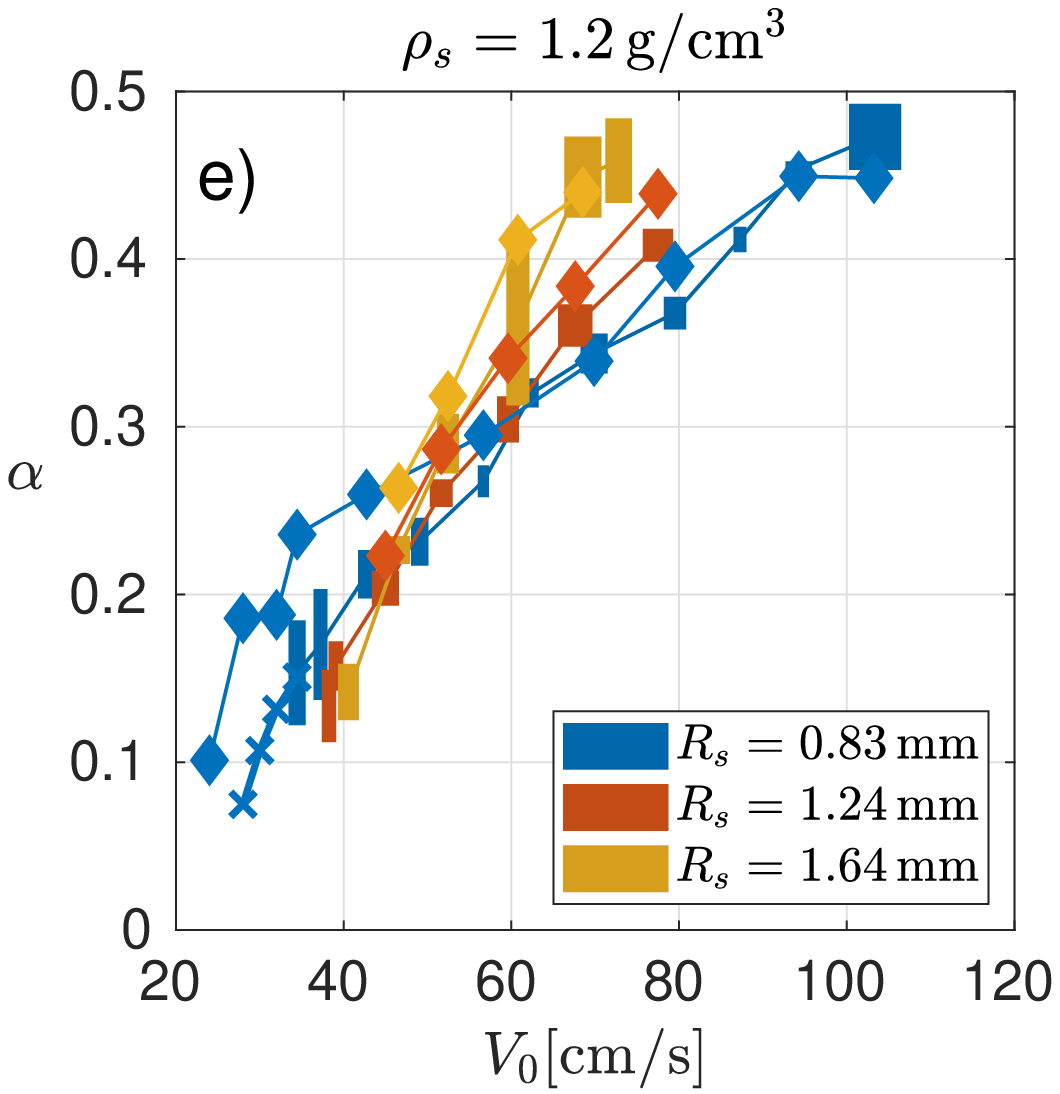}
     \\
     \\
     \includegraphics[width = .45\textwidth]{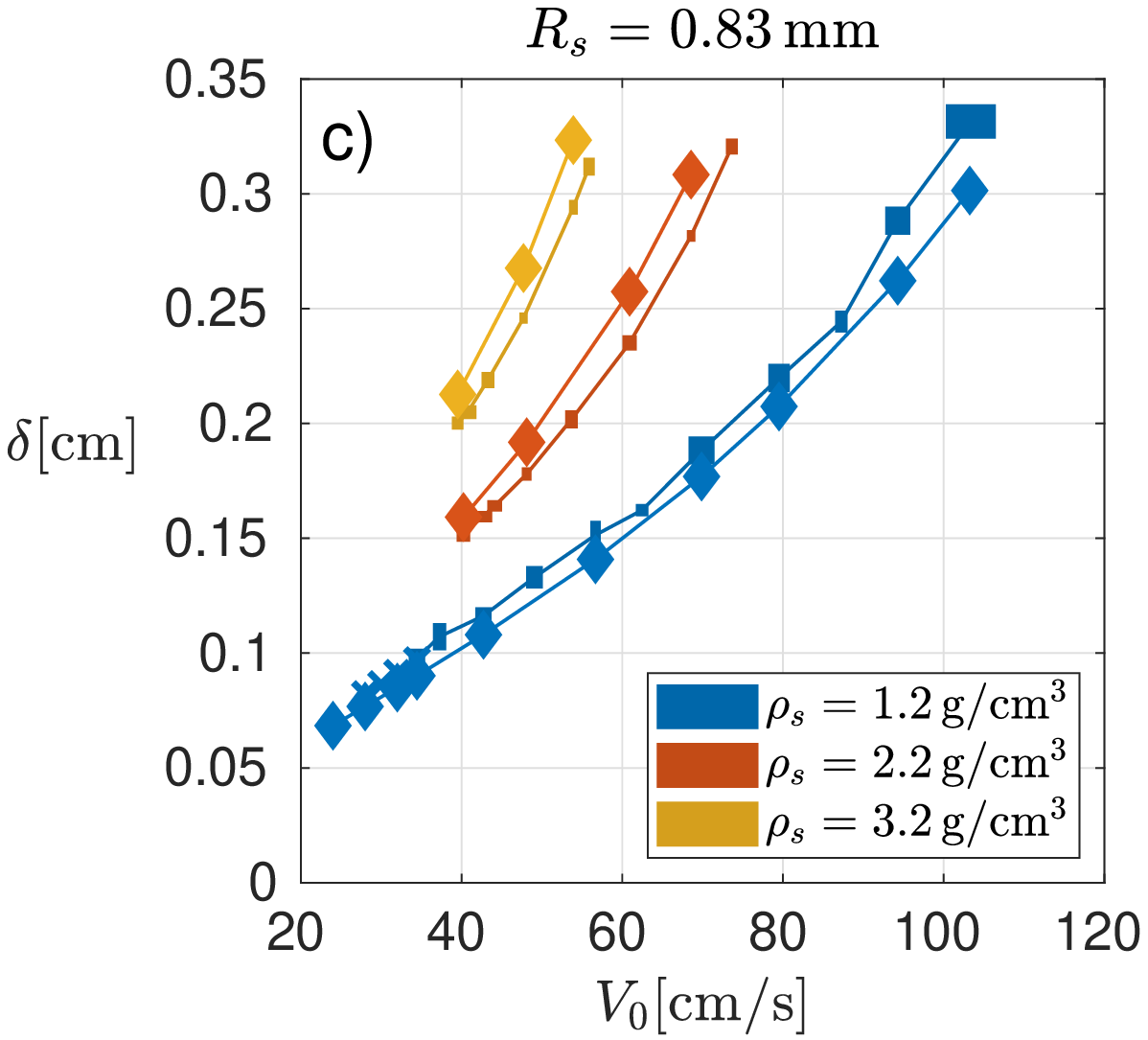} 
     &
     \includegraphics[width = .43\textwidth]{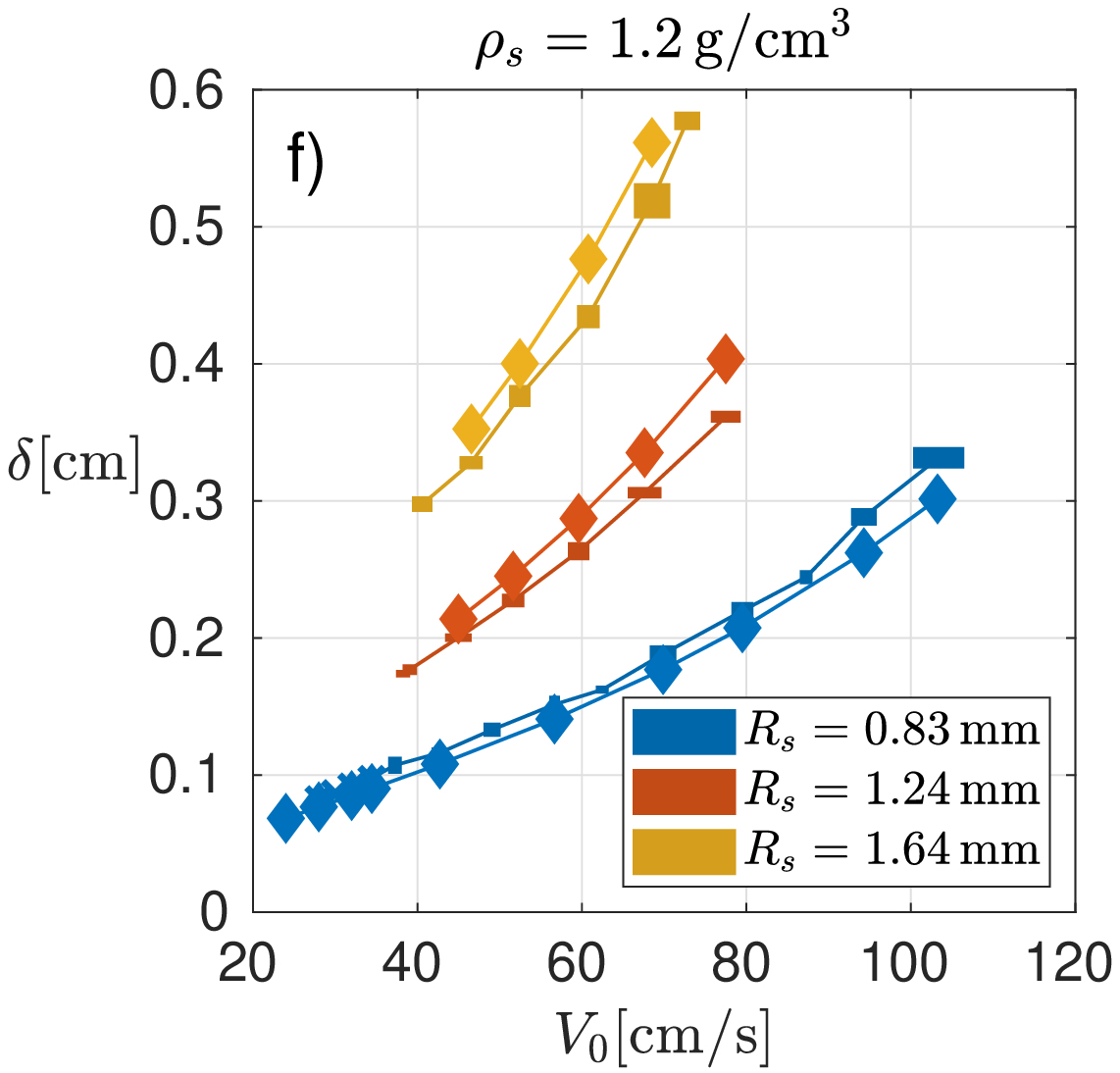}
    \end{tabular}
    \caption{Comparison of the contact time, coefficient of restitution and maximum penetration depth in experiments ($\blacksquare$), DNS ($\blacklozenge$) and KM ($\times$). The width and height of rectangular markers correspond to one standard deviation above and below the mean experimental values. All relevant parameters and notation are provided in Table~\ref{tab:parameters}.}
    \label{fig:SixPanels}
\end{figure}

\begin{figure}
    \centering
    \includegraphics[width = 13.5cm]{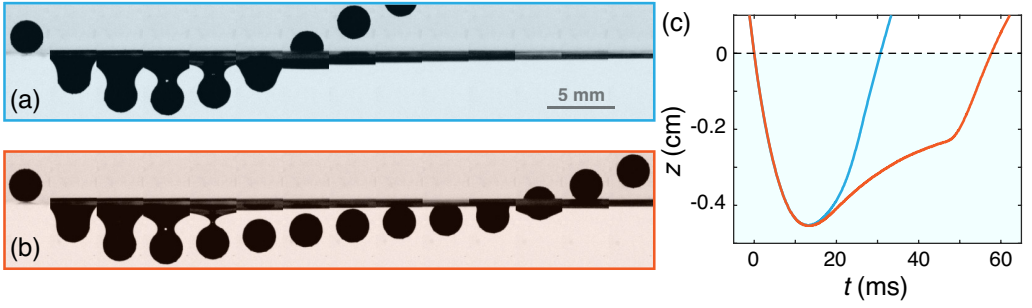}
    \caption{Two behaviours observed in experiment for nearly identical impact velocities for identical spheres with radius $R_s=0.124$ cm and density $\rho_s=1.2$ g/cm$^3$, just before sinking threshold. (a) Standard rebound, $V_0=86.7 \pm 1.7$ cm/s. (b) ``Resurrection'' phenomenon where cavity pinches off yet the sphere eventually resurfaces and rebounds completely, $V_0=87.4 \pm 3.5$ cm/s.  Images are evenly spaced in time by 4.8 ms, corresponding to 48 frames. (c) Trajectories associated with the images shown in parts (a) and (b). Videos corresponding to the trials shown in (a) and (b) are available as supplementary material.}
    \label{fig:res}
\end{figure}

\begin{figure}
    \centering
    \includegraphics[width = 13.5cm]{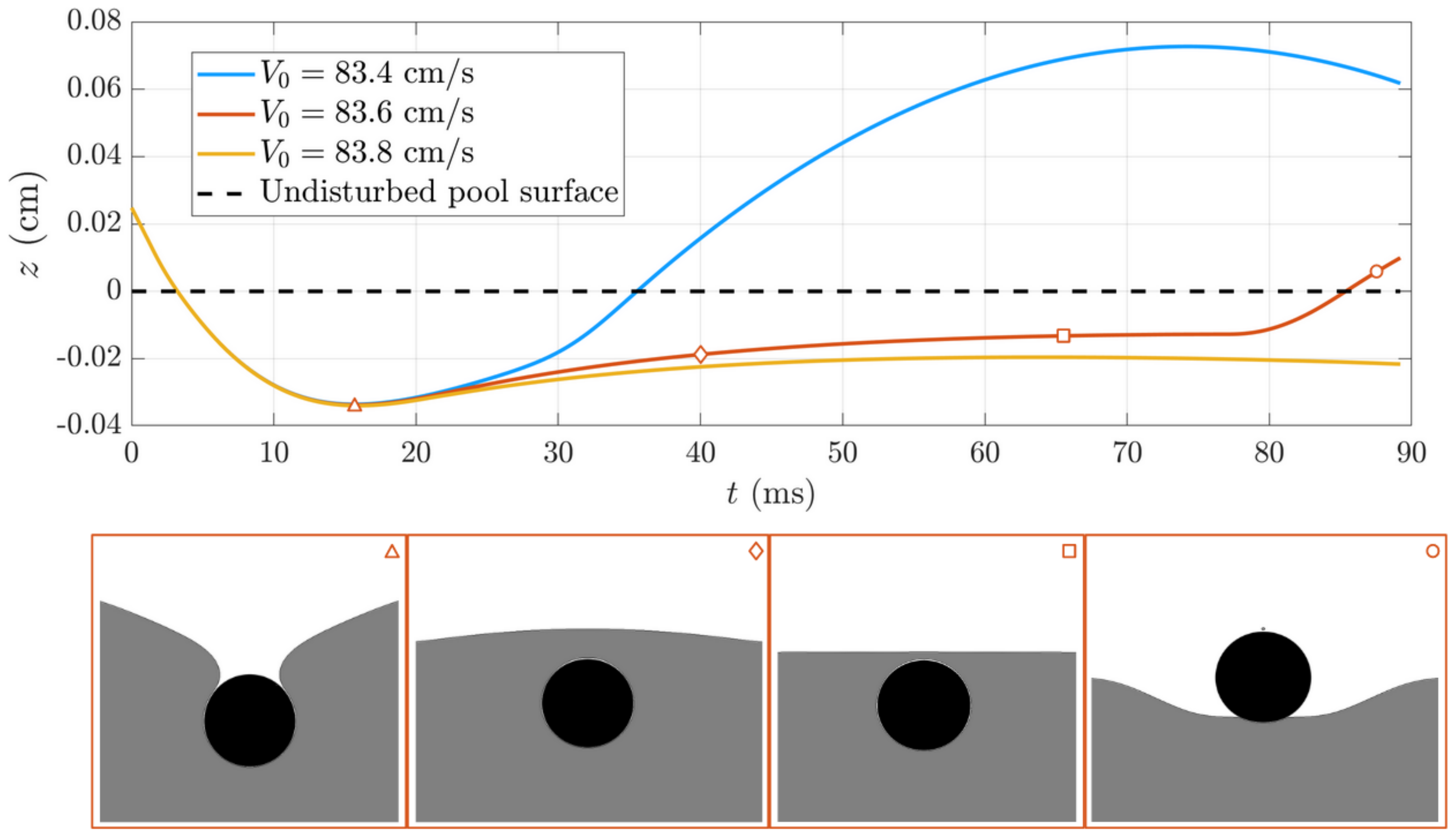}
    \caption{``Resurrection'' phenomenon observed using DNS for a pseudo-solid with radius $R_s=0.124$ cm and density $\rho_s=1.2$ g/cm$^3$, impacting with velocity $V_0 = 83.6$ cm/s. 
    Small impact velocity variations (of $\pm 0.2$ cm/s) result in either bouncing or sinking. Lines in the top panel represent the $z$-position of the centre of mass of the pseudo-solid as a function of time in each of these cases, while symbols indicate representative time steps in the flow evolution for the ``resurrection'' dynamics (illustrated in the bottom panel). A video summary contrasting these three scenarios is available as supplementary material.}
    \label{fig:resDNS}
\end{figure}

Figures \ref{fig:SixPanels}(a) and \ref{fig:SixPanels}(d) show that, for a given sphere (i.e. radius and density fixed, respectively), contact time is only weakly dependent on impact velocity.  The increase in contact time near the sinking threshold is presumably due to the highly nonlinear surface deformations observed in this regime.  This particular trend is apparent in the experimental trajectories presented in Figure \ref{fig:traj}(d), where nearly all rebounding trajectories intersect one another at a similar time, apart from the largest impact velocities, for which this tendency visibly diverges.  In fact, an entirely new exotic trajectory was observed just below the sinking threshold velocity, an example of which is documented in figure \ref{fig:res}.  We observe a new ``resurrection'' mode where the particle becomes completely submerged but is left with upward inertia following pinch-off, and ultimately completely de-wets and rebounds.  This surprising behaviour was observed in a very narrow regime of impact velocities and only for the lowest density spheres considered in our experiments: $\rho_s=1.2$ g/cm$^3$. To the authors' knowledge, this novel behaviour has not been previously reported for particles as dense or more dense than water. Guided by experimental insight, we were able to pinpoint and reproduce this type of dynamics computationally as well, an example of which we summarise in Figure~\ref{fig:resDNS}. 
This exploration shows that the ``resurrection'' is possible when the sphere has initiated its upward motion before the liquid bridge is formed over its north pole. Moreover, the small parameter window in which this peculiar phenomenon can be observed requires a delicate balance, wherein the gentle upward motion of the sphere overcomes the decelerating influences of both gravity and drag in order to pierce the liquid bridge above it.
We found that small variations ($\pm0.2$ cm/s) in the impacting velocity translate to either bouncing, if the penetration depth is sufficiently small to avoid the sphere becoming submerged, or sinking, in case the liquid bridge is sufficiently thick to successfully arrest the transient upward momentum of the sphere. A biological application that has some elements in common with the phenomenon of resurrecting spheres was reported in the work of \citet{KimEtAl2015}, in which they discuss the mechanics of plankton jumping out of water.

Returning to the broader parameter space, we find that the coefficient of restitution in figures \ref{fig:SixPanels}(b) and \ref{fig:SixPanels}(e) monotonically increases with impact speed for each sphere. The coefficient of restitution is more sensitive to the sphere's density than to its radius, with higher density spheres recovering relatively more energy during impact than otherwise equivalent lower density spheres.  Curiously, for the parameters studied here, we observed an approximate upper limit for the coefficient of restitution of around $\alpha=0.5$ in each case which occurred just below the sinking threshold.  Due to the relatively high Reynolds numbers considered in this work, the apparent {\it loss} of sphere energy during impact is in fact predominantly an energy {\it transfer} required to accelerate the bath fluid during impact.  In general, one can clearly observe that all trends present in the experimental curves are captured by the DNS. In particular, on panel \ref{fig:SixPanels}(e), we can see that smaller spheres show a higher coefficient of restitution ($\alpha$) at low velocities but a lower $\alpha$ at high velocities. The subtle trend is also present in the DNS results. 

Lastly, the penetration depth for all cases is shown in Figures \ref{fig:SixPanels}(c) and \ref{fig:SixPanels}(f), which monotonically increases with impact speed, sphere density, and radius.  These same trends are closely captured by the DNS.

Experiments and DNS show good agreement on the proposed metrics (figure \ref{fig:SixPanels}) over the full range accessible to experiments; namely, from velocities as low as to barely cause the rebounding sphere to recover past the initial impact height to impact velocities that cause the sphere to break through the surface and sink.  This fact strongly suggests that, for the parameters of interest, the non-wetting, pseudo-solid impactor is a very good approximation for the superhydrophobic sphere.  As discussed in the prior sections, the pseudo-solid approach simplifies the overall numerical model.  Remarkably, the data presented here thus suggests that the micro-scale roughness and dynamic contact line motion appear, at most, minimally relevant to the rebound metrics observed in experiments. All experimental and DNS trajectories that correspond to the points in figure \ref{fig:SixPanels} are presented in appendix \ref{app:trajects}.

There is a single experimental rebound for which the small surface slope assumption of the KM is satisfied. This case corresponds to $R_s = 0.83\,\mathrm{mm}$, $\rho = 1.2\, \mathrm{g/cm^3}$ and $V_0 = 34.45\,\mathrm{cm/s}$. KM simulations results are included for this case as well as for the same sphere with lower impact velocities in figure \ref{fig:SixPanels}. DNS results are also shown for this extension of the experimental regime. In this range of impact velocities, simulations smoothly extend the experimental results; however, a direct quantitative comparison is not possible with the current experimental setup, as the south pole of the sphere is obstructed for small rebound heights by the capillary wave field generated during the impact. Simulation results in the low Weber regime are expanded in the  following section.

\subsection{Model predictions}

We further explore the low Weber number regime, using the KM method. Specifically, we simulate the impact of spheres with radii smaller than those within the experimental range, densities below that of the materials tested in the experiments, and impact velocities including those that do not cause the sphere to fully rise above the $z=0$ level. Namely, we cover the range from the weakest impact velocity that is capable of producing a rebound to the highest impact velocity for which we satisfy $\|\nabla\eta\|_{\infty}\leq 1$.

\begin{figure}
    \centering
    \begin{tabular}{rr}
     \includegraphics[width = .495\textwidth]{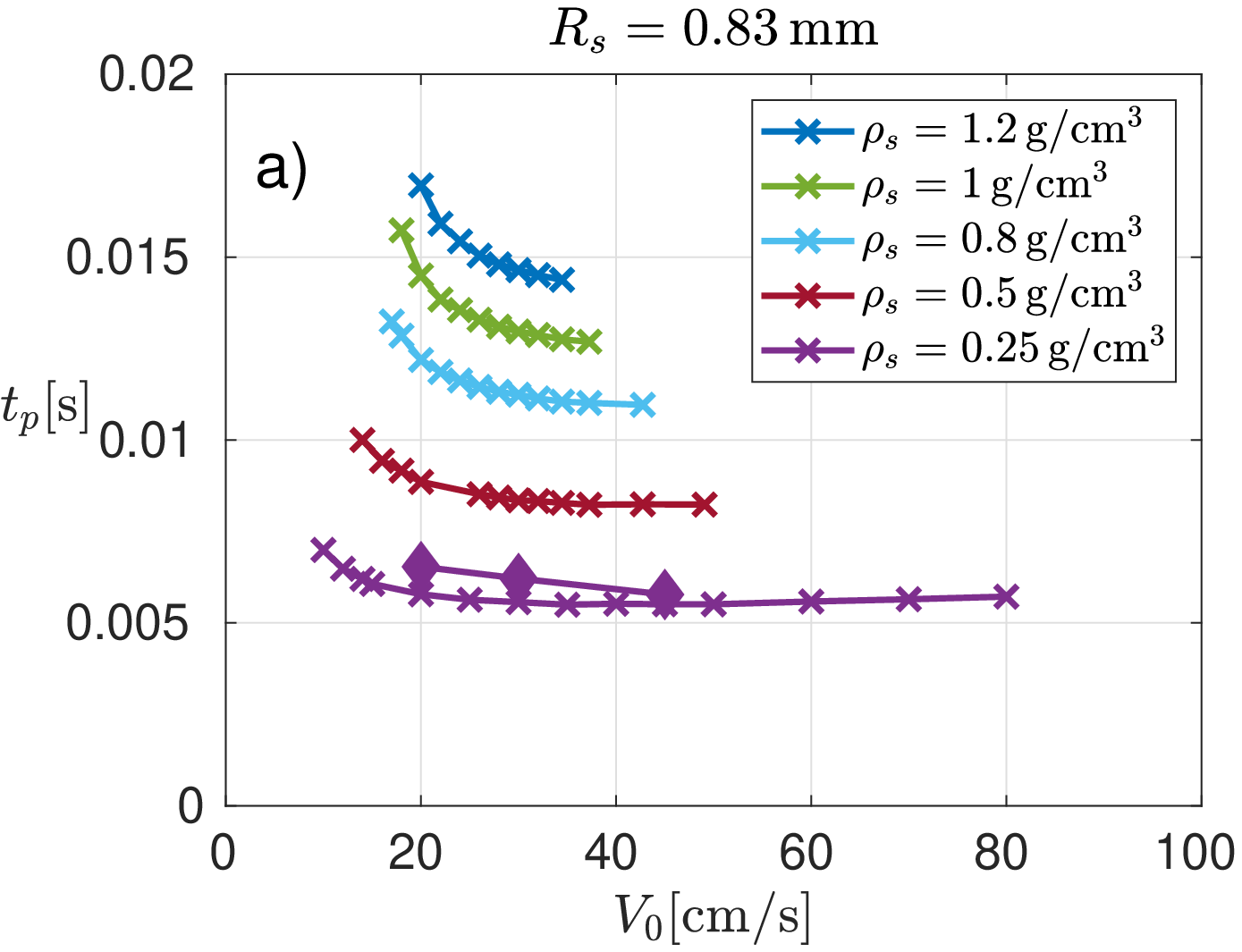}
     &
     \includegraphics[width = .495\textwidth]{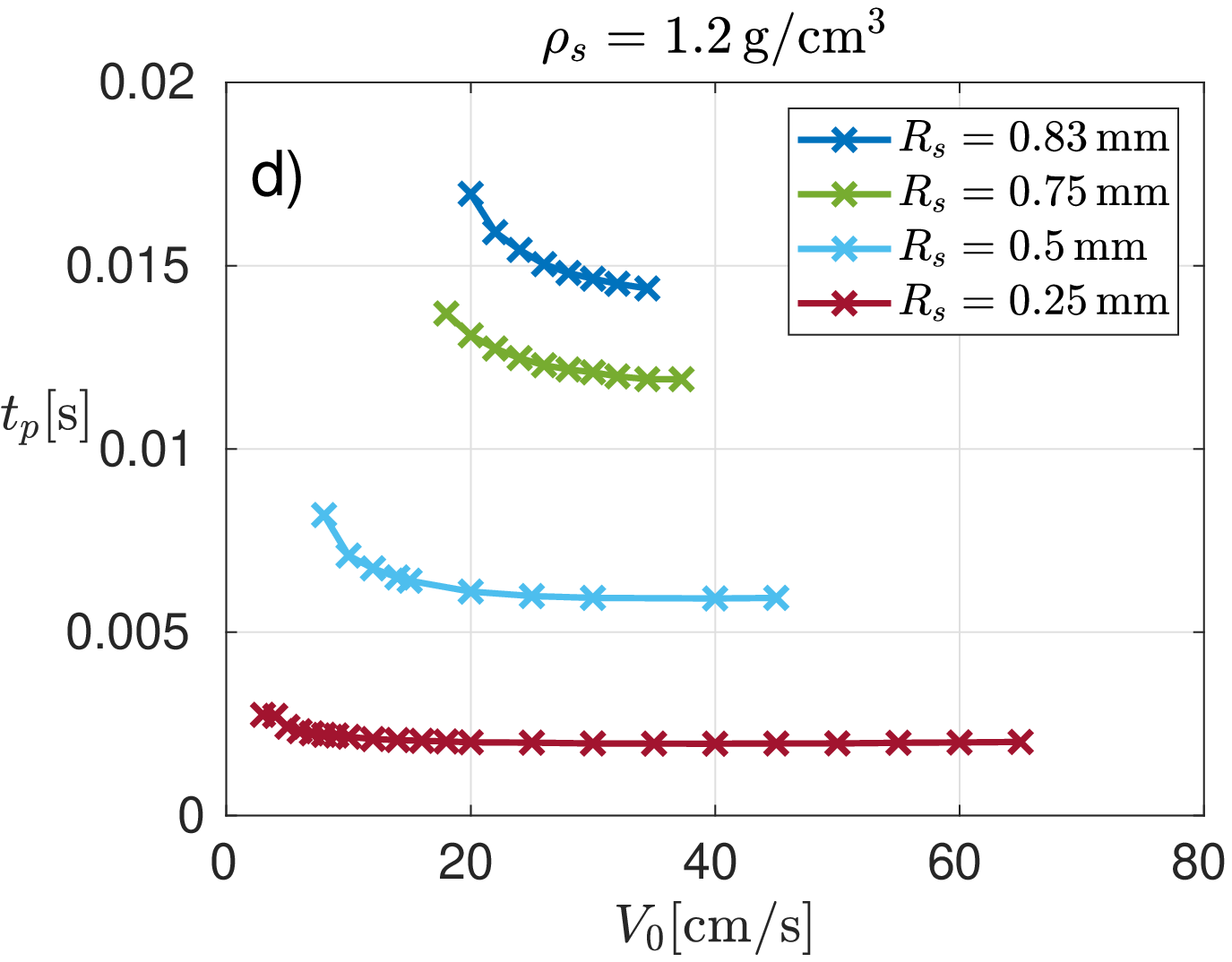}
     \\
     \\
     \includegraphics[width = .48\textwidth]{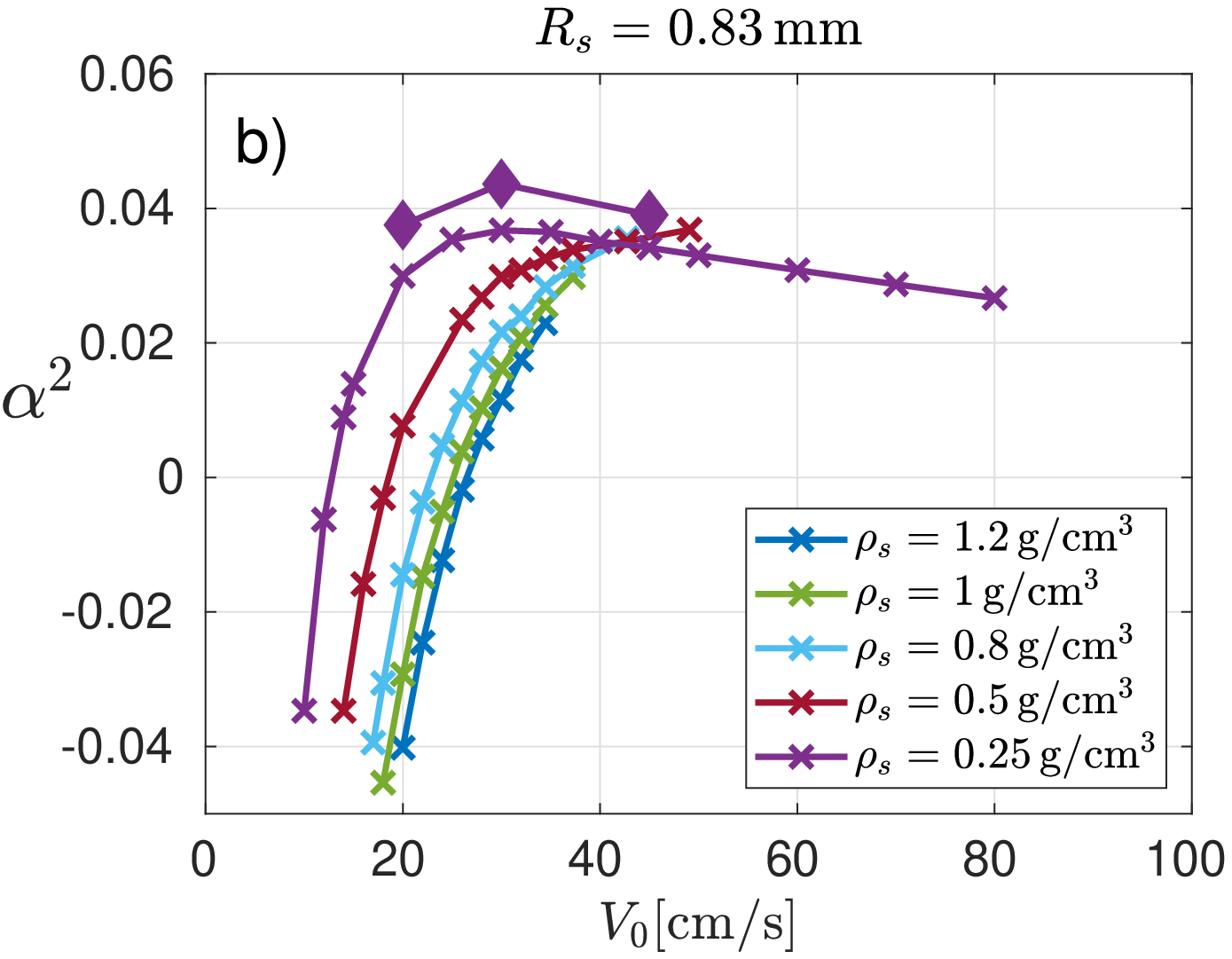}
     &
     \includegraphics[width = .48\textwidth]{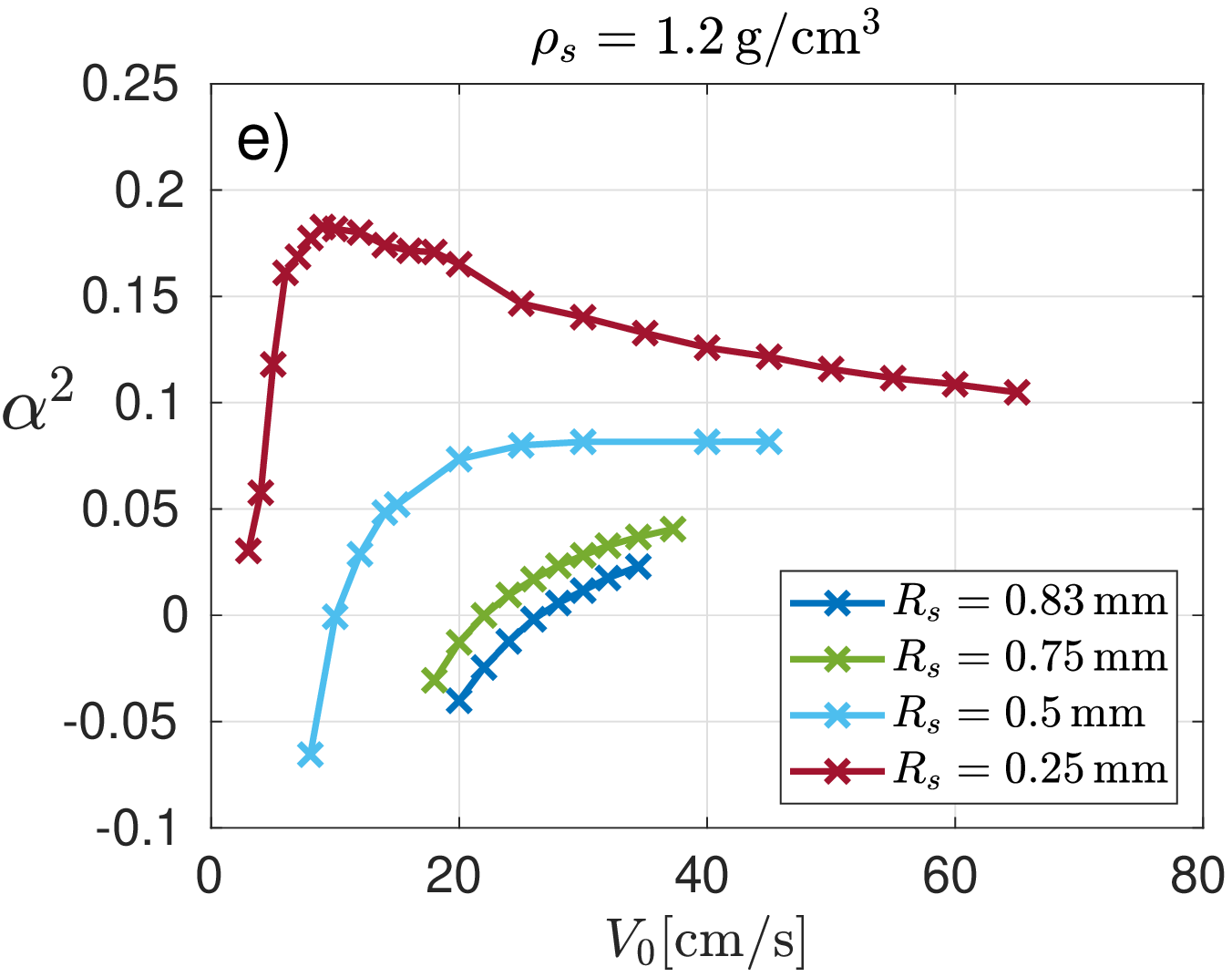}
     \\
     \\
     \includegraphics[width = .495\textwidth]{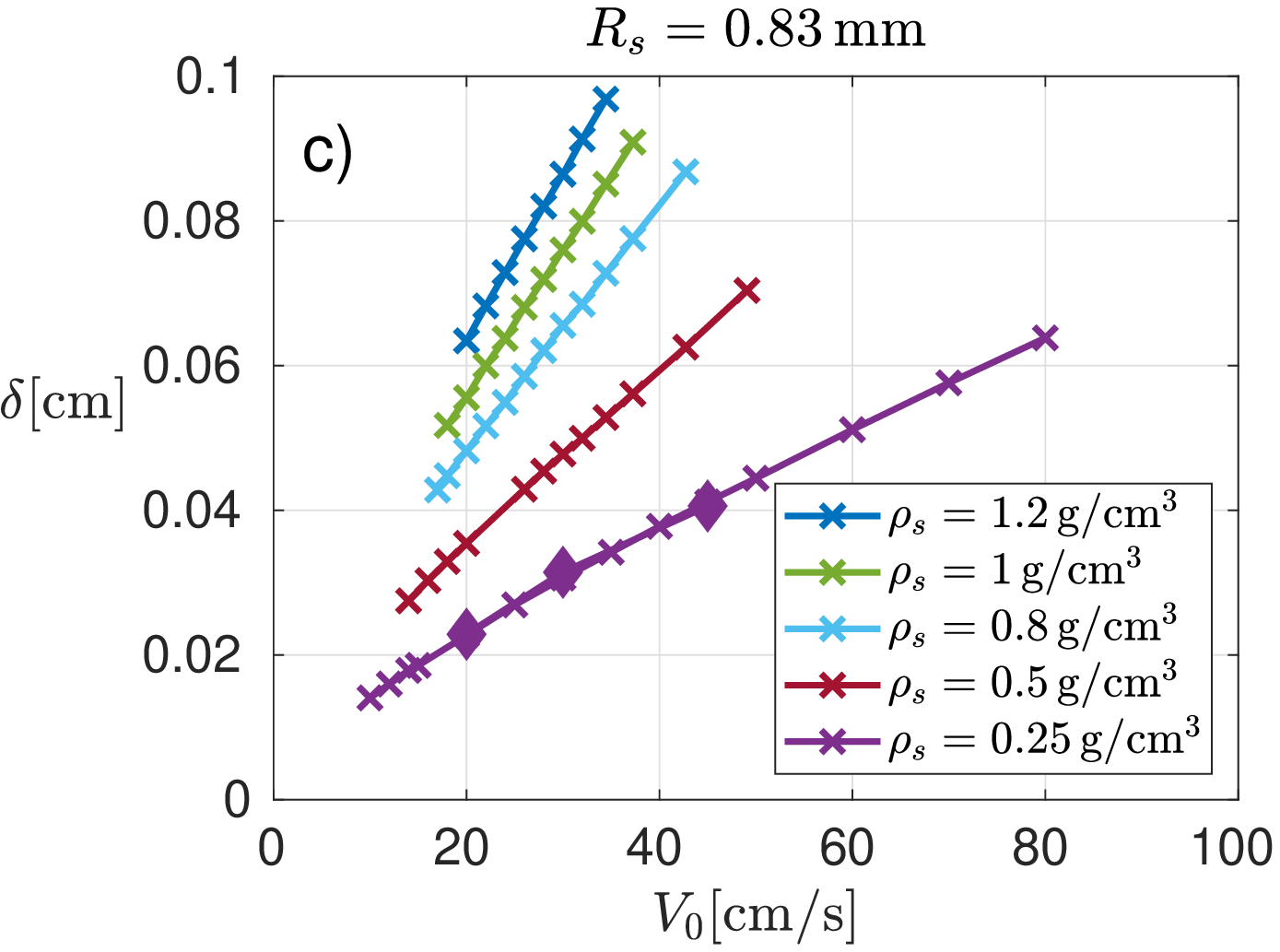} 
     &
     \includegraphics[width = .495\textwidth]{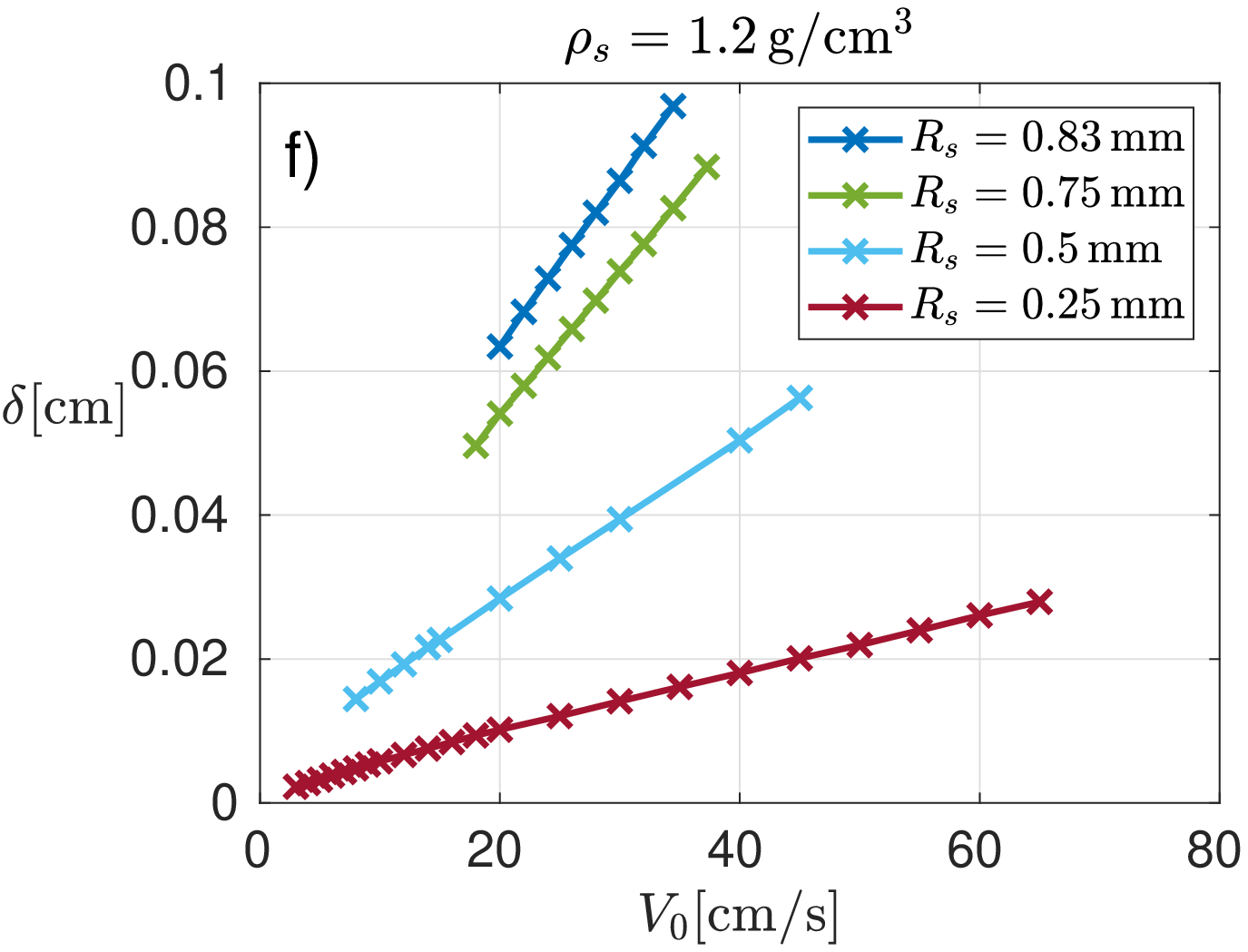}
     \end{tabular}
    \caption{Rebound metrics for weak impacts. Cross ($+$) markers correspond to KM predictions and diamond ($\blacklozenge$) markers to DNS predictions. In these impacts, different rebound metrics were used. These are the pressing time $t_p$, defined as the length of the time interval over which the south pole of the sphere is in direct contact with the fluid surface; and the coefficient of restitution squared $\alpha^2$, which can take negative values when the total energy transfer during the rebound is greater than the kinetic energy of the sphere as it starts its contact with the bath. All relevant notation and parameter values are provided in Table~\ref{tab:parameters}.}
    \label{fig:SixPanelsLinears}
\end{figure}

For this range of physical parameters, it is often the case that the bounce is so weak that the sphere does not recover enough mechanical energy to return to the impact height, thus rendering the definition of $t_c$ useless as a rebound metric. Instead, we define the time between touch-down and lift-off as the pressing time, $t_p$. For these cases, we also need to revisit the rebound metric $\alpha$.

When considering the normal impact of two rigid bodies, if one of the impacting masses reverses its direction following the impact, the standard definition for its coefficient of restitution 
$    \alpha 
    =
    -U_{\text{out}}/U_{\text{in}}
$ (i.e. minus the ratio of the outgoing velocity to the incoming velocity) can also be expressed as the square root of the ratio of its outgoing and incoming kinetic energies 
$
    \alpha 
    = 
    - U_{\text{out}}/U_{\text{in}}
    =
    \sqrt{
    E^k_{\text{out}}
    /
    E^k_{\text{in}
    }
    }.
$ If the impact takes place at the reference level for potential energy, this is also the ratio of their total mechanical energies (kinetic plus potential)
\begin{equation}
    \alpha 
    = 
    - \frac{U_{\text{out}}}{U_{\text{in}}}
    =
    \sqrt{
    \frac{
    E^k_{\text{out}}
    }{
    E^k_{\text{in}}
    }
    }
    =
    \sqrt{
    \frac{
    E^m_{\text{out}}
    }{
    E^m_{\text{in}}
    }
    }.
\end{equation}

This multiplicity of interpretations is possible when the impact is localised in time and space. In that scenario, external forces are unable to perform any work or exert any impulse on either of the impacting bodies. This is not the case in the impacts we study. As the free surface is allowed to deform, gravity does work and exerts an impulse on the sphere (in particular) over the duration of contact.

Variable $\alpha$ was in fact chosen to be the square root of the ratio of the outgoing mechanical energy to the incoming mechanical energy rather than minus the ratio of velocities at the start and end of contact, which (in general) take place at different heights. In the case when the sphere returns to the impact height, this is simply minus the ratio of the outgoing to the incoming velocity at the reference height (neglecting any losses from the moment the sphere lifts off to the moment when it crosses the reference level for the second time). However, near the lower limit of impact velocities the sphere transfers more than its initial kinetic energy to the bath. That is to say, it transfers all of the kinetic energy it had before impact plus some of its gravitational potential energy as it pushes down on the fluid. In these cases, though the sphere still reverses its direction of motion and detaches, it no longer reaches the impact height, i.e. $E^m_{\text{out}}$ is negative, thus turning $\alpha$ imaginary. To avoid introducing imaginary coefficients of restitution, we use $\alpha^2$ as our rebound metric near this regime, with the understanding that a negative value for $\alpha^2$ corresponds to the impactor losing more than its initial kinetic energy over the impact.

Despite $\alpha^2$ being a more general metric, we kept $\alpha$ as the parameter of choice for the other regimes, since in the study of impacts it is much more customary to consider the coefficient of restitution than its square.

The results of these low Weber number simulaitons are presented in figure \ref{fig:SixPanelsLinears}, where we identify behaviour that is qualitatively different from what was observed in the intermediate Weber number cases. We recall that in all cases shown above (see figure \ref{fig:SixPanels}), the $\alpha$ curve was always monotonic, whilst in the regime here considered, for a given sphere radius, it is possible to find a low enough density so as to produce a maximum in the coefficient of restitution $\alpha$ (or equivalently in $\alpha^2$). Similarly, for a given material density, we find a radius that is sufficiently small, so as to produce a non-monotonic curve for $\alpha$.

To the best of our knowledge, this is the first instance of a report of such behaviour for rebounding impactors on the free surface of a fluid. In order to independently verify these findings, we ran some selected cases in the DNS simulations. The results are presented in the three-point curve signalled with diamond ($\blacklozenge$) markers along with the kinematic match results. As can be seen in figure \ref{fig:SixPanelsLinears}, our DNS simulations verify the KM predictions. 

\subsection{Quasi-static approximation}\label{section:asymptotics}
\begin{figure}
    \centering
    \includegraphics[width = .5\textwidth]{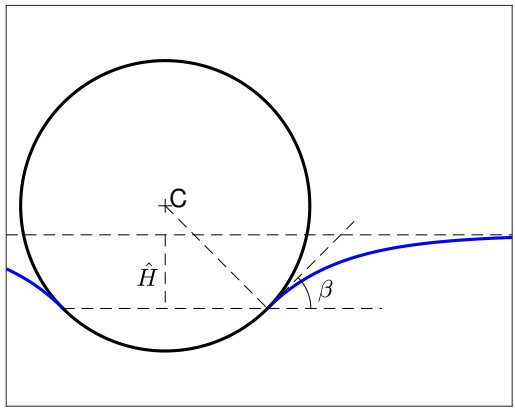}
    \caption{Schematic diagram for the quasi-static analysis. Point $C$ corresponds to the centre of the sphere, $\hat{H}$ is the depth of the boundary of the contact line, and $\beta$ is the angle formed between the horizontal and the free surface, where it meets the solid.}
    \label{fig:scheme}
\end{figure}

We use asymptotic analysis based on \cite{james1974meniscus} to derive a spring model which is able to collapse the curves for maximum penetration and contact time. A similar analysis has also been presented in \citet{Cooray2017}. Consider a sphere resting on the free surface of a quiescent bath. Buoyancy and surface tension effects result in a net vertical force given by
\begin{equation}\label{eqn:qs_balance}
    F_{\text{z}}
    =
    \underbrace{\rho g \pi R_s^2 \sin^2(\beta)\hat{H}+\frac{\rho g \pi R_s^3}{3}(2-3\cos(\beta)+\cos^3(\beta))}_{F_B}
    +
    \underbrace{2\pi\sigma R_s\sin^2(\beta)}_{F_T},
\end{equation}
where $\beta$ is the angle that the free surface makes with the horizontal direction at the boundary of the pressed area and $\hat{H}$ is the distance from the undisturbed free surface to the boundary of the pressed area (see figure \ref{fig:scheme}). The buoyancy force, $F_B$, is given by weight of the volume of fluid above the spherical cap that is in contact with the free-surface and the capillary force, $F_T$, is given by the vertical component of the surface tension acting along the contact line of the same spherical cap.

Taking $2\pi\sigma R_s$ as the unit of force and $R_s$ as the unit length, non-dimensionalising equation (\ref{eqn:qs_balance}) yields
\begin{equation}
    F 
    =
    \frac{F_{z}}{2\pi\sigma R_s} 
    =
    \sin^2(\beta)
    +
    \Bo
    \left[
    \frac{\sin^2(\beta)}{2}
    H
    +
    \frac{2-3\cos(\beta)+\cos^3(\beta)}{6}
    \right],
\end{equation}
where  $\Bo = \rho g R_s^2/\sigma$ is the Bond number and $H = \hat{H}/R_s$.

 We now consider the Young-Laplace equation for this set up,
 \begin{equation}\label{eqn:YL}
     \frac{1}{r}
     \partial_r
     \left[     \frac{r\partial_{r}\eta}{(1+\left(\partial_r\eta\right)^2)^{1/2}}
     \right]
     =
     \Bo\eta,
 \end{equation}
 subject to the boundary conditions
\begin{equation}
    \partial_r\eta_o(\sin(\beta)) 
    =
    \tan(\beta), \qquad
    \eta_o(\sin(\beta)) = -H,
\end{equation}
where $H$ is to be determined. We perform a boundary layer analysis in the limit of $\Bo \ll 1$. The region where curvature and surface deflection are $O(1)$ is the ``outer'' region (i.e. the boundary layer is at infinity), and the equation is approximated by neglecting the right-hand side of (\ref{eqn:YL}).

It follows that
\begin{equation}
    \eta_o(r)
    =
     \sin^2(\beta)
     \ln
     \frac{
    \left|r+\sqrt{r^2-\sin^4(\beta)}\right|
    }
    {
    \sin(\beta)\left(1+\cos(\beta)\right)
    }
    -H.
\end{equation}

For the ``inner'' solution, re-scaling $x = \sqrt{\Bo} \,r$, with $x=O(1)$, yields
\begin{equation}
    x\partial_{xx}\eta_i(x) + \partial_x \eta_i(x) - x\eta_i(x) 
    =
    0,
\end{equation}
subject to 
\begin{equation}
    \lim\limits_{x\to\infty}\eta_i(x) 
    =
    0;
\end{equation}
which implies
\begin{equation}
    \eta_i(r) = cK_0\left(\sqrt{\Bo}\,r\right),
\end{equation}
where $K_0$ is the modified Bessel function of the second kind and order $0$, and $c$ is an arbitrary constant.

\begin{figure}
    \centering
    \begin{tabular}{cc}
    \includegraphics[width = .49\textwidth]{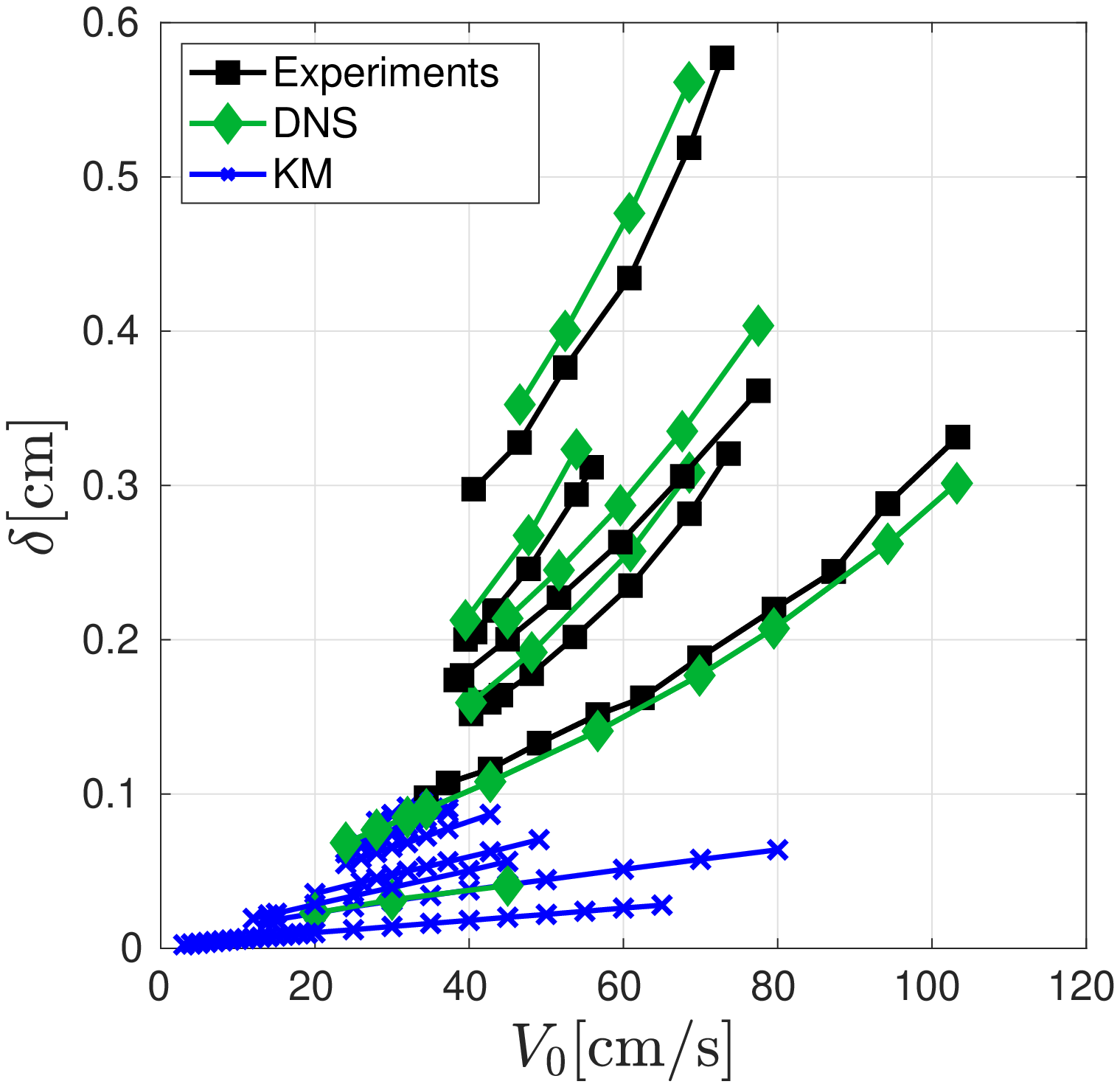} 
    &
    \includegraphics[width = .49\textwidth]{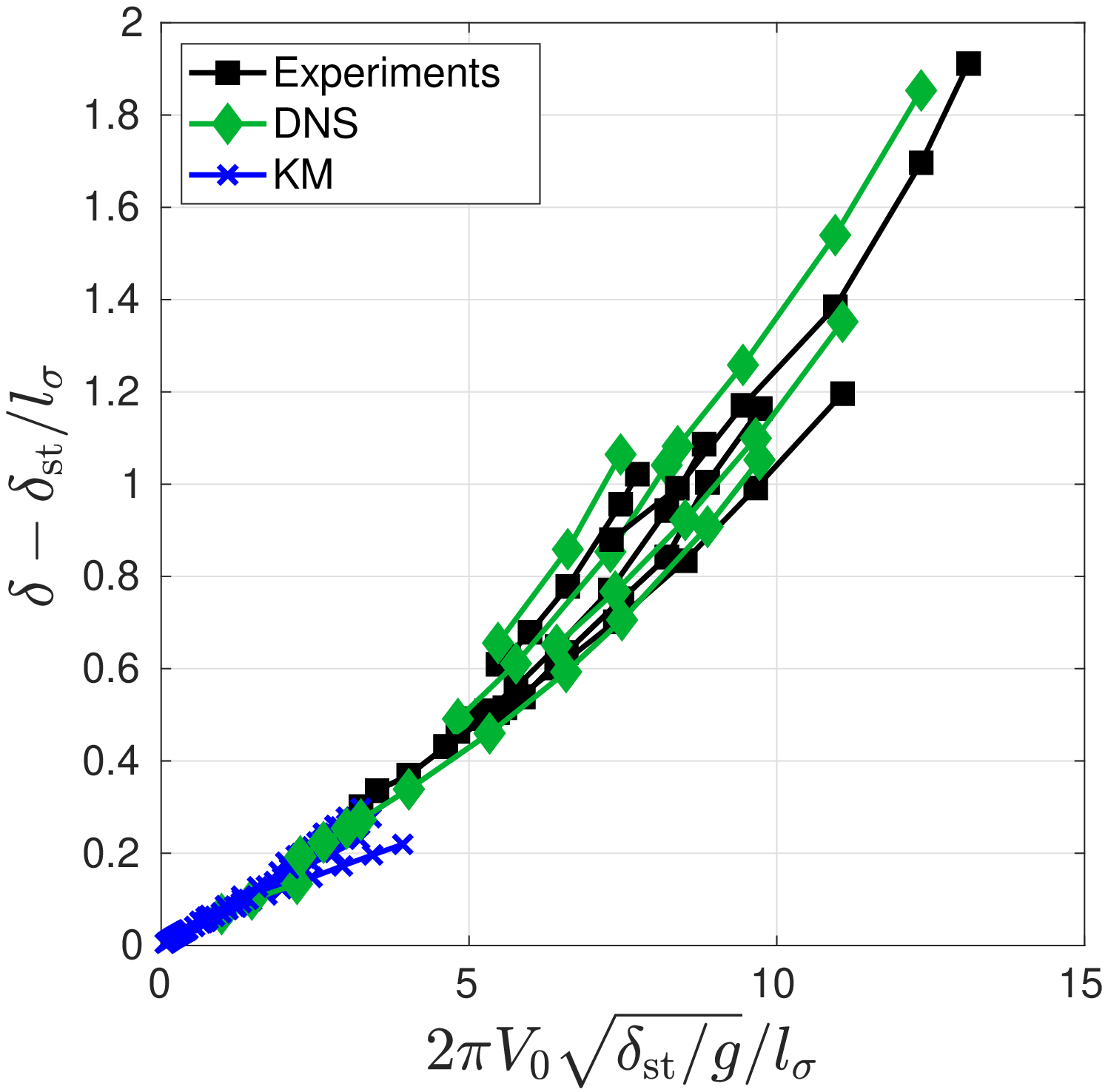}
    \end{tabular}
    \caption{Collapse of the maximum surface deflection on the basis of the non-linear spring model at the equilibrium deflection. Left panel: the full set of experimental and simulated data for which the sphere returns to the impact height. Right panel: the same data re-scaled using the variable suggested by the boundary layer analysis. The vertical axis on the right panel is normalised using the capillary length, $l_\sigma = \sqrt{\sigma/\rho g}$.}
    \label{fig:collapse_delta}
\end{figure}

In order to match the inner and outer solutions, we must consider the form of the inner solution for small $x$,
\begin{equation}
    \eta_i(r)
    \sim
    -c
    \left(
    \ln\left(r\right)
    +\ln\left(\frac{\sqrt{\Bo}}{2}\right)+\gamma\right),
\end{equation}
where $\gamma$ is the Euler-Mascheroni constant and the form of the outer solution for large values of $r$,
\begin{equation}
    \eta_o(r)
    \sim
    \sin^2(\beta)
    \left(\ln(r)
    +\ln(2)
    -
    \ln\left(\sin(\beta)\left(1+\cos(\beta)\right)\right)
    \right)
    -H.
\end{equation}
Thus we have $c = -\sin^2(\beta)$ and
\begin{equation} \label{eqn:H}
    H = 
    \sin^2(\beta)
    \ln
    \frac{4}{\sqrt{\Bo}e^\gamma\sin(\beta)\left(1+\cos(\beta)\right)}.
\end{equation}

In static equilibrium $F_{z}$ is equal to the mass of the sphere, hence $\frac{F_{z}}{2\pi\sigma R_s} = \frac{2}{3} \Dr \Bo$, and a small $\Bo$ and $\beta$ solution can be found with $\Bo \sim \beta^2$ (at leading order). Thus, we expect that this will continue to hold at small $We$, with
\begin{equation} \label{eqn:H}
    H 
    \sim
    \beta^2
    \ln
    \frac{
    2
    }
    {
    \sqrt{\Bo}e^\gamma\beta
    },
\end{equation}
and
\begin{equation}\label{eqn:F}
    F
    =
    \beta^2
    +
    \Bo\left(\frac{\beta^2}{2}H\right)
    +
    O(\beta^4).
\end{equation}

From equations (\ref{eqn:F}) and (\ref{eqn:H}), we have an approximate nonlinear ``interface spring'' stiffness given by the expression
\begin{equation}
    k 
    =
    \frac{F}{H} 
    \approx
    \left(
    \ln
    \frac{
    2
    }
    {
    e^\gamma
    \sqrt{\Bo F} 
    }
    \right)^{-1}.
\end{equation}

This spring model is now used to estimate the static deflection of the free surface due to the weight of the sphere (by taking $F$ in the argument of $\ln(\cdot)$ to be given by the aforementioned dimensionless weight of the sphere $F= 2\Dr\Bo/3$, yielding
\begin{equation}
    k_{\text{st}} 
    \approx
    \left(
    \ln
    \frac{
    \sqrt{6}
    }
    {
     e^\gamma
     \Bo
    \sqrt{\Dr}
    }
    \right)^{-1},
\end{equation}
and therefore
\begin{equation}
    \delta_{\text{st}} 
    \approx
    \frac{
    2\Bos_s 
    }
    {
    3}
    \left(
    \ln
    \frac{
    \sqrt{6}
    }
    {
     e^\gamma
     \Bo
    \sqrt{\Dr}
    }
    \right),
\end{equation}
where $k_{\text{st}}$ and $\delta_{\text{st}}$ are the stiffness and the deformation of the nonlinear spring at static equilibrium, respectively.
\begin{figure}
    \centering
    \begin{tabular}{cc}
    \includegraphics[width = .49\textwidth]{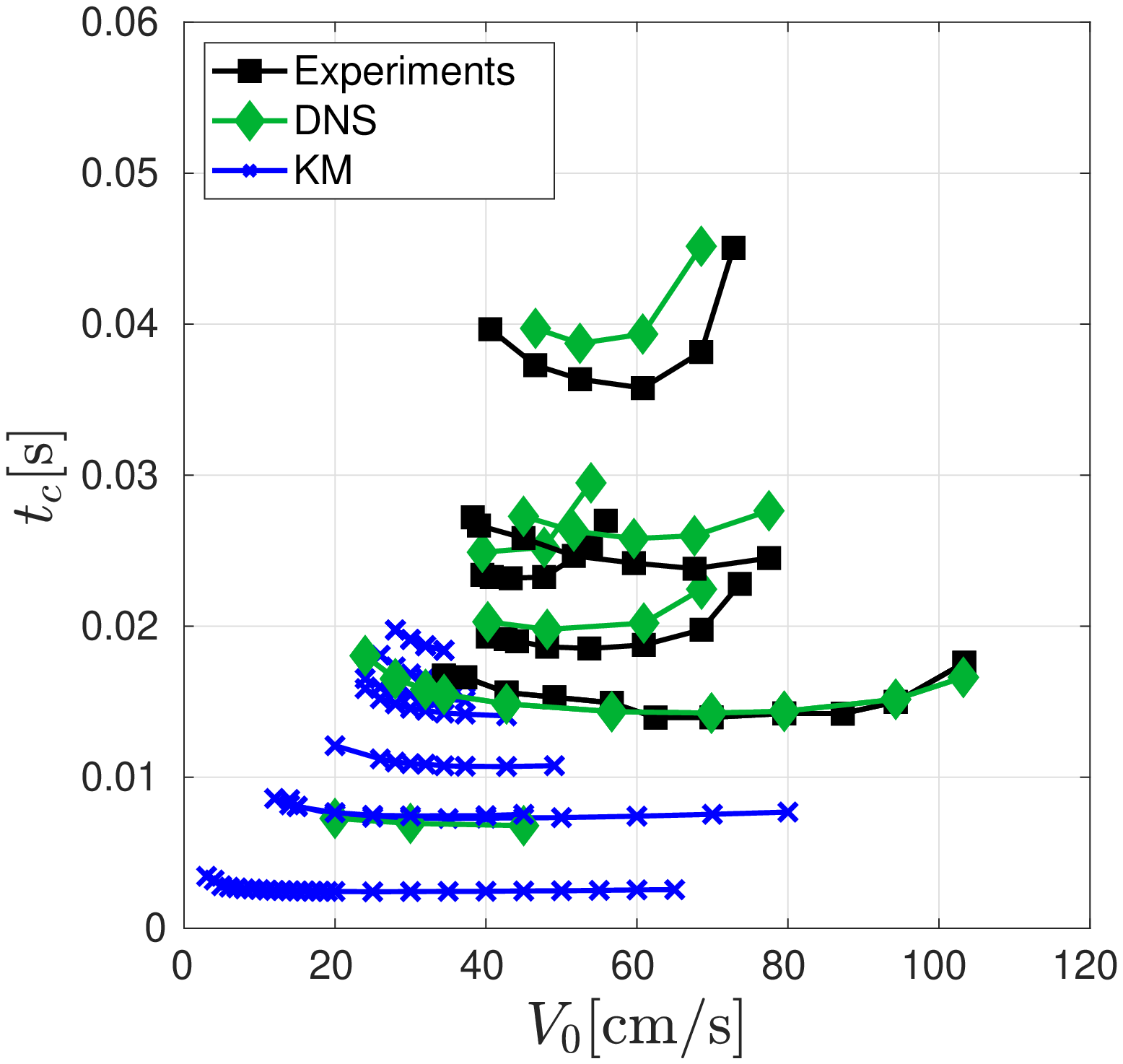} 
    &
    \includegraphics[width = .48\textwidth]{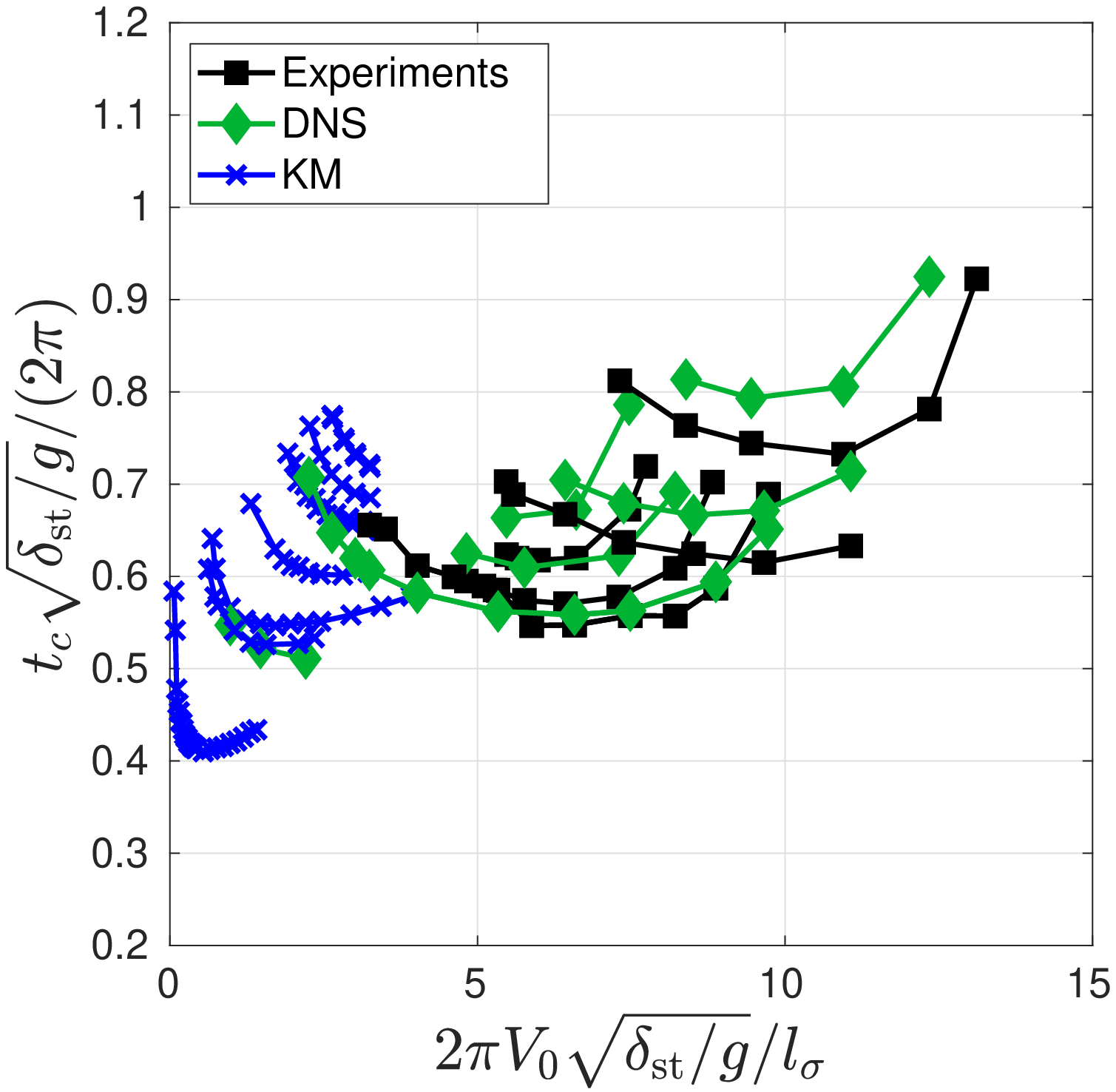}
    \end{tabular}
    \caption{Collapse of the contact time on the basis of the period of oscillation of the spring model. Left panel shows the full set of experimental and simulated data for which the sphere returns to the impact height. Right panel shows the same data re-scaled using the variables suggested by the boundary layer analysis.}
    \label{fig:collapse_tc}
\end{figure}

Figure \ref{fig:collapse_delta} shows the maximum surface deflection from all experimental, DNS and linearised fluid interface model data. The vertical axis measures maximum deflection with respect to the static deflection as estimated by the non-linear spring derived above. On the horizontal axis, velocity is given in units of capillary length over spring period. 

Figure \ref{fig:collapse_tc}, shows the contact time for all data, as a function of the period of oscillation of the spring. The clustering of the data around $0.6$ suggests that contact time can be interpreted as approximately half a period of oscillation of the spring. This is physically reasonable, as contact can be considered to occur during the negative-deflection part of an oscillation period. 

Figures \ref{fig:collapse_delta} and \ref{fig:collapse_tc} include all of our experimental and simulation points, with the exception of the simulation point for which the impactor never reaches the reference height following impact, as it is not possible to define $t_c$ for these points. 

Despite the assumptions, the collapse of the data is reasonable and suggests that the quasi-static asymptotic analysis and ``interface spring'' interpretation captures much of the dominant physics of the rebound.  This simple model however does not collapse the coefficient of restitution data.  This is not unexpected, as the asymptotic model does not include the {\it dynamic} effect of energy being transferred to the surface waves on the bath, which depend much more intricately on the physical parameters.

\section{Discussion}
The present work addresses a regime of impacts onto a free surface that had not hitherto received much attention, and reveals trends for the dependence of the contact time, the fraction of energy recovered by the impactor, and the maximum surface deformation. Moreover, carefully controlled experiments and modelling derived from first principles allow for the identification of new phenomena. Direct numerical simulations provide new insight into the dynamics and flow quantities that are difficult to measure. Moreover, the DNS also supply information and act as a validation test bed for the reduced-order model in appropriate regions of the parameter space that are challenging to investigate experimentally, thus acting as a bridge between the employed methods. Asymptotic analysis is used to derive a nonlinear spring approximation and provides a framework for the collapse and physical interpretation of data derived from all methods.

 Our experimental study spanned the range from intermediate to high impact velocities; namely from  impact velocities that cause the droplet to barely rise past the undisturbed free-surface height to the highest speeds for which the sphere bounces (higher velocities cause the sphere to sink). Moreover, the peculiar phenomenon of the ``resurrecting" sphere was uncovered in the experiments and captured by our DNS. Furthermore, robust trends in the contact time, coefficient of restitution, and penetration depth were established and compared directly with direct numerical simulations, and with the linearised model in the appropriate regime.

Direct numerical simulations are able to span the full range of experiments and reproduce the observed trends in contact time, coefficient of restitution and maximum surface deflection, and even capture the existence of a narrow parameter window where the new phenomenon of resurrection takes place. Our simulations also produce consistent predictions of the trajectory of the sphere throughout the range we study, allowing for the validation of KM results outside the experimental range.
Furthermore, DNS allow us to interrogate flow quantities of interest such as interfacial shapes, pressure and velocity field components in all flow phases, down to a scale of $\mathcal{O}(1)\ \mu$m. These will be reported in a subsequent publication. 

A linearised fluid model is used to efficiently explore the low Weber number limit. Since DNS are also used to explore the low velocity end of the bouncing regime, these provide a source of data for validation of the linearised model. Indeed, this model and the DNS coincide remarkably well when the linearity assumption holds. 
Despite its limitations to deal with higher Weber numbers, the linearised model remains useful, since it brings the obvious advantages of a much lower computational cost and relative simplicity. In particular, given that small spheres cause shorter (and therefore faster) capillary waves, their simulation becomes particularly costly when using DNS, as they require that the boundary of the numerical domain be far enough away to guarantee that waves are not being reflected and returning to influence the rebound. The linearised model is simpler and less computationally costly than the DNS; however, it remains far from trivial and there are a significant number of applications that could benefit from a further reduced mass-spring-damper model to predict rebounds on the free surface. For a given sphere radius and density, such a simplified model can be readily synthesised from the curves for contact time, coefficient of restitution and maximum penetration depth that are produced by the application of the kinematic match method to the linearised fluid model presented here, the code for which is made available as supplementary material. Furthermore, the kinematic match strategy \citep{GaleanoRiosEtAl2019} is not limited to a linearised free-surface model, nor to a fluid interface. Hence a similar study for impacts without linearising the free surface of the fluid, or impacts on flexible membranes and other deformable surfaces can be considered on the basis of the same modelling principles.

Agreement between the results of the linearised model and the DNS also reveals that flow in the air layer is unlikely to be a dominant element for rebound dynamics in the low Weber number regimes. Moreover, the pressure profiles that are predicted by the kinematic match method are in agreement with the existing literature \citep{HendrixEtAl2016}. Furthermore, \citet{HendrixEtAl2016} report that the maximum in pressure coincides with the annular region where the air layer is thinnest. Our air-free model thus provides a clear indication that this minimum in the width of the air layer is likely a consequence, rather than a cause, of the profile of the pressure distribution.

Exploring the weak impact end of the rebounding regime revealed that, for light enough spheres (in particular, lighter than the fluid), the dependence of the coefficient of restitution on impact velocity can be qualitatively different from what is seen for denser spheres. Specifically, the dependence of the coefficient of restitution as a function of the Weber number can have a {\it local} maximum in the interior of the Weber number spectrum, as opposed to at one end of it. Likewise, for a given sphere density (even if heavier than the fluid) we were able to observe a local maximum in the coefficient of restitution for a sufficiently small sphere. The latter observation is particularly interesting in light of the biological and bio-mimetic importance of surface impacts. If, for a given density and radius, there is an optimal velocity at which to impact the free surface so as to recover maximum energy, it is possible for some water-walking insect or mechanism to benefit from it.

The converse problem of a droplet impactor rebounding off a solid surface has been considered in several previous works, e.g. \citet{AndersEtAl1993, ClanetEtAl2004}. The dependence of the coefficient of restitution on the Weber number has previously been reported, e.g. \citet{BianceEtAl2006,AussillousAndQuere2006,GiletAndBush2012}. The trend observed in these studies is similar to what is found for low $\Dr$ and low $\Bo$ (see figure \ref{fig:SixPanelsLinears}), wherein the higher end of the $\We$ spectrum corresponds to a decrease in the coefficient of restitution with increasing $\We$. Our general results have greater similarities with the investigation of \citet{BianceEtAl2006}, in which they clearly found a growing coefficient of restitution (as a function of $\We$) for the low $\We$ regime, and a decreasing trend for higher $\We$. In our work, as we gradually increase $\We$ beyond the rebound threshold, we always find an increasing coefficient of restitution. This trend is sustained until we observe sinking of the sphere or, for low $\Dr$ and $\Bo$, reversal to a decreasing behaviour.

We have found that the regime diagram reported in figure 7 of \citet{LeeAndKim2008} does not capture the behaviour of the simulations considered here. In particular, our experiments and simulations consistently indicate that the scaling for the bouncing threshold reported in the respective study is unlikely to provide a collapse. \citet{LeeAndKim2008} propose that, for a given density ratio $\Dr$, the minimum $\We$ for bouncing increases as the $\Bo$ decreases. The opposite relation is found in our work.

Our boundary layer analysis provides a nonlinear spring model, which yields a framework for the collapse and interpretation of the maximum penetration depth and contact time data from the three methods. Moreover, a collapse based on a linear spring model was attempted but resulted in very limited success. This is, to some extent, in contrast with what was found in similar systems, for example those of droplets bouncing on a fluid trampoline \citep{GiletAndBush2009JFM}, and it indicates that the interaction of the impactor with the underlying flow adds significant complexity to this problem.

It is worth mentioning that other non-linear spring models which have been successfully used in similar (though not identical) contexts are available in the literature. In particular, we highlight the model presented in \citet{GiletAndBush2009PRL} and \citet{GiletAndBush2009JFM}. It is also quasi-static; however, it differs from ours in that there is no fluid bulk underneath the interface, hence the model in question does not need to account for the effect of hydro-static pressure. Moreover, the presence of a trampoline rim in the works of Gilet and Bush, impose a different set of boundary conditions for their resulting Young-Laplace equation. Other similar models include the work of \citet{MolacekBush2012,MolacekAndBush2013a} and \citet{TerwagneEtAl2013}. These studies present spring models derived from energy principles and include the storage of energy in the deformation of the impactor as a key element in the dynamics.

Our work combines experiments, DNS, linearised free-surface models and asymptotics to span the full range of the topic at hand. 
We use each of these approaches within 
their respective ranges of validity
and cross-compare 
the results where they overlap. 
This articulation of different methods allowed us to uncover the general trends in rebound metrics, collapse the curves for contact time and penetration depth, efficiently explore the low Weber number regime with the appropriate metrics, and identify the new phenomenon of ``resurrecting'' spheres.

\vspace{5mm}
C.A.G.-R. and P.A.M. gratefully acknowledge the support of EPSRC project EP/N018176/1.  D.M.H. acknowledges the financial support of the Brown OVPR Seed Award and the UTRA Undergraduate Research program, the preliminary experimental work by undergraduate John Edmonds (UNC-CH), and laboratory space generously loaned by K. Breuer which allowed the authors to rapidly establish the experimental component of the project at Brown. R.C. is grateful for the resources and continued support of the Imperial College London Research Computing Service. All authors would like to thank the referees for their constructive suggestions.

Declaration of interests. The authors report no conflict of interest.

\appendix
\section{Trajectories}\label{app:trajects}

\begin{figure}
    \centering
    \begin{tabular}{cc}
     \includegraphics[width = .39\textwidth]{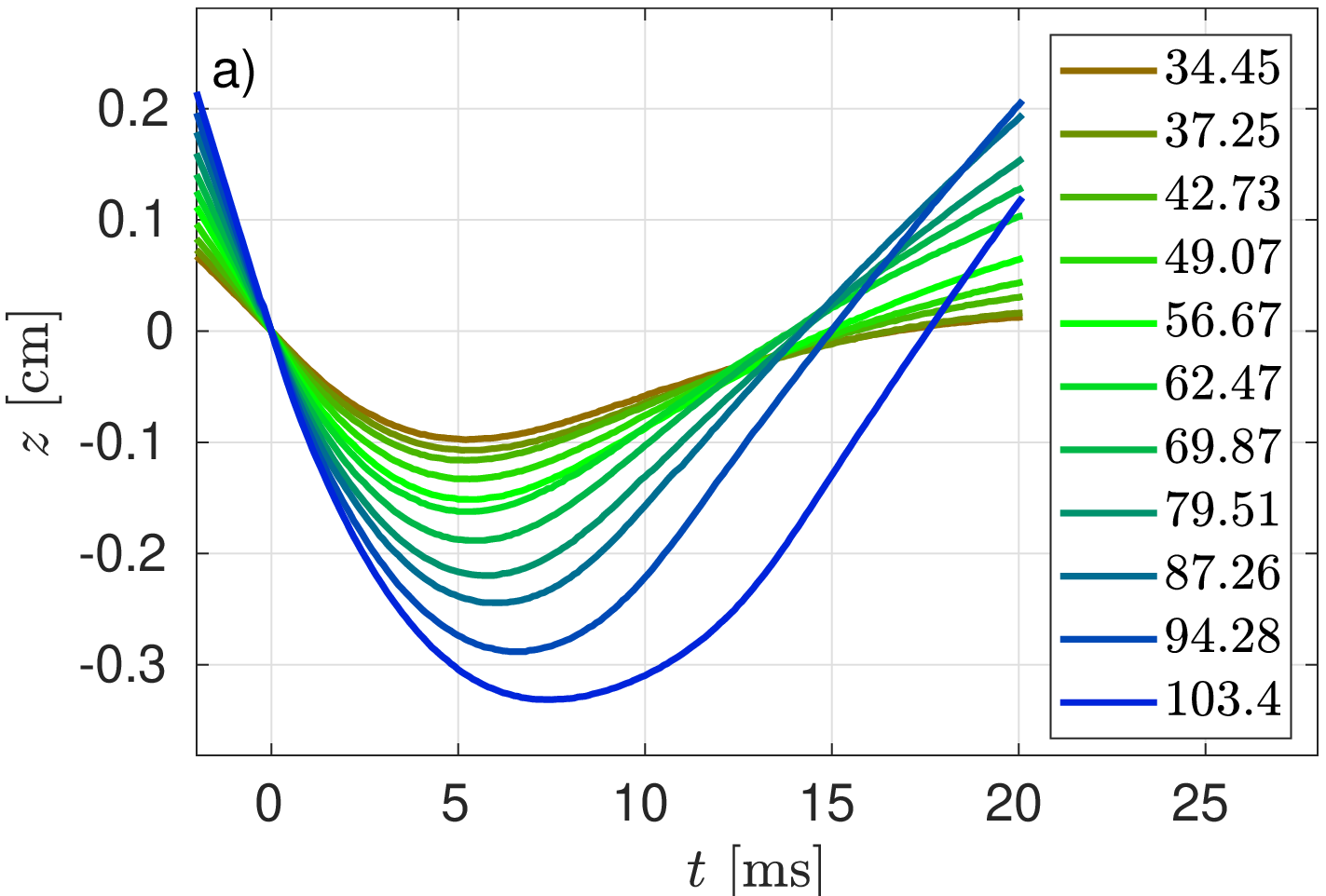}
     &
     \includegraphics[width = .39\textwidth]{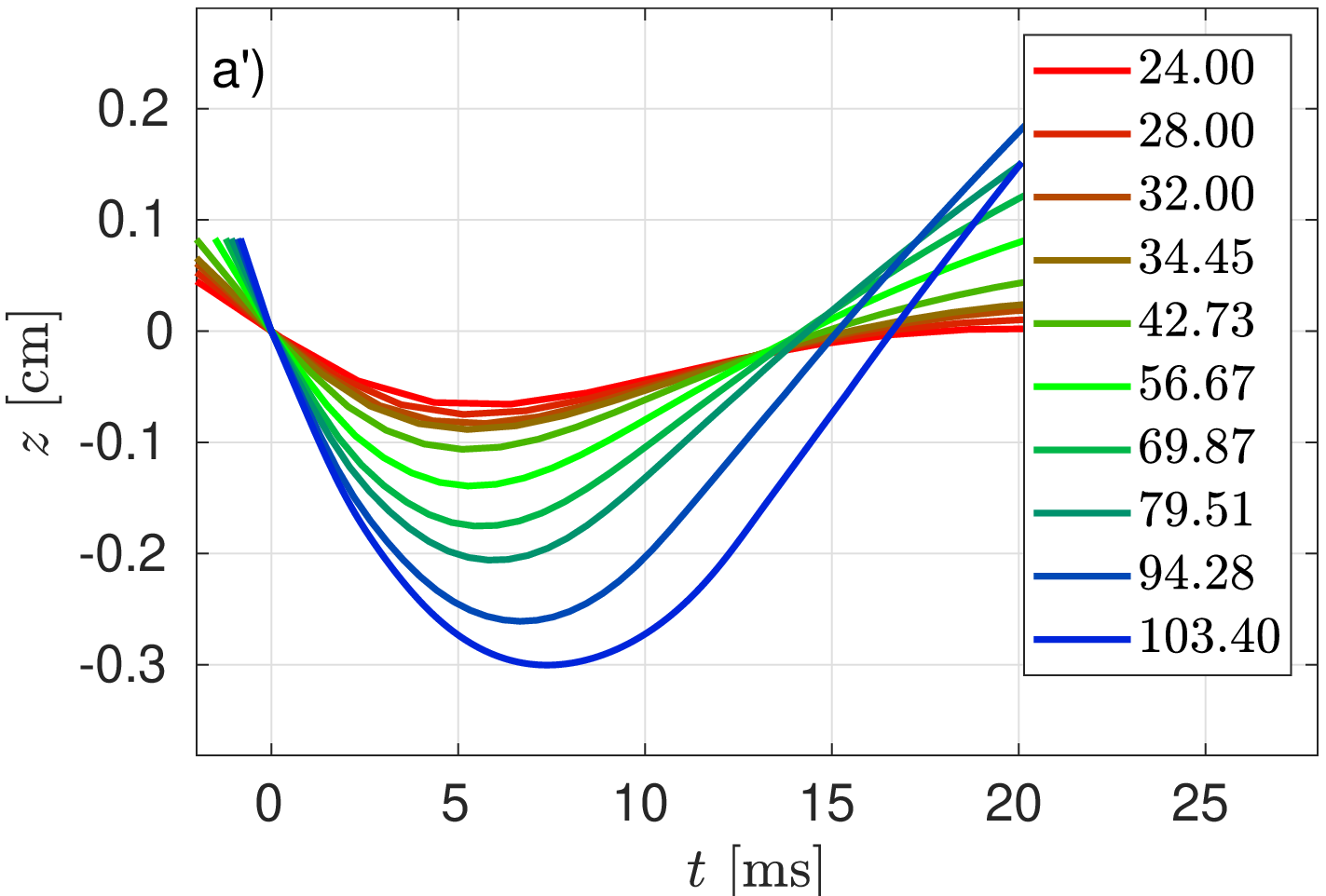}
     \\
     \includegraphics[width = .39\textwidth]{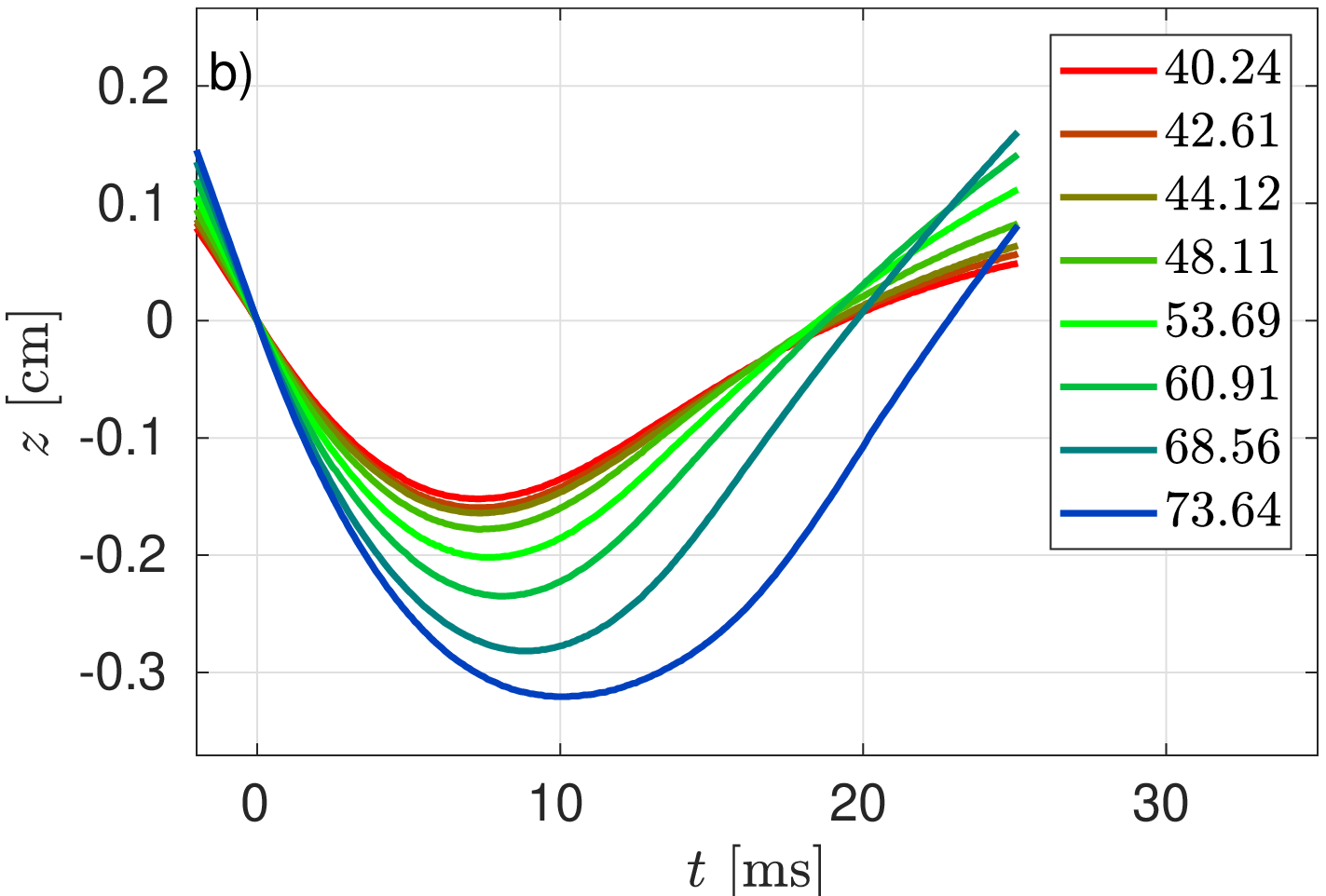}
     &
     \includegraphics[width = .39\textwidth]{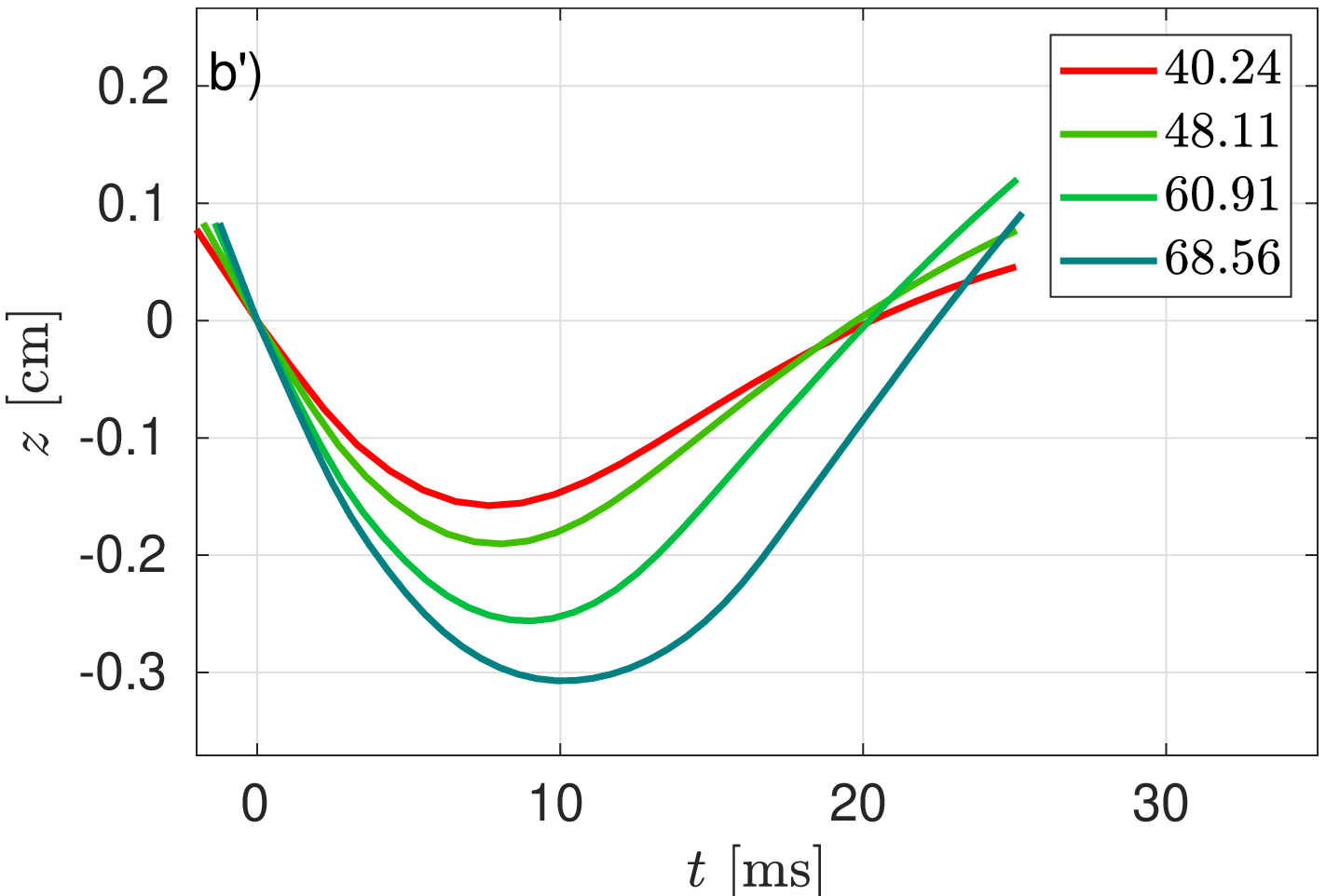}
     \\
     \includegraphics[width = .39\textwidth]{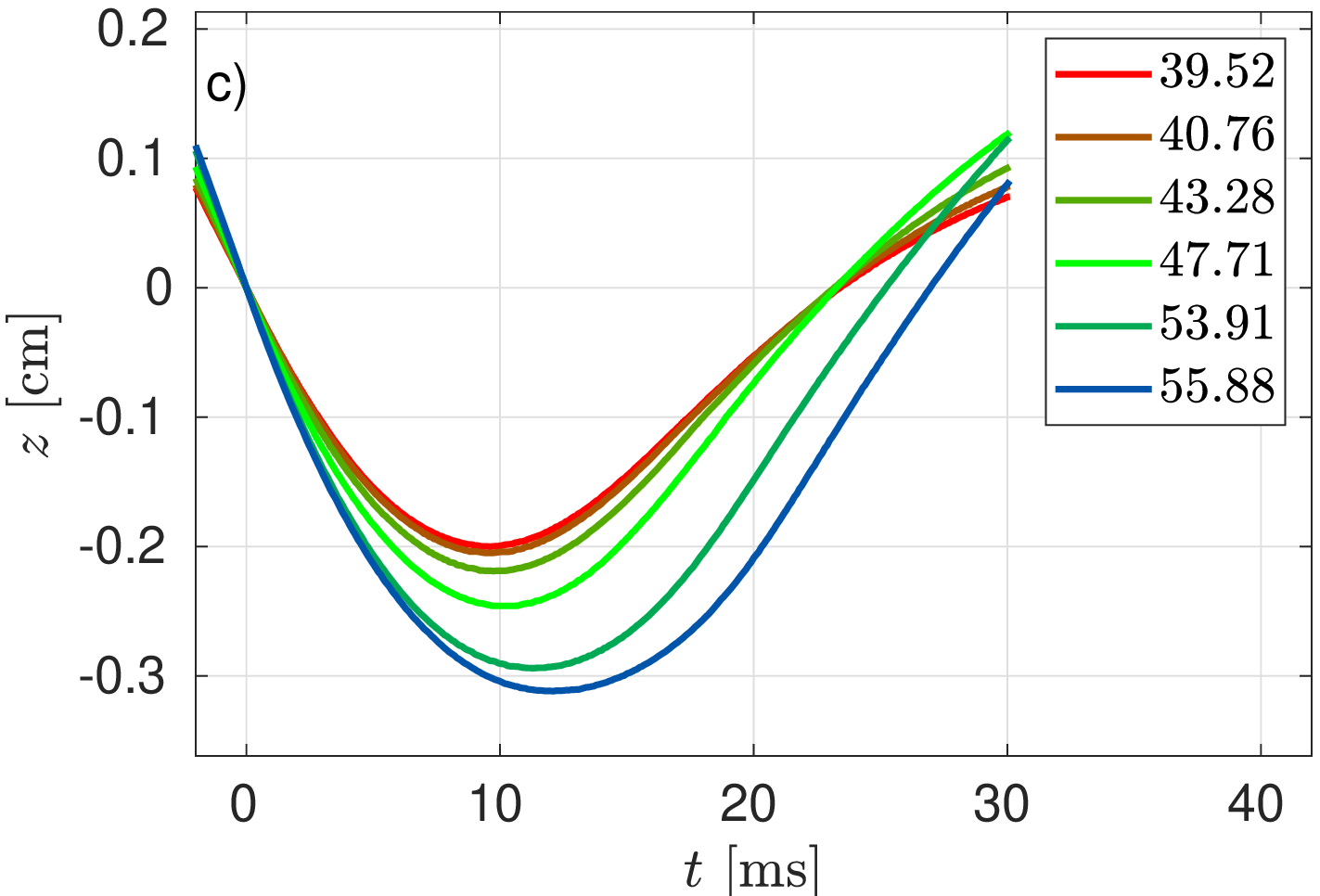} 
     &
     \includegraphics[width = .39\textwidth]{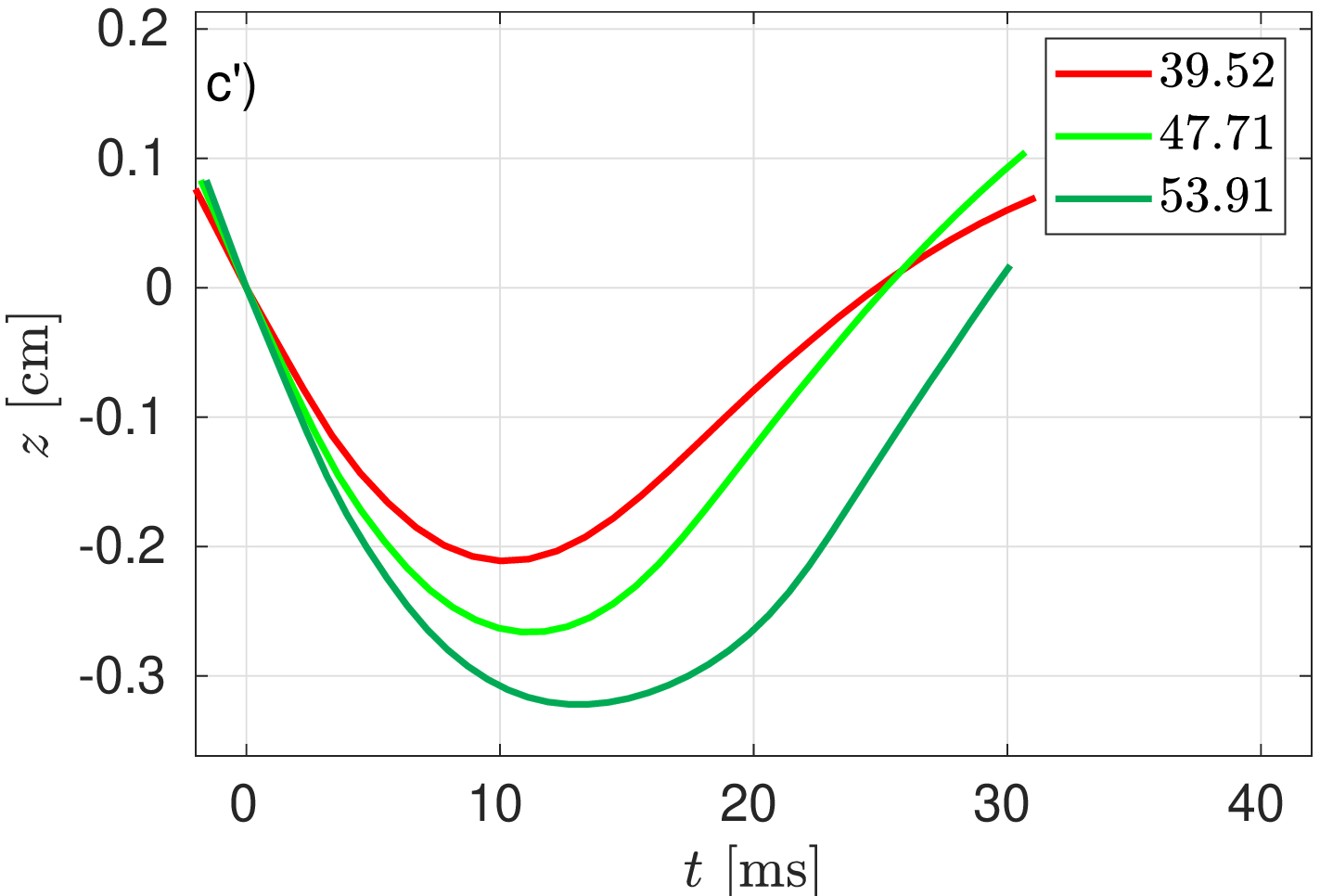}
     \\
     \includegraphics[width = .39\textwidth]{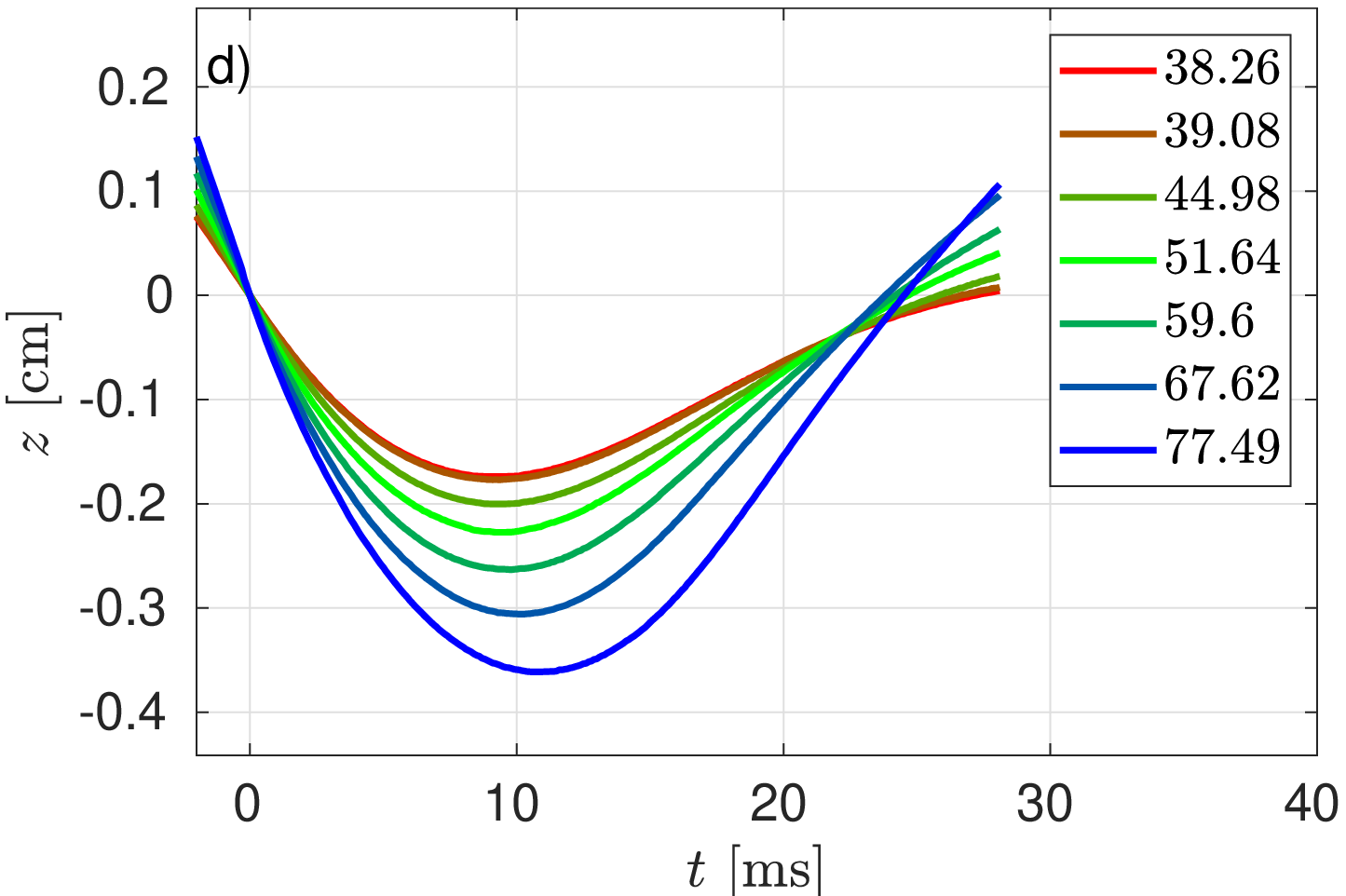} 
     &
     \includegraphics[width = .39\textwidth]{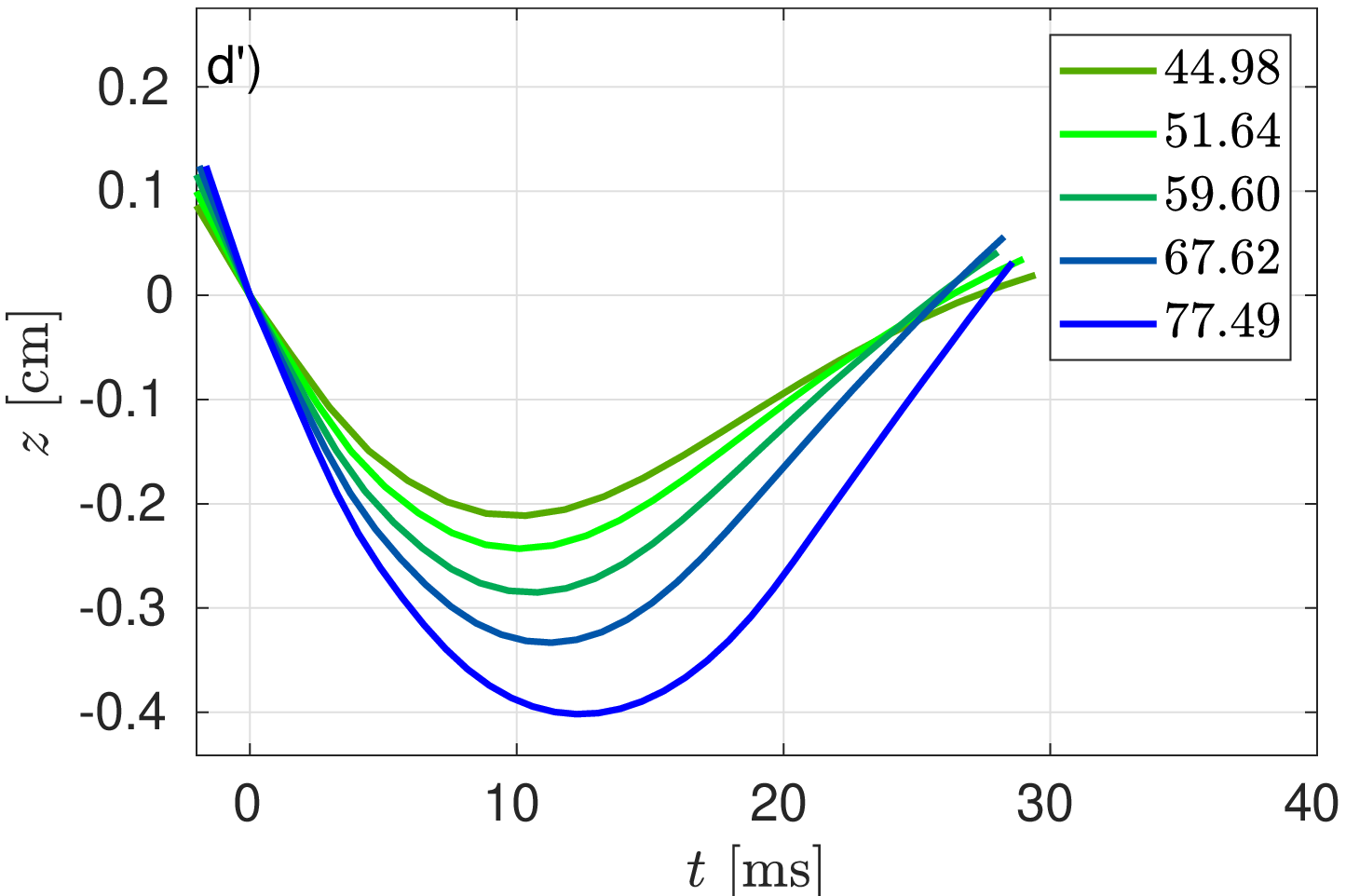}
     \\
     \includegraphics[width = .39\textwidth]{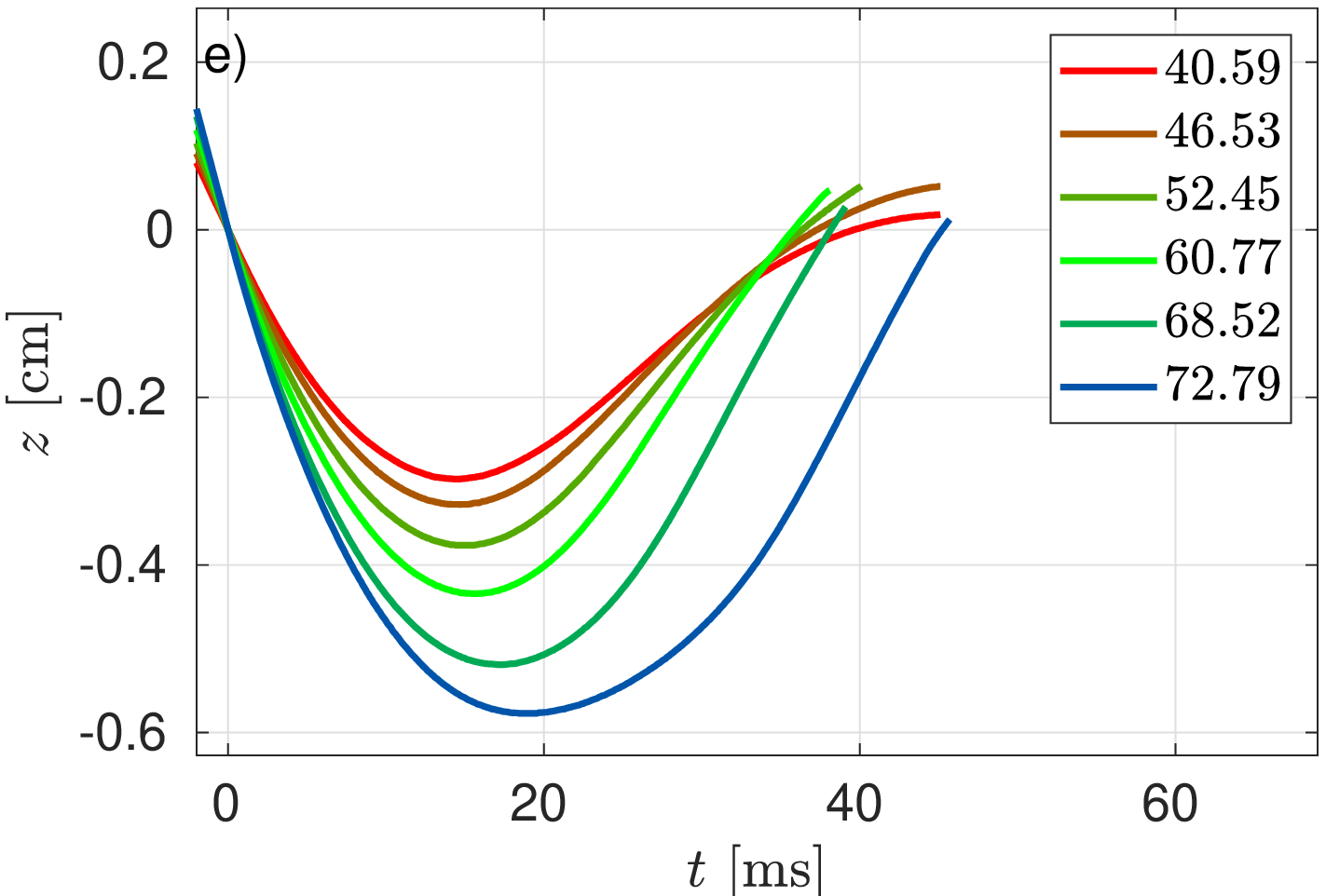} 
     &
     \includegraphics[width = .39\textwidth]{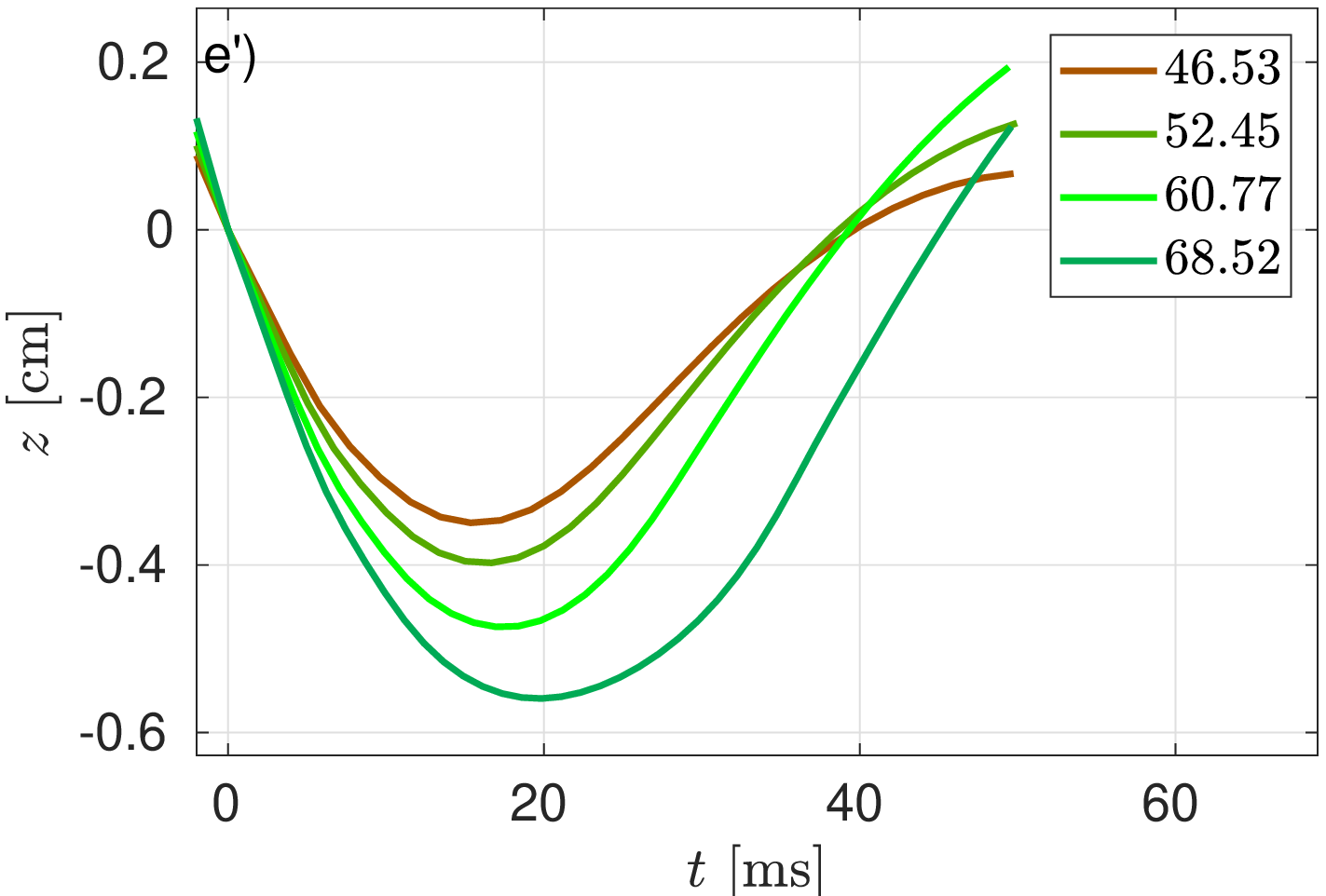}
     \end{tabular}
    \caption{Average south pole trajectories for each sphere in experiments (a-e) and DNS south pole trajectories for the corresponding pseudo-spheres (a'-e'). Trajectories are colour coded by impact speed as indicated in the legends (in cm$/$s). (a-a') $\rho_s = 1.2\,$g$/$cm$^3$, $R_s = 0.83\,$mm, (b-b') $\rho_s = 2.2\,$g$/$cm$^3$, $R_s = 0.83\,$mm,  (c-c') $\rho_s = 3.2\,$g$/$cm$^3$, $R_s = 0.83\,$mm, (d-d') $\rho_s = 1.2\,$g$/$cm$^3$, $R_s = 1.24\,$mm, (e-e') $\rho_s = 1.2\,$g$/$cm$^3$, $R_s = 1.64\,$mm.}
    \label{fig:trajects} 
\end{figure}

We present the average south pole trajectories for each set of physical parameters used in the experiments. Panels \ref{fig:trajects}.a-\ref{fig:trajects}.e illustrate these datasets, each panel corresponding to one hydrophobic sphere used. The corresponding panels \ref{fig:trajects}.a'-\ref{fig:trajects}.e' show the south pole trajectories obtained using DNS for corresponding modelled pseudo-solid spheres within the same parameter regimes. We highlight that the choice of DNS cases was not intended to represent a one-to-one map of the experiments; instead we aimed to cover a similar range of impact velocities in order to verify the trends in rebounds metrics, as presented in figure \ref{fig:SixPanels}.

In experiments, the cut-off for impact velocities at the low end corresponds to rebounds for which the sphere does not return to the initial impact height, and at the high end to sinking of the sphere. In the DNS, the lower end cut off was ignored for one sphere, in order to provide some more trajectories for validation of the KM method used on the linearised free-surface model. These three trajectories correspond to the lowest velocities in panel \ref{fig:trajects}a'. 

DNS results also accurately predict the cut-off at the high end of impact velocities. At times, the pseudo-spheres sink and coalesce slightly below the maximum velocity for sinking of the sphere in the experiments. Indeed, panels \ref{fig:trajects}b', \ref{fig:trajects}c' and \ref{fig:trajects}e' lack the trajectory for the highest impact velocity precisely because the value used in the experiments caused the pseudo-sphere to sink and coalesce by falling just slightly short of recovering and bouncing back. 

\bibliographystyle{jfm}
\bibliography{BiblioCGR}

\end{document}